\documentclass[acmsmall, screen]{acmart}

\renewcommand\footnotetextcopyrightpermission[1]{}
\usepackage{booktabs}
\usepackage{subcaption}

\usepackage[outline]{contour}

\usepackage{kotex}
\usepackage{fancyvrb}
\usepackage{nameref}
\usepackage{hyperref}
\usepackage{xspace}
\usepackage{wrapfig}
\usepackage{lipsum}
\usepackage{fancyvrb}
\usepackage{balance}
\usepackage{mathpartir}
\usepackage{stmaryrd}
\usepackage{cprotect}
\usepackage{pbox}
\usepackage[normalem]{ulem}
\usepackage{dashbox}
\usepackage{soul}
\usepackage{xcolor}
\usepackage{multirow}
\usepackage{framed}
\usepackage{enumitem}
\usepackage{outlines}
\usepackage[super]{nth}
\usepackage[draft]{fixme}
\usepackage{xstring}
\usepackage[all]{xy}

\usepackage[]{inputenc}
\usepackage[T1]{fontenc}
\usepackage[color]{coqdoc}
\usepackage{amsmath}
\usepackage{url}
\usepackage{kotex}
\usepackage{cancel}

\newbool{showcomments}
\boolfalse{showcomments}
\newbool{showoldtext}
\boolfalse{showoldtext}

\usepackage{amsthm}
\usepackage{bm}
\usepackage{arydshln}
\usepackage{cleveref}
  \Crefformat{chapter}{\S#2#1#3}
  \Crefformat{section}{\S#2#1#3}
  \Crefformat{figure}{Fig. #2#1#3}
  \Crefformat{appendix}{\S#2#1#3}
  \Crefformat{axiom}{Axiom #2#1#3}
\usepackage{tikz-cd}
\usepackage{setspace}
\usetikzlibrary{matrix, arrows.meta, positioning, calc, decorations.pathreplacing}
\usepackage{url}
\usepackage{multicol}
\usepackage{seqsplit}

\setcopyright{none}

\begin{document}
\newtheorem*{remark}{Remark}
\newtheorem{axiom}{Axiom}

\newcommand{\ie}{\textit{i.e.,} }
\newcommand{\cf}{\textit{cf.} }
\newcommand{\eg}{\textit{e.g.,} }
\newcommand{\etal}{\textit{et al.}}

\newcommand{\todo}[1]{{\textcolor{purple!80!black}{\textbf{TODO: #1}}}}

\newcommand{\code}[1]{\texttt{#1}\xspace}

\newcommand{\originaltxt}[1]{}
\newcommand{\hide}[1]{}
\newcommand{\unhide}[1]{#1}

\newcommand{\kor}[1]{}

\newcommand{\myparagraph}[1]{\paragraph{\hspace*{-3.8mm}\bfseries{#1}}}
\definecolor{grey}{rgb}{0.5,0.5,0.5}
\newcommand{\caselabel}[1]{\begin{small}{\color{grey}{(\textsc{#1})}}\end{small}}
\newcommand{\mycomment}[1]{\begin{small}{\vspace{1.5mm}\color{grey}{(*****\;#1\;*****)}}\end{small}}

\newcommand{\defeq}{\ensuremath{\triangleq}}

\newcommand{\option}{\code{option}}
\newcommand{\some}{\code{Some}}
\newcommand{\none}{\code{None}}

\newcommand{\setofz}[1]{\{#1\}}
\newcommand{\setof}[1]{\{\, #1 \,\}}
\newcommand{\suchthat}{\,|\,}
\newcommand{\metaspec}[1]{\overline{#1}}

\newcommand{\unitset}{()}

\newcommand{\bfparagraph}[1]{\paragraph{\textbf{#1}}}

\newcommand{\ang}[1]{\langle{#1}\rangle}
\newcommand{\Type}[0]{\mathbf{Type}}
\newcommand{\Prop}[0]{\mathbf{Prop}}
\newcommand{\cons}[0]{\mathtt{scons}}
\newcommand{\head}[0]{\mathtt{shead}}
\newcommand{\tail}[0]{\mathtt{stail}}
\newcommand{\node}[0]{\mathtt{node}}
\newcommand{\val}[0]{\mathtt{val}}
\newcommand{\sub}[0]{\mathtt{sub}}
\newcommand{\natS}[0]{\mathtt{S}}
\newcommand{\stream}[0]{\mathtt{stream}}
\newcommand{\streamP}[0]{\mathtt{streamR}}
\newcommand{\streamC}[0]{\mathtt{streamC}}
\newcommand{\streamR}[0]{\mathtt{streamR}}
\newcommand{\bintree}[0]{\mathtt{bintree}}
\newcommand{\bintreeP}[0]{\mathtt{bintreeR}}
\newcommand{\bintreeC}[0]{\mathtt{bintreeC}}
\newcommand{\bhead}[0]{\mathtt{bhead}}
\newcommand{\bleft}[0]{\mathtt{bleft}}
\newcommand{\bright}[0]{\mathtt{bright}}
\newcommand{\bcons}[0]{\mathtt{bcons}}
\newcommand{\lang}[0]{\mathtt{lang}}
\newcommand{\langP}[0]{\mathtt{langR}}
\newcommand{\langC}[0]{\mathtt{langC}}
\newcommand{\lcons}[0]{\mathtt{lcons}}
\newcommand{\llhead}[0]{\mathtt{lhead}}
\newcommand{\ltail}[0]{\mathtt{ltail}}
\newcommand{\covector}[0]{\mathtt{covector}}
\newcommand{\covectorP}[0]{\mathtt{covectorR}}
\newcommand{\covectorC}[0]{\mathtt{covectorC}}
\newcommand{\bunfold}[0]{\mathtt{bunfold}}
\newcommand{\bfold}[0]{\mathtt{bfold}}
\newcommand{\cvhead}[0]{\mathtt{cvhead}}
\newcommand{\cvtail}[0]{\mathtt{cvtail}}
\newcommand{\cvcons}[0]{\mathtt{cvcons}}
\newcommand{\cvnil}[0]{\mathtt{cvnil}}
\newcommand{\smap}[0]{\mathtt{smap}}
\newcommand{\smapS}[0]{\mathtt{smapS}}
\newcommand{\rmap}[0]{\mathtt{rmap}}
\newcommand{\lmap}[0]{\mathtt{lmap}}
\newcommand{\self}[0]{\mathtt{self}}
\newcommand{\ra}[0]{\!\rightarrow{}\!}
\newcommand{\Ra}[0]{\Rightarrow{}}
\newcommand{\la}[0]{\leftarrow{}}
\newcommand{\La}[0]{\Leftarrow{}}
\newcommand{\lra}[0]{\leftrightarrow{}}
\newcommand{\rla}[0]{\leftrightarrow{}}
\newcommand{\Lra}[0]{\Leftrightarrow{}}
\newcommand{\Rla}[0]{\Leftrightarrow{}}
\newcommand{\nat}[0]{\mathbb{N}}
\newcommand{\hnat}[0]{\mathbb{N}^{\infty}}
\newcommand{\natsucc}[0]{\texttt{succ}}
\newcommand{\natpred}[0]{\texttt{pred}}

\newcommand{\depfun}[2]{(#1) \ra #2}
\newcommand{\deppair}[2]{(#1) \times #2}
\newcommand{\id}[0]{\mathbftt{id}}
\newcommand{\fst}[0]{\mathbftt{fst}} 
\newcommand{\snd}[0]{\mathbftt{snd}} 
\newcommand{\inl}[0]{\mathbftt{inl}}
\newcommand{\inr}[0]{\mathbftt{inr}}

\newcommand{\Container}{\mathrm{Container}}
\newcommand{\vctn}{\mathcal{D}}
\newcommand{\fctnS}{\mathtt{shp}}
\newcommand{\fctnD}{\mathtt{deg}}
\newcommand{\vspf}{\mathcal{F}}
\newcommand{\fspf}{\mathtt{F}}
\newcommand{\fspm}{\mathtt{map}}
\newcommand{\fspr}{\mathtt{rel}}
\newcommand{\fsprone}{\mathtt{rel1}}
\newcommand{\fspe}{\mathtt{emb}}
\newcommand{\fspsem}[1]{\widehat{#1}}

\newcommand{\SPF}{\mathrm{SPF}}

\newcommand{\SPE}{\mathrm{SPE}}
\newcommand{\SPEI}{\mathrm{SPEI}}
\newcommand{\vspe}{\mathcal{E}}
\newcommand{\fold}{\mathbftt{fold}}
\newcommand{\unfold}{\mathbftt{unfold}}
\newcommand{\scofix}[1]{\overline{#1}}
\newcommand{\gop}[1]{~\hat{#1}~}
\newcommand{\gdepfun}[2]{(#1) \gop{\ra} #2}
\newcommand{\gdeppair}[2]{(#1) \gop{\times} #2}
\newcommand{\cop}[1]{~\tilde{#1}~}
\newcommand{\cdepfun}[2]{(#1) \cop{\ra} #2}
\newcommand{\cdeppair}[2]{(#1) \cop{\times} #2}
\newcommand{\spfcompose}[0]{\underline{\circ}}
\newcommand{\cttcompose}{\gop{\circ}}
\newcommand{\sttcompose}{\cop{\circ}}

\newcommand{\sem}[1]{{\llbracket #1 \rrbracket}}
\newcommand{\semf}[1]{{\llbracket #1 \rrbracket}_\mathtt{map}}
\newcommand{\semr}[1]{{\llbracket #1 \rrbracket}_\mathtt{rel}}

\newcommand{\powset}[1]{\mathbb{P}(#1)}

\newcommand{\equptoc}[1]{=_{#1}}
\newcommand{\equpto}[1]{\equiv_{#1}}
\newcommand{\Nat}{\mathbb{N}}
\newcommand{\Integer}{\mathbb{Z}}
\newcommand{\level}{\mathtt{level}}
\newcommand{\leveleachtxt}{\mathtt{level\_each}}
\newcommand{\leveleach}[1]{\widebar{#1}}
\newcommand{\levelA}[1]{A_{#1}}
\newcommand{\levelfun}[1]{{{#1}_{\shortuparrow}}}
\newcommand{\productivity}[2]{\powset{\level(#1)\times\level(#2)}}
\newcommand{\stdunif}{\mathscr{u}}
\newcommand{\stdunifZ}{\stdunif_\Integer}
\newcommand{\stdcase}{\mathscr{c}}
\newcommand{\allproductivity}[0]{\mathscr{U}}
\newcommand{\hlevel}{\mathtt{level}_{h}}

\newcommand{\CTT}{\mathrm{CTT}}
\newcommand{\STT}{\mathrm{STT}}
\newcommand{\vctt}{\mathcal{C}}
\newcommand{\vstt}{\mathcal{S}}
\newcommand{\cttf}{\mathcal{F}}
\newcommand{\ctta}{\texttt{A}}

\newcommand{\FamT}{\mathrm{FamT}}
\newcommand{\vfamt}{\vec{\vtcon}}
\newcommand{\FamC}{\mathrm{Fam}}
\newcommand{\vfamc}{\vec{\vspe}}
\newcommand{\ffamI}{\mathtt{I}}
\newcommand{\ffamE}{\mathtt{E}}
\newcommand{\ffami}{\mathtt{idx}}
\newcommand{\tconop}[1]{{{[}{#1}{]}}}
\newcommand{\twoset}{\setofz{1,2}}

\newcommand{\tuple}[1]{{\langle #1 \rangle}}
\newcommand{\dollar}[1]{{\${#1}}}
\newcommand{\identCTT}[1]{\tuple{#1}}
\newcommand{\constCTT}[1]{\gop{#1}}
\newcommand{\selfCTT}[1]{\tuple{\dollar{#1}}}
\newcommand{\constSTT}[1]{\cop{#1}}
\newcommand{\identSTT}[1]{(\!-_{#1}\!)}

\newcommand{\Bool}{\mathbb{B}}
\newcommand{\ite}[3]{#1 \;{?}\; #2 : #3}
\newcommand{\cbite}[3]{#1 \;{\bm{?}}\; #2 \bm{:} #3}
\newcommand{\btrue}{\texttt{true}}
\newcommand{\bfalse}{\texttt{false}}
\newcommand{\cany}[3]{\bm{[}{#1}\bm{]}^{#2}_{#3}}
\newcommand{\band}{\&\&}
\newcommand{\bor}{||}

\newcommand{\False}{\textrm{False}}

\newcommand{\Empty}{{\emptyset}}

\newcommand{\Unit}{\textrm{Unit}}
\newcommand{\unit}{\texttt{tt}}

\newcommand{\Const}{\textrm{Const}}

\newcommand{\List}{\mathrm{List}}
\newcommand{\Stream}{\mathrm{Stream}}

\newcommand{\Id}{\mathrm{Id}}

\newcommand{\Arity}{\mathrm{Arity}}
\newcommand{\varity}{\mathcal{A}}

\newcommand{\lord}{\le}
\newcommand{\hord}{\le_{h}}
\newcommand{\hpord}{\le_{hp}}

\newcommand{\productive}[1]{\mathfrak{prd}(#1)}
\newcommand{\productiveat}[2]{\mathfrak{prd}(#1,#2)}
\newcommand{\prd}[1]{{#1\textrm{-}}}
\newcommand{\eqpa}[2]{\approxeq_{#1}^{#2}\!}
\newcommand{\soproductive}[1]{\mathfrak{prd}^2 #1}
\newcommand{\soproductiveJ}[1]{\mathfrak{prd}^2_{I} #1}
\newcommand{\soproductivepJ}[1]{\mathfrak{prd}^2_{pI} #1}

\newcommand{\cFun}[2]{\sem{\vctt_#1}\ra\sem{\vctt_#2}}
\newcommand{\cFFun}[4]{(\cFun{#1}{#2})\ra(\cFun{#3}{#4})}
\newcommand{\cFunJ}[3]{\depfun{j\!:\!#1}\sem{\vec{\vctt_#2}(j)}\ra\sem{\vec{\vctt_#3}(j)}}
\newcommand{\cFFunJ}[6]{(\cFunJ{#1}{#3}{#4})\ra(\cFunJ{#2}{#5}{#6})}
\newcommand{\cFFunpJ}[4]{(\cFunJ{J}{#1}{#2})\ra(\cFun{#3}{#4})}

\newcommand{\CTTFunctionType}[0]{\sem{\vctt_1} \ra \sem{\vctt_2}}

\newcommand{\por}[0]{\cup}
\newcommand{\pand}[0]{\cap}
\newcommand{\wfs}{\mathtt{well\!\!-\!\!founded\!\!-\!\!s}}
\newcommand{\asc}{asc}
\newcommand{\acc}{\mathtt{Acc}}
\newcommand{\sacc}{\mathtt{SAcc}}
\newcommand{\infsat}[2]{\mathtt{infsat}_{#1}(#2)}

\newcommand{\mathbftt}[1]{\bm{\mathtt{#1}}}
\newcommand{\const}{\mathbftt{const}}
\newcommand{\fcompose}{\bm{\circ}}
\newcommand{\justcompose}{\circ}
\newcommand{\fpair}{\bm{\otimes}}
\newcommand{\fproduct}{\boxtimes}
\newcommand{\ceapp}{\mathbftt{ceapp}}
\newcommand{\cgeapp}{\mathbftt{cgeapp}}
\newcommand{\cgecurry}{\mathbftt{cgecurry}}
\newcommand{\cecurry}{\mathbftt{cecurry}}
\newcommand{\ceswap}{\mathbftt{ceswap}}
\newcommand{\ceproj}{\mathbftt{ceproj}}
\newcommand{\fcopair}{\bm{\oplus}}
\newcommand{\fcoproduct}{\bm{\boxplus}}
\newcommand{\fmap}{\mathbftt{fmap}}
\newcommand{\convone}{{\mathbftt{conv}}_{1}}
\newcommand{\convtwo}{{\mathbftt{conv}}_{2}}
\newcommand{\helper}{\mathtt{inner}}
\newcommand{\comm}{\mathbftt{comm}}
\newcommand{\assoc}{\mathbftt{assoc}}
\newcommand{\antiassoc}{\mathbftt{antiassoc}}
\newcommand{\cnswap}{\mathbftt{cnswap}}
\newcommand{\cnif}{\mathbftt{cnif}}
\newcommand{\cnproj}{\mathbftt{cnproj}}
\newcommand{\curry}{\mathbftt{curry}}
\newcommand{\uncurry}{\mathbftt{uncurry}}
\newcommand{\cepair}{\mathbftt{cepair}}

\newcommand{\pcompose}{\ang{\circ}}
\newcommand{\ppair}{\ang{,}}
\newcommand{\pproduct}{\ang{\times}}
\newcommand{\pcopair}{\ang{|}}
\newcommand{\pcoproduct}{\ang{+}}

\newcommand{\sfcompose}{\fcompose^2}
\newcommand{\sfindep}{\mathbftt{\lambdabar}}
\newcommand{\sfself}{\mathbftt{sfself}}
\newcommand{\sfpair}{\fpair^2}
\newcommand{\sfproduct}{\fproduct^2}
\newcommand{\sfcopair}{\fcopair^2}
\newcommand{\sfcoproduct}{\fcoproduct^2}
\newcommand{\sfmap}{\mathbftt{sfmap}}
\newcommand{\sfecurry}{\mathbftt{sfecurry}}

\newcommand{\spifcompose}{\sfcompose_{pI}}
\newcommand{\spifindep}{\sfindep_{pI}}
\newcommand{\spifself}{\mathbftt{spifself}}
\newcommand{\spifpair}{\sfpair_{pI}}
\newcommand{\spifproduct}{\sfproduct_{pI}}
\newcommand{\spifcopair}{\sfcopair_{pI}}
\newcommand{\spifcoproduct}{\sfcoproduct_{pI}}
\newcommand{\spifmap}{\mathbftt{spifmap}}

\newcommand{\strcons}{::}
\newcommand{\szero}{\mathtt{s0}}
\newcommand{\sone}{\mathtt{s1}}
\newcommand{\stwo}{\mathtt{s2}}
\newcommand{\sthr}{\mathtt{s3}}
\newcommand{\sfour}{\mathtt{s4}}
\newcommand{\sfive}{\mathtt{s5}}
\newcommand{\foo}{\mathtt{foo}}
\newcommand{\gfun}[1]{{F_{#1}}}
\newcommand{\Fsone}{\gfun{\mathtt{s1}}}
\newcommand{\Fstwo}{\gfun{\mathtt{s2}}}
\newcommand{\Fsthr}{\gfun{\mathtt{s3}}}
\newcommand{\Fsfour}{\gfun{\mathtt{s4}}}
\newcommand{\Fsfive}{\gfun{\mathtt{s5}}}
\newcommand{\Ffoo}{\gfun{\mathtt{foo}}}

\newcommand{\zeros}{\mathtt{zeros}}
\newcommand{\ones}{\mathtt{ones}}
\newcommand{\growing}{\mathtt{growing}}
\newcommand{\pingpong}{\mathtt{pingpong}}
\newcommand{\filter}{\mathtt{filter}}
\newcommand{\listappend}{+\!+}

\newcommand{\bisim}{bisim}
\newcommand{\bisimF}{bisimF}
\newcommand{\fib}{\mathtt{fib}}
\newcommand{\zip}{\mathtt{zip}}
\newcommand{\zipp}{\mathtt{zip}_{+}}
\newcommand{\zipL}{\mathtt{zipL}}
\newcommand{\bfs}{\mathtt{bfs}}
\newcommand{\bfsone}{\mathtt{bfs}_{1}}
\newcommand{\bfstwo}{\mathtt{bfs}_{2}}
\newcommand{\roundrobin}{\mathtt{roundrobin}}
\newcommand{\distribute}{\mathtt{distribute}}
\newcommand{\alternatingsub}{\mathtt{alternatingsub}}
\newcommand{\bmap}{\mathtt{bmap}}
\newcommand{\sbt}{\mathtt{sbt}}
\newcommand{\sbtsuc}{\mathtt{sbtsuc}}
\newcommand{\sbtinv}{\mathtt{sbtinv}}
\newcommand{\lplus}{\mathtt{lplus}}
\newcommand{\append}{\mathtt{append}}

\newcommand{\cancelred}[1]{\color{red}\cancel{\color{black}#1}\color{black}}
\newcommand{\minus}{\text{-}}

\newcommand{\coco}{\texttt{Coco}\xspace}


\title{Coco: Corecursion with Compositional Heterogeneous Productivity}
\subtitle{Extended Version}

\author{Jaewoo Kim}
\orcid{0009-0002-8789-5996}
\affiliation{
  \institution{Seoul National University}
  \country{Korea}
}
\email{jaewoo.kim@sf.snu.ac.kr}

\author{Yeonwoo Nam}
\authornote{The second author, Yeonwoo Nam, made significant contributions comparable to those of the first author.}
\orcid{0009-0004-0837-4573}
\affiliation{
  \institution{Seoul National University}
  \country{Korea}
}
\email{yeonwoo.nam@sf.snu.ac.kr}

\author{Chung-Kil Hur}
\orcid{0000-0002-1656-0913}
\affiliation{
  \institution{Seoul National University}
  \country{Korea}
}
\email{gil.hur@sf.snu.ac.kr}

\setlength{\floatsep}{0.3em}
\setlength{\textfloatsep}{1em}
\setlength{\abovecaptionskip}{2pt}

\begin{abstract}
Contemporary proof assistants impose restrictive syntactic guardedness conditions that reject many valid corecursive definitions.
Existing approaches to overcome these restrictions present a fundamental trade-off between coverage and automation.

We present Compositional Heterogeneous Productivity (CHP), a theoretical framework that unifies high automation with extensive coverage for corecursive definitions.
CHP introduces heterogeneous productivity applicable to functions with diverse domain and codomain types, including non-coinductive types.
Its key innovation is compositionality: the productivity of composite functions is systematically computed from their components,
enabling modular reasoning about complex corecursive patterns.

Building on CHP, we develop Coco, a corecursion library for Rocq that provides extensive automation for productivity computation and fixed-point generation.
\end{abstract}



\makeatletter
\def\@ACM@printacmref{}
\fancyfoot[L]{}
\fancyfoot[C]{}
\fancyfoot[R]{}
\makeatother

\maketitle

\thispagestyle{empty}
\makeatletter
\let\@ACM@printacmref\@empty
\makeatother
\fancypagestyle{firstpagestyle}{
  \fancyhf{}
  \renewcommand{\headrulewidth}{0pt}
  \renewcommand{\footrulewidth}{0pt}
}

\section{Introduction} 
\label{Introduction}

Coinductive datatypes and predicates play an essential role in formal verification,
enabling reasoning about infinite or potentially infinite structures and behaviors. 
While traditional methodologies model infinite behaviors through functions over infinite domains,
coinduction enables more elegant proofs and natural definitions.
This capability is particularly useful for
expressing and reasoning about potentially nonterminating programs.
Notable examples include interaction trees \cite{xia19}, CompCert \cite{leroy09} and
conditional contextual refinement \cite{song23}.

Despite their importance, coinductive datatypes and predicates suffer from
limited or inadequate support in contemporary proof assistants
such as Rocq (formerly known as Coq) \cite{rocqwebsite}, Lean \cite{leanwebsite} and Isabelle/HOL \cite{nipkow02}.
For example, Rocq's underlying kernel, based on the Calculus of
(Co)Inductive Constructions, employs syntactic guardedness checking to
prevent ill-formed cyclic definitions (\ie those that do not have a unique solution),
but this approach has significant
practical limitations. First, it overly restricts valid coinductive
proofs and corecursive definitions, rejecting constructions that are
inherently sound. Second, the syntactic approach lacks
compositionality, making it difficult to build larger terms and proofs
from smaller components.
Finally, the guardedness condition cannot be expressed within the logic itself,
preventing the modular reasoning that would enable more flexible coinductive
programming and verification patterns.

While theoretical and practical innovations \cite{sangiorgi98,hur13,pous16,schafer17} have
resolved the limitations of coinductive reasoning about coinductive predicates,
progress in corecursion over coinductive datatypes has been more limited.
The challenge lies in simultaneously achieving two objectives:
maximizing the \emph{coverage} of sound
corecursive definitions that systems can accept, and maintaining
\emph{automation} of the acceptance process to minimize manual user
intervention.
Currently, two state-of-the-art approaches offer
improvements over traditional syntactic guardedness checking but with
complementary trade-offs between coverage and
automation. Chargu\'eraud's work \cite{chargueraud10} introduces a
fixed point theory that achieves theoretical completeness on
coverage---it can handle any corecursive definition provided that an
appropriate underlying equivalence structure is specified and a
corresponding semantic guardedness condition is proved by the user.
However, this approach's lack of compositionality significantly
hampers automation efforts. Conversely, AmiCo, an Isabelle/HOL tool
for corecursion, leverages the concept of \emph{friends} to achieve
compositionality in guardedness analysis, thereby enabling better
automation support \cite{blanchette17}. Unfortunately, the
friend-based approach has significant coverage limitations that
prevent it from handling various valid corecursive definitions.

\subsection{AmiCo and Chargu\'eraud's approach}

We now examine a series of motivating examples that illustrate the contrasting strengths 
in coverage and automation of the aforementioned approaches.
We first briefly explain coinductive datatypes and corecursive definitions before presenting the examples.

An example of a \emph{coinductive datatype} is $\stream$,
consisting of an infinite stream of natural numbers defined with a
single constructor:
\[
\begin{minipage}{\textwidth}
\begin{coqdoccode}
\coqdocnoindent
\coqdockw{CoInductive} \coqdef{x.stream}{stream}{\coqdocinductive{stream}} := \coqdef{x.scons}{scons}{\coqdocconstructor{scons}} (\coqdef{x.hd:3}{hd}{\coqdocbinder{hd}}: \coqexternalref{nat}{http://coq.inria.fr/doc/V8.20.1/stdlib//Coq.Init.Datatypes}{\coqdocinductive{nat}}) (\coqdef{x.tl:4}{tl}{\coqdocbinder{tl}}: \coqref{x.stream:1}{\coqdocinductive{stream}}).\coqdoceol
\end{coqdoccode}  
\end{minipage}
\]
It is equipped with constructor $\cons : \nat \ra \stream \ra \stream$ and
destructors $\head : \stream \ra \nat$ and $\tail : \stream \ra \stream$.
More generally, coinductive datatypes include trees with infinite
branches and possibly infinitely-branching nodes. Their definitions
may be mutually recursive with other coinductive datatypes or with
inductive datatypes (known as mixed inductive-coinductive
datatypes).

A \emph{corecursive definition} defines an infinite value in a coinductive
type through a cyclic definition.
For example, a stream $\szero$ is cyclically defined by
$\szero = \cons \ 0 \ \szero$, which yields an infinite stream of $0$s.
Similarly, the corecursive function $\smapS : \stream \to \stream$ is cyclically defined by
$\smapS \ (\cons \ x \ s) = \cons \ (S \ x) \ (\smapS \ s)$
, which adds $1$ to every element of an input stream.
More generally, corecursive functions may be mutually defined
with recursive functions (known as mixed recursion-corecursion).
Note that not every cyclic definition is well-formed:
for instance, $\szero = \szero$ is ill-formed because it does not uniquely
define a stream.

\bfparagraph{Problem with Rocq}
Consider the following corecursive definition for stream $\sone$:
$$\sone = \cons \ 0 \ (\smapS \ \sone)$$
This definition is inherently well-defined, as demonstrated by repeated substitution,
writing $\strcons$ for $\cons$ for brevity:
\[\begin{array}{@{}r@{\ }l@{}}
    \sone =& 0 \strcons \smapS \ \sone = 0 \strcons \smapS \ (0 \strcons \smapS \ \sone) = 0 \strcons 1 \strcons \smapS^2 \ \sone \\ 
    =& 0 \strcons 1 \strcons \smapS^2 (0 \strcons 1 \strcons \smapS^2 \ \sone) = 0 \strcons 1 \strcons 2 \strcons \smapS^3 \ \sone = \cdots
  \end{array}\]
The resulting stream $\sone$ satisfies $(\sone)_i = i$ for $i \geq 0$.
Although this definition is semantically well-guarded (\ie
producing $n+1$ elements by consuming $n$ elements), Rocq's kernel
rejects it because the syntactic guardedness checker cannot analyze
the guardedness properties of $\smapS$.

\bfparagraph{Solutions of AmiCo and Chargu\'eraud's approach}
We now examine how AmiCo and the Chargu\'eraud's approach handle this example.

In AmiCo, registering a function as a \emph{friend} is key to
accepting nontrivial corecursive definitions. Intuitively, a friend is
a function that produces $n$ elements in its output whenever it
consumes $n$ elements from its input. For example, $\smapS$ qualifies as a
friend, but $\tail$ does not. Friend functions preserve
guardedness because they do not consume the guard provided by a
constructor, enabling AmiCo to freely permit friend functions within
corecursive definitions---a flexibility not afforded by Rocq's kernel. In
this example, $\smapS$ can be registered as a friend,
which AmiCo accepts automatically via a syntactic analysis. Once registered, the
definition of $\sone$ is allowed since it remains effectively
guarded even with the friend function present.

Chargu\'eraud's approach relies on the mathematical notion of
\emph{contractivity} to capture a semantic notion of guardedness.
A corecursively-defined element of set $X$
is characterized as the unique fixed point of a generating function $F: X \ra X$.
The existence and uniqueness of this fixed point are
guaranteed when the generating function is proved contractive with
respect to an equivalence structure on $X$.  In
the above example, $\sone$ is the unique fixed point of $\Fsone$
(\ie the unique solution of the fixed point equation $x = \Fsone \ x$),
where $\Fsone \ x = \cons \ 0 \ (\smapS \ x)$.
To prove $\Fsone$ contractive,
we first choose a suitable equivalence structure $=_{i}$ on $\stream$:
let $x \equptoc{i} y$ for streams $x, y$ if their first $i$ elements are equal.
A function $F : \stream \ra \stream$ is contractive if for all $x, y$ and
$i \in \nat$, whenever $x =_{j} y$ for all $j < i$, then $F \ x =_{i} F \ y$.
We then manually prove $\Fsone$ contractive:
for $i \ge 1$, if $x =_{i-1} y$, then $\smapS \ x =_{i-1} \smapS \ y$, and
consequently $\Fsone \ x =_{i} \Fsone \ y$.
The fixed point $\sone$ follows from this contractivity proof.

\bfparagraph{Problem with AmiCo and Chargu\'eraud's approach}

These two methods, however, exhibit respective weaknesses, which we illustrate through two examples $\stwo$ and $\sthr$.

First, AmiCo more effectively handles $\stwo$ below than Chargu\'eraud's approach:
\[ \stwo = \cons \ 0 \ (\smapS \ (\smapS \ \stwo)) \]
Once $\smapS$ is registered as a friend, AmiCo requires no
additional user effort to validate $\stwo$, despite the multiple
occurrences of $\smapS$ in the definition. The friend information for
$\smapS$ exists independently of $\sone$ and can be reused for
$\stwo$, demonstrating how compositionality facilitates corecursive
definitions through assembly of known functions. On the other hand,
Chargu\'eraud's approach requires the user to
construct a new contractivity proof for the generating function $\Fstwo$ of $\stwo$.
Each new corecursive definition, even if only slightly modified, necessitates its own explicit contractivity proof,
with complexity growing proportionally to the intricacy of the functional combinations.
This method thus suffers from a lack of automation.

Second, AmiCo cannot handle $\sthr$ below while Chargu\'eraud's approach can:
\[ \sthr = \cons \ 0 \ (\cons \ 1 \ (\tail \ \sthr)) \quad \cdots(1) \]
This definition is sound because one occurrence of $\tail$ is
compensated by two occurrences of $\cons$, making the overall
definition guarded.  However, AmiCo only supports the notion of
guardedness preservation---it cannot utilize the information that
$\tail$ consumes only one element, leading to rejection of $\sthr$.
This demonstrates AmiCo's limited coverage.  Chargu\'eraud's approach
does not suffer from this limitation: $\sthr$ can be defined
through contractivity of the generating function $\Fsthr$ of $\sthr$,
and this contractivity can be proven similarly as done for $\Fsone$.

It is important to note that $\sthr$ has an equivalent definition
$\sthr = \cons \ 0 \ \ones$, where $\ones = \cons \ 1 \ \ones$.
While AmiCo accepts this alternative definition, 
users who prefer the equation (1) must expend additional effort to
prove it separately using coinductive methods.
In contrast, our tool directly accepts the original equation,
expanding coverage for such use cases.

\subsection{Challenges and Our Approach}

The preceding analysis underscores the need for a corecursive theory
that combines high automation with extensive coverage. To handle
the previous examples in a modular and automated fashion, the theory must
$(i)$ associate each component function with \emph{productivity}
information detailing how many elements it produces or consumes, and
$(ii)$ provide a means of systematically calculating the productivity of
composite functions based on that information. Furthermore, to account
for guardedness compensation in $\sthr$, this notion of
productivity must be quantitative, in contrast to the friend approach
that merely checks whether guardedness is preserved.

An intuitive, though ultimately na\"ive, strategy for such a
compositional and quantitative system is to represent productivity as
an integer value. Concretely, we can record $(+n)$ if a function
produces $n$ elements, and $(-n)$ if it consumes $n$ elements. The
expectation is that a generating function with productivity of at
least $(+1)$ yields a unique fixed point. In the case of $\stwo$,
$\cons$ has $(+1)$ productivity producing one element, and $\smapS$ has $(0)$ productivity producing no elements.  
Hence, the system can compute that $\Fstwo$, the composition of those component
functions (\ie one $\cons$ and two $\smapS$), has $(+1)$
productivity because adding those productivities yields $1$ (\ie $1 + 0 + 0 = 1$).
In the case of $\sthr$, $\tail$ has $(-1)$ productivity consuming one
element and thus $\Fsthr$ has $(+1)$ productivity by adding the
productivities of its components (\ie two $\cons$ and one $\tail$).
Since the compositionality of this system is realized by addition, we
term it the \emph{addition mechanism}.

However, this na\"ive method fails in more general cases. For example,
the function $\texttt{zip}\,f: \stream \times
\stream \to \stream$, found in stream processing
languages such as \cite{oleg17}, applies $f:\texttt{nat}\times\texttt{nat}\to\texttt{nat}$
to each pair of elements from the two input streams. Since this zip function
takes two streams as input, we may need to track productivity for each
stream separately to achieve more complete coverage. Moreover, it is
possible to generalize the zip function into
$\texttt{zipL}\,f$ of type $\texttt{list}\,\stream\to\stream$,
which requires a more complex notion of productivity.

To illustrate the subtlety in generalizing the notion of productivity,
we consider the stream $\sfour$ defined as follows:
\[ \sfour = \cons \ (\head \ \sfour) \ (\cons \ 1 \ (\tail \ \sfour)) \]
The definition of $\sfour$ is similar to $\sthr$: it consists of two occurrences of $\cons$ and
one occurrence of $\tail$. Despite the similarity, unlike $\sthr$, $\sfour$ is not well-defined
since even the first element of $\sfour$ cannot be determined from the definition.
Given that the presence of $\head$ in the definition
constitutes the difference, a tempting approach is to assign $\head$ a productivity of $(-1)$,
setting aside the deeper issue of how to define productivity for $\head:\stream\ra\nat$
whose codomain $\nat$ is not even a coinductive type.
This additional $(-1)$ would bring the productivity of the generating function $\Fsfour$ of $\sfour$ to $(0)$,
which matches the fact that it does not produce an element.

This intuition is, however, inaccurate, given that $\sfive$ below with a similar definition is well-defined:
\[ \sfive = \cons \ 0 \ (\cons \ (\head \ \sfive) \ (\tail \ \sfive)) \]
Repeated substitution
reveals that $\sfive$ is a stream where each element is $0$%
\footnote{While the definition can be rewritten as $\sfive = \cons \ 0 \ \sfive$ in this case,
we present this example to better illustrate the intuition about the limitations of na\"ive productivity counting.
More realistic examples that cannot be simplified in this way would exhibit the same fundamental issue.}:
\[\begin{array}{@{}r@{\ }l@{}}
    \sfive =& 0 \strcons (\head \ \sfive) \strcons (\tail \ \sfive) = 0 \strcons (\head \ (0 \strcons (\head \ \sfive) \strcons (\tail \ \sfive))) \strcons (\tail \ \sfive) \\
    =& 0 \strcons 0 \strcons \tail \ \sfive = 0 \strcons 0 \strcons \tail \ (0 \strcons 0 \strcons \tail \ \sfive) = 0 \strcons 0 \strcons 0 \strcons \tail \ \sfive = \cdots
  \end{array}\]
Since the definition of $\sfive$ has the same number of occurrences of $\cons$, $\head$, and $\tail$ as $\sfour$, 
a simplistic approach based merely on summing productivities 
would either reject both $\sfour$ and $\sfive$ or accept both, none of which are desirable.

This analysis highlights two primary challenges. First, characterizing
productivity for functions with non-coinductive domains or codomains
is problematic. A function like $\head$ with codomain~$\nat$ cannot be
characterized using the simple notion of producing or
consuming elements, which is straightforward for functions operating
on coinductive types like $\stream$ but becomes unclear for
non-coinductive types.  Second, the addition mechanism fails to handle
complex patterns of function combination. The component functions in
$\sfour$ and $\sfive$ are combined in ways more
intricate than simple function composition. In the case of
$\sfour$, the outer $\cons$ takes two arguments
from the respective outputs of $\head$ and the inner
$\cons$, representing a structural pattern of combination
that cannot be captured by merely summing individual
productivities. Therefore, to differentiate between the distinct
validity of these cases, a more refined notion of productivity is
required---one that can accurately capture both the behavior of
functions with non-coinductive domains or codomains and
the various ways functions can be structurally combined.

Based on the preceding discussion, three fundamental questions naturally arise:
\begin{enumerate}
\item How can the notion of productivity be generalized to apply to functions over various domain and codomain types, particularly when those types are not themselves coinductive?
  Specifically, what does the productivity of $\head : \stream \ra \nat$ mean?
\item How can a compositional calculus be established to compute the productivity of a function from its components when it is constructed from a diverse set of combinators?
\item How can productivity information be leveraged to guarantee the existence of a unique fixed point for a given generating function?
\end{enumerate}

We address these questions through a novel theoretical framework named
\textbf{Compositional Heterogeneous Productivity (CHP)}.
CHP introduces a robust notion of \emph{heterogeneous} productivity
applicable to functions $f : A \ra B$, even when the domain $A$ and
codomain $B$ differ, or when one or both are not coinductive types.
This encompasses, for instance, functions of type
$\stream \ra \nat$, $\nat \times \stream \ra \stream$ or even $\stream + \bintree \ra (\stream\times\nat) + \mathtt{list}(\bintree)$.
A cornerstone of CHP is its \emph{compositionality}: the productivity of a composite function,
formed by integrating its constituents through multiple combinators,
is systematically determined from the productivities of its
components. Productivity information accumulates as new functions are
registered within the system. Eventually, the fixed point of a
generating function is obtained by proving it possesses sufficiently
positive productivity.

An additional advantage of CHP is that productivity is expressed within Rocq.
For example, consider defining a corecursive function $\foo_{l,f} : \stream \ra \stream$
for $l : \List \ \nat$ and $f : \stream \ra \stream$ satisfying the
equation $\foo_{l,f}(x) = l \listappend \ f(x)$,
where $l \listappend \ s$ denotes appending a list $l$ to a stream $s$.
CHP accepts this corecursive definition provided $f$ is $(-n)$-productive%
\footnote{The $(-n)$-productivity here
reflects the intuition from the 
addition mechanism sketched earlier. The rigorous definition within CHP
is via $\stdunifZ$, defined later.}
for some $n < \text{length } l$.
This ability to quantify over productivity
assumptions by treating productivity as a mathematical property
distinguishes CHP from other approaches such as AmiCo.

Note that CHP focuses on defining coinductive datatypes and corecursive functions,
because established tools \cite{sangiorgi98,hur13,pous16,schafer17} already provide 
powerful support for coinductive reasoning about coinductive predicates.

Building upon CHP, we have developed \textbf{Coco} (Compositional
Corecursion), a corecursion library for the Rocq proof assistant. Coco
implements the CHP framework, enabling extensive automation for computation
of productivities and generation of fixed points for a large class of
corecursive definitions covering most examples in the literature on the topic.
When automation fails, users retain the option of proving
productivity properties manually.
Notably, Coco is a user-level library
that does not modify Rocq's underlying kernel,
though it relies on standard classical and choice axioms (see \Cref{Axioms}).

A caveat is that Coco currently requires more boilerplate code,
while AmiCo offers superior engineering with minimal boilerplate.
This stems from our implementation choice: writing tactics in Ltac within Rocq 
rather than writing Rocq plugins in OCaml.
AmiCo's automation for mixed recursive-corecursive definitions is 
particularly noteworthy:
it requires only termination proofs for recursive components,
whereas Coco currently requires users to prove semi-well-foundedness of productivity conditions,
demanding greater proof effort (detailed in \Cref{Fixed Point}).

\bfparagraph{Outline}

\Cref{Key idea} presents the key ideas underlying CHP theory.
\Cref{Coinductive Type} reviews the formalization of coinductive types, serving as the basis for the subsequent sections.
\Cref{First-order Productivity} formally defines productivity and establishes the fixed point principle and combination principles.
\Cref{Second-order Productivity} introduces second-order productivity, 
useful for the derivation of productivity for functions such as $\smapS$ that are themselves corecursively defined.
\Cref{Examples} demonstrates the coverage of CHP through various illustrative examples.
\Cref{Implementation and Automation} describes how Coco functions as a system providing automation and extensibility.
\Cref{Related Work} discusses related work in the field.
\Cref{Future Work} concludes the paper with directions for future research.

\section{Key Ideas} 
\label{Key idea}

This section outlines the key ideas of our CHP framework 
using the motivating examples $\Fsfour$ and $\Fsfive$ described in \Cref{Introduction}. 
We begin by introducing the concept of heterogeneous productivity in terms of \emph{levels} and 
illustrate the productivities of the constructor ($\cons$) and destructors ($\head$, $\tail$). 
Subsequently, we demonstrate how our framework guarantees a unique fixed point $\sfive$ of $F_\sfive$ 
through composition of productivities while rejecting the definition of $\sfour$.
We conclude the section by discussing our approach to automating this analysis.

In our theory, the concept of levels provides the foundation for heterogeneous productivity.
To clarify this notion, we revisit the integer-based productivity in the addition mechanism, informally described in \cref{Introduction}.
For instance, we assign a productivity (+1) to $\cons \ 0 : \stream \ra \stream$ in this mechanism, since the function produces one element.
This notion of ``producing one element'' can be captured by a mathematical property:
fixing the $n$-prefix of the input determines the $(n+1)$-prefix of the output.

Generally, we can formalize this concept in terms of \emph{equality-up-to} relations based on \emph{levels}.
The degree of similarity between two streams corresponds to the length of their common prefix.
We therefore introduce a hierarchy of levels for the type $\stream$, where each level $n \in \nat$ defines a corresponding equivalence relation.
More precisely, let streams $s_1$ and $s_2$ be equal up to $n$ if the first $n$ elements of $s_1$ and $s_2$ are identical.
Then, for all $n$, $\cons \ 0 \ s_1$ and $\cons \ 0 \ s_2$ are equal up to $n+1$ if $s_1$ and $s_2$ are equal up to $n$.
This relationship between the level $n$ in domain and the level $n+1$ in codomain is what the integer productivity $(+1)$ captures.
In this sense, the addition mechanism can be interpreted as imposing $\nat$ as a level set on all involved domains and codomains.

However, this uniform level structure is insufficient for analyzing functions over more complex types, 
as demonstrated by the examples of $\sfour$ and $\sfive$. 
In particular, in the analysis of $\cons$, assigning $\nat$ as the level set causes the loss of information about the first argument of $\cons$.
This loss prevents the addition mechanism from distinguishing the productivity of $\Fsfour$ from that of $\Fsfive$.
To resolve this limitation, we must allow different types to be associated with distinct, type-specific level sets.

\bfparagraph{Appropriate level sets}

We illustrate how type-specific level set assignment enables accurate
analysis by re-examining $\cons$ and $\head$ using new level
sets. While these are manually selected, we will show how
to systematically and automatically derive those level sets in
\Cref{First-order Productivity}.

For $\cons : \nat \times \stream \ra \stream$, we use the refined level set $\Bool \times \nat$ for its domain $\nat \times \stream$ 
($\Bool$ is the boolean type with elements $\btrue$ and $\bfalse$).
In detail, for $(b, m) \in \Bool \times \nat$, let $x_1 = (n_1, s_1)$ and $x_2 = (n_2, s_2)$ be equal up to $(b, m)$ if
(i) $b$ is $\bfalse$ or $n_1 = n_2$,
and (ii) $s_1$ and $s_2$ are equal up to $m$.
Then $\cons$ satisfies the property: for all $m > 0$ and $x_1, x_2 : \nat \times \stream$, $\cons \ x_1$ and $\cons \ x_2$ are equal up to $m$ whenever $x_1$ and $x_2$ are equal up to $(\btrue, m\minus1)$.
The boolean component of the level successfully encodes the requirement on the first argument of $\cons$.

Similarly, for $\head : \stream \ra \nat$, we assign the level set $\Bool$ to its codomain $\nat$, 
with the equality-up-to relation defined as follows:
$n_1$ and $n_2$ are equal up to $b$ if $b$ is $\bfalse$ or $n_1 = n_2$.
Then for $s_1$ and $s_2$, $\head \ s_1$ and $\head \ s_2$ are equal up to $\btrue$ whenever $s_1$ and $s_2$ are equal up to $1$.
The level $\bfalse$ in the codomain does not require any level in the domain.

This reveals two key insights.
First, the levels for $\stream$ depend on the coinductive structure of the type.
Second, the level set for $\nat \times \stream$ represents a combination of the level sets for $\nat$ and $\stream$.
Our framework embodies this compositional approach:
we first assign level sets to coinductive types, then extend these assignments to their \emph{mixtures}.
The systematic construction of this assignment process is presented in \Cref{CTT} and \Cref{Equality-up-to}.

\bfparagraph{Productivities of constructor and destructors}
Since level sets are no longer uniform (\ie not always $\nat$),
the notion of productivity must capture more complex relationships between potentially disparate level sets.
This demands a more expressive formulation than simple integer values such as $(+1)$.
While one might consider representing productivity as a function $p: \level(B) \to \level(A)$ for a function $f: A \to B$,
this approach lacks sufficient expressiveness. 
In certain cases, productivity should associate a single level in codomain with multiple levels in domain.

We therefore formalize productivity as a relation, represented as an element of $\productivity{A}{B}$.
We say that $f: A \ra B$ is $p$-productive for $p \in \productivity{A}{B}$, 
if for any $l \in \level(B)$ and $a_1, a_2 \in A$, 
$f \ a_1$ and $f \ a_2$ are equal up to $l$ whenever $a_1$ and $a_2$ are equal up to all levels related to $l$ by $p$.
This approach is highly adaptable, 
as the selection of appropriate level sets 
preserves all information required for productivity reasoning across diverse types.

We now derive the heterogeneous productivities of $\tail$, $\cons$ and $\head$ in the relation form.
The productivity of $\tail$ is $\{ (n\!+\!1,\ n)\ |\ n > 0 \}$, effectively capturing the property that $\tail \ s_1$ and $\tail \ s_2$ are equal up to $n$ if $s_1$ and $s_2$ are equal up to $n+1$.
Note that no input level is related to the output level $0$ because any two output streams are equal up to $0$ vacuously.
The productivity of $\cons$ is $\{ ((\btrue, n\minus1),\ n)\ |\ n > 0 \}$: 
for $\cons(v_1,s_1)$ and $\cons(v_2,s_2)$ to be equal at level $n>0$, their first arguments $v_1$ and $v_2$ must be identical and the second arguments be equal up to $n\minus1$.
Following the same line of reasoning, we determine the productivity of $\head$ to be $\{ (1,\ \btrue) \}$.

A productivity relation can also be represented equivalently by its defining constraint.
For example, the productivity of $\tail$ in constraint-based form is $\textcolor{blue}{n_\texttt{I} = n_\texttt{O} + 1 \land n_\texttt{O}>0}$ where $n_\texttt{I}, n_\texttt{O} \in \nat$ are the domain (input) and codomain (output) levels, respectively. 
The productivity of $\head$ is $\textcolor{red}{n_\texttt{I} = 1 \land b_\texttt{O}=\btrue}$ for domain level $n_\texttt{I} \in \nat$ and codomain level $b_\texttt{O} \in \Bool$. 
The productivity of $\cons$ is $\textcolor{teal}{b_\texttt{I} = \btrue \land n_\texttt{I}+1=n_\texttt{O}}$ for domain level $(b_\texttt{I},n_\texttt{I}) \in \Bool\times\nat$ and codomain level $n_\texttt{O}\in\nat$.
We mainly use the constraint-based form throughout the remainder of this section for intuitive explanation.

\bfparagraph{Analysis via productivity composition}
We proceed to illustrate how productivity composition enables our framework to accept the definition of $\sfive$ while rejecting the definition of $\sfour$.
To achieve this, we decompose $\Fsfour$ and $\Fsfive$ into component functions using \emph{function combinators},
allowing compositional application of the productivities derived in previous paragraphs.

We employ three function combinators to construct $\Fsfour$ and $\Fsfive$.
The most elementary is \emph{function composition} $\fcompose$, defined by $(g \fcompose f)(x) = g(f(x))$.
The second is \emph{function pairing} $\fpair$.
Given two functions $f : D \ra C_1$ and $g : D \ra C_2$, 
their pairing yields a new function $f\fpair g : D \ra C_1 \times C_2$,
defined as $(f \fpair g) (x) = (f(x), g(x))$.
The third is the \emph{constant} combinator $\const$, where $\const(a) = \lambda \_. a$.
Using these combinators and three basic block functions ($\cons$, $\head$, $\tail$),
$\Fsfour$ and $\Fsfive$ can be expressed as follows, 
avoiding the confusion between partial application $\cons\ 0 : \stream \ra \stream$ and the original form $\cons : \nat \times \stream \ra \stream$. 
\[\begin{array}{@{}r@{\ }l@{}}
  \Fsfour = \cons \fcompose (\head \fpair (\cons \fcompose (\const(1) \fpair \tail))) \\
  \Fsfive = \cons \fcompose (\const(0) \fpair (\cons \fcompose (\head \fpair \tail)))
\end{array}\]

We now derive the productivity of $\Fsfive$ through
detailed compositional analysis. The step-by-step process of composing
productivities is shown in \Cref{Fs5 productivity}, where each step
presents a function in combinator form on the upper line, with its
corresponding productivity constraint on the lower line. In this
derivation, for $f \fpair g$ we take the
disjunction of the productivity constraints for $f$ and $g$;
for $f \fcompose g$ we transitively compose the
productivity constraints for $f$ and $g$ by existentially
quantifying the connecting variables; and for $\const(x)$ we simply take $\False$.
The precise combination principles for more combinators
are formally presented in \Cref{Combination of Productivity}.

\begin{figure}[t]
\begin{equation*}
\begin{array}{@{}l@{}}
  \hline
  \textcolor{red}{\head} \fpair \textcolor{blue}{\tail}: \\
  \hfill (\textcolor{red}{n_\texttt{I}=1 \land b_\texttt{O}=\btrue}) \lor (\textcolor{blue}{n_\texttt{I} = n_\texttt{O} + 1 \land n_\texttt{O}>0}) \\
  \hline
  \textcolor{teal}{\cons} \fcompose (\textcolor{red}{\head} \fpair \textcolor{blue}{\tail}): \\
  \hfill \exists n_1,b_1,\ (\textcolor{teal}{b_1=\btrue \land n_1+1=n_\texttt{O}}) \land ((\textcolor{red}{n_\texttt{I}=1 \land b_1=\btrue}) \lor (\textcolor{blue}{n_\texttt{I} = n_1 + 1 \land n_1>0})) \\
  \hline
  \const(0) \fpair (\textcolor{teal}{\cons} \fcompose (\textcolor{red}{\head} \fpair \textcolor{blue}{\tail})): \\
  \hfill \False \lor (\exists n_1,b_1,\ (\textcolor{teal}{b_1=\btrue \land n_1+1=n_\texttt{O}}) \land ((\textcolor{red}{n_\texttt{I}=1 \land b_1=\btrue}) \lor (\textcolor{blue}{n_\texttt{I} = n_1 + 1 \land n_1>0}))) \\
  \hline
  \textcolor{teal}{\cons} \fcompose (\const(0) \fpair (\textcolor{teal}{\cons} \fcompose (\textcolor{red}{\head} \fpair \textcolor{blue}{\tail}))): \\
  \hfill \exists n_2,b_2,\ (\textcolor{teal}{b_2=\btrue \land n_2+1=n_\texttt{O}})\ \land \\
    \hspace{0pt} (\False \lor (\exists n_1,b_1,\ (\textcolor{teal}{b_1=\btrue \land n_1+1=n_2}) \land ((\textcolor{red}{n_\texttt{I}=1 \land b_1=\btrue}) \lor (\textcolor{blue}{n_\texttt{I} = n_1 + 1 \land n_1>0})))) \\
  \hline
\end{array}
\end{equation*}
\caption{Process of composing productivities for $\Fsfive$}
\label{Fs5 productivity}
\end{figure}

The CHP fixed point theorem then establishes the unique fixed point of $\Fsfive$ from its derived productivity.
Specifically, the theorem guarantees the unique fixed point,
provided the productivity constraint for $\Fsfive$ implies the special productivity constraint $\textcolor{purple}{n_\texttt{I}+1 \le n_\texttt{O}}$ (see \Cref{Fs5 condition}).
To verify such conditions, our library Coco provides a tactic-based automatic solver, utilizing Rocq's standard nonlinear arithmetic solver \texttt{nia} \cite{rocqdoc} for arithmetic reasoning.
Our solver automatically proves the well-definedness condition for $\Fsfive$.
Indeed, one can see that the productivity constraints for $\Fsfive$ is simplified to
$(n_\texttt{I} = 1 \land n_\texttt{O} = 3) \lor (n_\texttt{I} > 1 \land n_\texttt{O} = n_\texttt{I} + 1)$,
which implies $\textcolor{purple}{n_\texttt{I}+1 \le n_\texttt{O}}$.

\begin{figure}[t]
\begin{equation*}
\begin{array}{@{}l@{\ }l@{}}
  \forall n_\texttt{I},n_\texttt{O},~(\exists n_2,b_2,\ (b_2=\btrue \land n_2+1=n_\texttt{O})\ \land \\
  \ (\False \lor (\exists n_1,b_1,\ (b_1=\btrue \land n_1+1=n_2) \land ((n_\texttt{I}=1 \land b_1=\btrue) \lor (n_\texttt{I} = n_1 + 1 \land n_1>0))))) \\
  \ra\ {\textcolor{purple}{n_\texttt{I}+1 \le n_\texttt{O}}}
\end{array}
\end{equation*}
\vspace{-2mm}
\caption{Well-definedness condition for $\sfive$ (satisfied)}
\label{Fs5 condition}
\end{figure}

Similarly, we can derive the productivity of $\Fsfour$ and its well-definedness condition, which is shown in \Cref{Fs4 condition}.
In this case, our solver fails to verify the well-definedness condition because it has a counterexample $(n_\texttt{I},n_\texttt{O}) = (1, 1)$.
Through this method, CHP successfully distinguishes between well-definedness of $\sfour$ and $\sfive$.
The general and systematic methodology underlying this process is presented throughout this paper.

\begin{figure}[t]
\begin{equation*}
\begin{array}{@{}l@{\ }l@{}}
  \forall n_\texttt{I},n_\texttt{O},~(\exists n_2,b_2,\ (b_2=\btrue \land n_2+1=n_\texttt{O})\ \land \\
  \ ((n_\texttt{I}=1 \land b_2=\btrue) \lor (\exists n_1,b_1,\ (b_1=\btrue \land n_1+1=n_2) \land (\False \lor (n_\texttt{I} = n_1 + 1 \land n_1>0))))) \\
  \rightarrow\ {\textcolor{purple}{n_\texttt{I}+1 \le n_\texttt{O}}}
\end{array}
\end{equation*}
\vspace{-2mm}
\caption{Well-definedness condition for $\sfour$ (not satisfied)}
\label{Fs4 condition}
\end{figure}

\bfparagraph{Second Order Productivity and Automation}
To achieve high automation in the productivity analysis,
we automatically derive the following: $(i)$ productivity of constructors and destructors of coinductive datatypes
and $(ii)$ productivity of composite functions from the productivities of their constituent functions.
Although we can cover many examples by combining these constructions,
we still have important functions whose productivity cannot be derived in such a way.

The function $\smapS$ serves as such an example.
Since it is used as a component of $\Fsone$, finding and proving the productivity of $\smapS$ is important.
Note that $\smapS$ is defined by $F_{\smapS} = \lambda f.\ \lambda x.\ \cons \ (1+ (\head \ x),\ f(\tail \ x))$
and the productivity of $F_{\smapS}$ can be derived from the productivities of its constituent functions such as $\cons$.
However, the fixed point principle introduced in this section guarantees only the existence and uniqueness of the fixed point $\smapS$,
not that $\smapS$ is $0$-productive.
Of course, it is always possible for the user to manually prove the productivity for $\smapS$.

However, for better automation, we address this problem by introducing the notion of \emph{second-order productivity}.
This provides a new fixed point principle that not only establishes the fixed point but also validates its productivity.
Second-order productivities can also be proved automatically just as (first-order) productivity, as detailed in \Cref{Second-order Productivity}.

\bfparagraph{Implementation Challenges}

Transitioning from the theory of CHP to its practical implementation in the Rocq library raises
several challenges regarding automation, usability, and extensibility.
Specifically, we address three key questions.
First, how is a productivity proof, once established, automatically leveraged by the system?
Second, how can corecursive functions defined via Coco be made accessible to users unfamiliar with the underlying CHP theory?
Third, how can the framework accommodate user-defined type combinators?
Our approach to these implementation challenges is presented in \Cref{Implementation and Automation}.

\section{Coinductive Type}
\label{Coinductive Type}

In this section, we provide a formal treatment of coinductive types, establishing the necessary background for later discussions of levels and productivity.
We begin by presenting \emph{semantic polynomial functors}, a specialized class of functors used to construct coinductive types.
We then review coinductive types as \emph{final coalgebras} derived from these functors.

Throughout the paper, we use the following notation conventions for readability.
The notations $x : T$ and $x \in T$ are used interchangeably.
For $T_1 \in \Type$ and $T_2 \in T_1 \ra \Type$,
we denote $\prod_{x \in T_1} T_2(x)$ by $\depfun{x\!:\!T_1}{T_2(x)}$ and
$\sum_{x\in T_1}T_2(x)$ by $\deppair{x\!:\!T_1}{T_2(x)}$.
For $X, Y \in I \ra \Type$, we write $X \Ra Y$ for $\depfun{i:I}{X(i)\ra Y(i)}$.
We omit implicit parameters $X,Y$ when clear from context.
$\powset{A}$ denotes the power set of $A$.
$\option(A)$ is the option type with constructors $\some(a)$ for $a : A$ and $\none$.
$\unitset$ is the unit type with element $\unit$.
The notation $\ite{\_}{\_}{\_}$ represents the ternary if-then-else expression.

\subsection{Semantic Polynomial Functor}
\label{Semantic Polynomial Functor}

A coinductive type $C$ can be understood as the greatest fixed point of a type functor $F_C$ \cite{moss97, rutten00}.
For instance, $F_{\mathtt{s}} \ X = \nat \times X$ is the type functor for the stream type (\ie $\stream = \nu F_{\mathtt{s}}$).
However, some type functors such as $F \ X = (X \ra \Bool)$ do not possess a fixed point.
Consequently, Rocq's $\mathtt{CoInductive}$ accepts only syntactically \emph{strictly positive} functors, which are known to exhibit fixed points \cite{rocqdoc}.
Since we develop a user-level library, we must semantically capture the notion of strict positivity within Rocq.
To achieve this, we employ the concept of \emph{polynomial functor} \cite{moerdijk00}, adopting the terminology from the \emph{container} formulation \cite{abbott05}.

A polynomial functor $F : \Type \ra \Type$ is characterized by a shape set $\fctnS : \Type$ and a degree function $\fctnD : \fctnS \ra \Type$,
which maps a type $X$ to the type $F \ X = (s : \fctnS) \times (\fctnD (s) \ra X)$.

\begin{wrapfigure}{r}{0.6\textwidth}
\vspace*{-2mm}
\begin{minipage}{\textwidth}
\begin{subfigure}{0.3\textwidth}
\hspace{20pt}
\begin{tikzpicture}
    [level distance=6mm,
    every node/.style={rectangle,inner sep=0pt},
    level 1/.style={sibling distance=17mm,nodes={}},
    level 2/.style={sibling distance=6mm,nodes={}},
    level 3/.style={sibling distance=5mm,nodes={}},
    level 4/.style={sibling distance=2mm,nodes={}}]
    \node {$s_0$}[grow=down]
      child {node {$s_1$}
        child {node {$s_2$}
          child {node {$s_4$}
            child{node {$\vdots$}}
            child{node {$\vdots$}}
            child{node {$\vdots$}}
            child[missing]
          }
          child {node {$s_1$}
            child {node {$\vdots$}}
          }
          child {node {$s_5$}
            child {node {$\vdots$}}
            child {node {$\vdots$}}
          }
          child {node {$\cdots$}}
          child[missing]
        }
      }
      child {node {$s_2$}
        child {node {$s_3$}}
        child {node {$s_0$}
          child {node {$s_0$}
            child{node {$\vdots$}}
            child{node {$\vdots$}}
          }
          child {node {$s_3$}}
        }
        child {node {$s_1$}
          child[missing]
          child {node {$s_0$}
            child{node {$\vdots$}}
            child{node {$\vdots$}}
          }
        }
        child{node {$\cdots$}}
      };
\end{tikzpicture}
\caption{General coinductive type}
\label{coinductive-type-tree}
\end{subfigure}
\begin{subfigure}{0.11\textwidth}
\begin{center}
\begin{tikzpicture}
  [level distance=6mm,
    every node/.style={rectangle,inner sep=0.5pt},
    level 1/.style={sibling distance=15mm,nodes={}},
    level 2/.style={sibling distance=7.5mm,nodes={}},
    level 3/.style={sibling distance=3mm,nodes={}}]
  \node {$1$}[grow=down]
    child {node {$2$}
    child {node {$3$}
    child {node {$8$}
    child {node {$\vdots$}
    }}}};
  \end{tikzpicture}
\end{center}
\caption{$\stream$}
\label{stream-tree}    
\end{subfigure}
\begin{subfigure}{0.2\textwidth}
\begin{tikzpicture}
    [level distance=6mm,
    every node/.style={rectangle,inner sep=0.5pt},
    level 1/.style={sibling distance=12mm,nodes={}},
    level 2/.style={sibling distance=7mm,nodes={}},
    level 3/.style={sibling distance=4mm,nodes={}}]
    \node {$2$}[grow=down]
        child {node {$5$}
            child {node {$1$}
                child {node {$\vdots$}}
                child {node {$\unit$}}
            }
            child {node {$\unit$}}
        }
        child {node {$3$}
            child {node {$0$}
                child {node {$\vdots$}}
                child {node {$\vdots$}}
            }
            child {node {$1$}
                child {node {$\vdots$}}
                child {node {$\unit$}}
            }
        };
\end{tikzpicture}
\caption{$\bintree_\bot$}
\label{bintree_bot-tree}
\end{subfigure}
\end{minipage}
\end{wrapfigure}
The coinductive type $\nu F$ constructed by $F$ can be conceptualized as a collection of infinite trees with varying number of branches.
Each node within a tree in $\nu F$ is determined by a shape $s \in \fctnS$ and its corresponding degree $\fctnD(s)$ (\ie possessing $|\fctnD(s)|$ subtrees), as illustrated in the figure (a) above.
For example, as depicted in the figure (b) above, $F_{\mathtt{s}}$ is equivalent to 
a polynomial functor $F_\stream = \lambda X. (\_ : \nat) \times (\unitset \ra X)$. 
Likewise, as depicted in the figure (c) above, $F_{\bintree_\bot} = \lambda X. (i : \unitset + \nat) \times ((\ite{i \!=\! \unit}{\Empty}{\Bool}) \ra X)$ 
is a polynomial functor corresponding to $\bintree_\bot$,
the coinductive type of binary trees with possibly infinite depth.

To accommodate indexed coinductive types, we use the indexed variant of polynomial functors.
We provide formal definitions of these indexed functors and their corresponding coinductive types.

\begin{figure}[t]
\[
\begin{array}{@{}r@{\ \ }c@{\ \ }l@{}}
\vctn \in \Container(I) &\defeq& \setof{ (\fctnS,\fctnD) \suchthat \fctnS\in\Type \land \fctnD\in \fctnS \ra I \ra \Type }
\\
\sem{\vctn}(X) &\defeq& \deppair{s:\vctn.\fctnS}{(\vctn.\fctnD(s) \Ra X)}\\
&&\text{where } \sem{\vctn} \in (I\ra \Type)\ra\Type
\\
\semf{\vctn}(f)(s, x) &\defeq& (s, \lambda i.~\lambda d.~ f(i)(x(i)(d)))\\
&&\text{where } \semf{\vctn} \in (X \Ra Y) \ra \sem{\vctn}(X) \ra \sem{\vctn}(Y) 
\\
\semr{\vctn}(r) &\defeq& \setof{ ((s,x),(s,y)) \suchthat \forall\, (i\in I)\, (d \in \vctn.\fctnD(s)(i)),\; (x(i)(d), y(i)(d)) \in r(i)}\\
&&\text{where } \semr{\vctn} \in (\depfun{i:I}{\powset{X(i) \times Y(i)}}) \ra \powset{\sem{\vctn}(X) \times \sem{\vctn}(Y)}
\\
\end{array}
\]
\caption{Definition of container}
\label{definition of container}
\end{figure}

A \emph{container} is a structure that encodes a polynomial functor
via shape $\fctnS$ and degree function $\fctnD$ (\Cref{definition of container}).
These determine the polynomial functor $\sem{\vctn}$,
along with its \emph{map function} $\sem{\vctn}_\fspm$ and \emph{relator} $\sem{\vctn}_\fspr$.
The map function lifts functions while preserving structure.
Analogously, the relator \cite{rutten98,traytel12}
lifts an indexed relation between $X$ and $Y$ to a relation between $\sem{\vctn}(X)$ and $\sem{\vctn}(Y)$ for $X, Y : I \ra \Type$.
For example, let $\vctn_0 = (\nat, \lambda\_.\unitset) \in \Container(\unitset)$
so that $\sem{\vctn_0}(X) = F_\stream(X\ \unit)$. 
Then $\semr{\vctn_0}(\lambda\_.r)$ relates $x' : F_{\stream} (X\ \unit)$ and $y' : F_{\stream} (Y\ \unit)$ 
if and only if
$\exists n \in \nat, x \in X(\unit), y \in Y(\unit),\ x' = (n, \lambda \_ .x) \land y' = (n, \lambda \_ .y) \land r(x, y)$.

For greater flexibility,
we work with functors \emph{semantically isomorphic} to polynomial functors defined by containers,
which we term \emph{semantic polynomial functors} (SPFs), formalized in \Cref{definition of SPF}.
An SPF is a triple of a functor, a map function, and a relator,
together with an isomorphism to a polynomial functor.
For $\vspf \in \SPF(I)$, we simply write $\vspf(X)$ for $\vspf.\fspf(X)$.
We present examples of SPFs in \Cref{SPF examples}, with $\fspm$, $\fspr$ and bijections ($\fspe, \fspe'$) omitted for brevity.
The notation $(x=y):\Type$ denotes the type defined as (if $x=y$ then $\unitset$ else $\Empty$).

\begin{figure}
\[
\begin{array}{@{}r@{\ \ }l@{~}l@{}}  
  \vspf \in \SPF(I) &\defeq \setof{ (\fspf, \fspm, \fspr, \vctn, \fspe, \fspe') \suchthat
    \fspf \in (I\ra\Type)\ra\Type ~\land \\
    & \fspm \in (X \Ra Y) \ra \fspf(X) \ra \fspf(Y) ~\land \\
    & \fspr \in (\depfun{i:I}{\powset{X(i) \times Y(i)}}) \ra \powset{\fspf(X) \times \fspf(Y)} ~\land{} \\
    & \vctn \in \Container(I)\ \land\ \fspe \in \fspf \Ra \sem{\vctn}\ \land\ \fspe' \in \sem{\vctn} \Ra \fspf\ \land\\
    & (\forall\, X\, Y\, (f \in X \Ra Y)\, (x \in \fspf(X)),~
       \fspe(\fspm(f)(x)) = \semf{\vctn}(f)(\fspe(x))) \land{} \\
    & (\forall\, X\, Y\, r \, (x : \fspf(X))\, (y : \fspf(Y)),~
       (x,y)\in\fspr(r) \!\iff\! (\fspe(x),\fspe(y)) \in \semr{\vctn}(r)) \land{}\\
    & (\forall\, X\,(x \in \fspf(X)),~ \fspe'(\fspe(x)) = x)\ \land\\
    & (\forall\, X\,(x' \in \sem{\vctn}(X)),~ \fspe(\fspe'(x')) = x')
  }
  \\
\end{array}
\]
\caption{Definition of SPF}
\label{definition of SPF}
\end{figure}

\begin{figure}
\[
\begin{array}{@{}r@{\ \ }l@{\quad }l@{}}
  \tconop{\times} \in \SPF(\Bool) \defeq& \lambda X.~ (X\ \btrue) \times (X\ \bfalse), 
  &\vctn\ = (\unitset,\ \lambda\_.\lambda \_. \unitset)
  \\
  \tconop{+} \in \SPF(\Bool) \defeq& \lambda X.~ (X\ \btrue) + (X\ \bfalse),
  &\vctn\ = (\Bool,\ \lambda b_1.\lambda b_2.\ (b_1=b_2):\Type)
  \\
  \tconop{A\ra} \in \SPF(\unitset) \defeq& \lambda X.~ A\ra (X\ \unit),
  &\vctn\ = (\unitset,\ \lambda\_.\lambda \_. A)
  \\
  \tconop{\depfun{a:A}{}} \in \SPF(A) \defeq& \lambda X.~ \depfun{a:A}{X(a)},
  &\vctn\ = (\unitset,\ \lambda\_.\lambda a.\ \deppair{a':A}{((a=a'):\Type)})
  \\
  \tconop{\deppair{a:A}{}} \in \SPF(A) \defeq& \lambda X.~ \deppair{a:A}{X(a)},
  &\vctn\ = (A,\ \lambda a_1.\lambda a_2.\ (a_1=a_2):\Type)
  \\
  \tconop{\List} \in \SPF(\unitset) \defeq& \lambda X.~ \List~(X\ \unit),
  &\vctn\ = (\Nat,\ \lambda n.\lambda \_. \setof{0,1,\cdots,n\!-\!1})
  \\
  \tconop{\Id} \in \SPF(\unitset) \defeq& \lambda X.~ X\ \unit,
  &\vctn\ = (\unitset,\ \lambda\_.\lambda\_. \unitset)
  \\
  \tconop{\Id_{a:A}} \in \SPF(A) \defeq& \lambda X.~ X\ a,
  &\vctn\ = (\unitset,\ \lambda\_.\lambda a'.\ (a=a'):\Type)
  \\
  \tconop{B} \in \SPF(I) \defeq& \lambda \_.~ B \quad\text{ for } I,B \in \Type,
  &\vctn\ = (B,\ \lambda\_.\lambda \_. \Empty)
\end{array}
\]
\caption{Examples of SPF}
\label{SPF examples}
\end{figure}

An $\SPF$ $\vspf$ can be composed with a family of $\SPF$s $\vec{\vspf}$ to yield a new $\SPF$ $\vspf \spfcompose\ \vec{\vspf}$ (\Cref{SPF composition}).

\begin{figure}
\[\begin{array}{@{}l@{\ \ }l@{\quad }l@{}}
  \quad\ \ \ \vspf \spfcompose\ \vec{\vspf} \in \SPF(I_1) \defeq \lambda X.~\vspf(\lambda i_2.~ \vec{\vspf}(i_2)(X)),\qquad
  \vctn\ = (\fctnS, \fctnD)\\
  \qquad\ \text{for }\ \vspf\in \SPF(I_2),\ \vec\vspf\in I_2\ra\SPF(I_1), \\
  \qquad\ \fctnS = \deppair{s:\vspf.\vctn.\fctnS}{(\depfun{i_2:I_2}{(\vspf.\vctn.\fctnD(s)(i_2))\ra (\vec\vspf(i_2)).\vctn.\fctnS})},\\
  \qquad\ \fctnD = \lambda (s,p_s).~ \lambda i_1.~ \deppair{i_2:I_2}{\ \deppair{d: \vspf.\vctn.\fctnD(s)(i_2)}{\ ((\vec\vspf(i_2)).\vctn.\fctnD\ (p_s(i_2)(d))\ (i_1))}}
\end{array}\]
\caption{Composition of SPFs}
\label{SPF composition}
\end{figure}

The concept of SPF is similar to that of \emph{quotient polynomial functor (QPF)} \cite{avigad19}.
The key difference is that QPFs additionally incorporate quotients over the structure of polynomial functors. 
We do not foresee any fundamental difficulty in generalizing from SPFs to 
QPFs within the CHP framework.

\subsection{Final Coalgebra}
\label{Final Coalgebra}

To derive coinductive types from polynomial functors,
we introduce the notion of \emph{semantic polynomial endofunctor} (SPE) from SPF.
An SPE $\vspe \in \SPE(I)$ is an indexed SPF, interpreted as an endofunctor $\fspsem{\vspe}$.
An SPE admits the greatest fixed point $\nu\fspsem{\vspe} \in I \ra \Type$
with indexed functions $\fold_\vspe \in \fspsem{\vspe}(\nu\fspsem{\vspe}) \Ra \nu\fspsem{\vspe}$ and $\unfold_\vspe \in \nu\fspsem{\vspe} \Ra \fspsem{\vspe}(\nu\fspsem{\vspe})$, 
satisfying the universal property stated in \Cref{indexed final coalgebra}.
\[
\begin{array}{@{}r@{\ }l@{\ \ }c@{\ \ }l@{}}
  \SPE(I) \in& \Type &\defeq& I \ra \SPF(I)
  \\
  \fspsem{\vspe} \in& (I\ra\Type)\ra I \ra \Type & \defeq & \lambda X.~ \lambda i.~ \vspe(i).\fspf(X)
  \\
\end{array}
\]

\begin{minipage}[m]{0.54\textwidth}
\begin{theorem}[Final Coalgebra]\label{indexed final coalgebra}
  For a coalgebra $X \in I \ra \Type$ with $\alpha \in X \Ra \fspsem{\vspe}(X)$,
  there uniquely exists $\scofix{\alpha}\in X \Ra \nu\fspsem{\vspe}$ such that
  the diagram on the right commutes for all $i$ : 
\end{theorem}
\end{minipage}
\begin{minipage}[m]{0.45\textwidth}
\[
\xymatrix@C=5PC@R=1.4PC{
  \fspsem{\vspe}(X)(i)  
  \ar[r]^-{\vspe(i).\fspm(\scofix{\alpha})}  
  &
  \fspsem{\vspe}(\nu\fspsem{\vspe})(i)  
  \\
  X(i)
  \ar[u]^-{\alpha(i)}  
  \ar[r]^-{\scofix{\alpha}(i)}  
  &
  \nu\fspsem{\vspe}(i)
  \ar[u]_-{\unfold_\vspe(i)}
}
\]
\end{minipage}

The resulting $\nu\fspsem{\vspe}\in I\ra\Type$ obtained from \Cref{indexed final coalgebra} represents the general form of indexed coinductive types.

\begin{remark}
  $\forall i,\ \fold_\vspe(i) \fcompose \unfold_\vspe(i) = \lambda x.~x$,\ \ $\unfold_\vspe(i) \fcompose \fold_\vspe(i) = \lambda x.~x$.
\end{remark}

For non-indexed coinductive types (\ie those without mutual dependencies),
the construction reduces to the fixed point of $F\in\Type\ra\Type$.
We adopt simplified notation below for these cases.
The subsequent sections primarily address these non-indexed cases.
The general indexed version is presented in the appendix (see \Cref{CHP for Indexed Coinductive Types}).

\begin{flalign*}
  &\vspe\in\SPE\defeq \setof{(\vspf,\fspf,\fspm,\fspr)\ |\ \ \vspf\in\SPF(\unitset), &&\\
  &\hspace{100pt}\fspf = \lambda X.\ \vspf.\fspf(\lambda\_.X) \in \Type\ra\Type,\ &&\\
  &\hspace{100pt}\fspm = \lambda f.\ \vspf.\fspm(\lambda\_.f) \in (X\ra Y)\ra F(X)\ra F(Y),\ &&\\
  &\hspace{100pt}\fspr = \lambda r.\ \vspf.\fspr(\lambda\_.r) \in \powset{X\times Y}\ra\powset{F(X)\times F(Y)} } &&
\end{flalign*}
\noindent $\fspsem{\vspe}\defeq \lambda X.~\vspe.\fspf(\lambda\_.X) \in \Type\ra\Type$.\ \ $\nu\fspsem{\vspe}\in\Type$.
\noindent $\fold_\vspe : \fspsem{\vspe}(\nu\fspsem{\vspe}) \ra \nu\fspsem{\vspe}$, $\unfold_\vspe : \nu\fspsem{\vspe} \ra \fspsem{\vspe}(\nu\fspsem{\vspe})$

\subsection{Bisimulation}
\label{Bisimulation}

Bisimulation is a coinductive equivalence relation on coinductive types that serves as the natural notion of equality.
In Rocq, however, identifying bisimulation with the standard inductively defined equality requires an axiom.

\begin{definition}\label{def_bisim}
  Let $\bisim$ be the greatest relation satisfying $\bisim = \bisimF\ \bisim$ where\\
\indent\quad$\bisimF : \powset{\nu\fspsem{\vspe} \times \nu\fspsem{\vspe}} \ra \powset{\nu\fspsem{\vspe} \times \nu\fspsem{\vspe}}$, 
  $\bisimF \ r \defeq \{ (c_1, c_2) \ | \ \vspe.\fspr \ r \ (\unfold_\vspe(c_1), \unfold_\vspe(c_2)) \}$.\\
\indent\quad We write $c_1 \equiv c_2$ for $(c_1, c_2) \in \bisim$.
  
\end{definition}

\begin{axiom}\label{bisim_eq}
  $\forall c_1,c_2\in\nu\fspsem{\vspe}$, if $c_1\equiv c_2$, then $c_1=c_2$.
\end{axiom}
\section{First-order Productivity}
\label{First-order Productivity}

In this section, we formally define equality-up-to relations and productivity, along with their mathematical properties.
We begin by defining the notion of \emph{coinductive-leafed type trees} to model mixture types and endow them with leveled equivalence relations.
Upon this foundation, we then define productivity, establish its combination principles, and present the fixed point theorem.
Also, we introduce \emph{slot-leafed type trees} to systematically assign the productivities of constructors and destructors.
To distinguish productivity in this section from the second-order productivity in \Cref{Second-order Productivity}, we often refer to it as \emph{first-order productivity}.

\subsection{Coinductive-leafed Type Tree}
\label{CTT}
For a general definition of heterogeneous productivity,
equality-up-to relations must apply not only to simple coinductive types but also to their mixtures.
These mixture types can involve complex combinations, such as $\nat \times \stream$ and $(\stream\times\nat) + \mathtt{list}(\bintree)$.
We achieve this generalization by decomposing mixture types into a tree-based structure, which we term the \emph{coinductive-leafed type tree} (CTT), 
defined inductively as follows:
\[\begin{array}{@{}r@{}r@{\ \ }c@{\ \ }l@{~}l@{}}
  \text{(Leaf case)}& \qquad \identCTT{\vspe} &\in& \CTT \qquad &\text{for } \vspe\in\SPE\\
  \text{(Internal node case)}& \qquad \cttf \cttcompose \vec{\vctt}  &\in& \CTT \qquad &\text{for } \ctta\in\Type,\ \cttf\in\SPF(\ctta),\ \vec\vctt\in \ctta\ra\CTT
\end{array}\]

The first constructor $\identCTT{\cdot}$ represents the leaf case,
and the second constructor $\cttcompose$ represents the internal node case with $\ctta$-indexed children $\vec\vctt$.

We recursively define the corresponding type $\sem\vctt$ for each $\CTT$ $\vctt$, 
which realizes the mixture type that the $\CTT$ was designed to represent.
The leaf represents a coinductive type,
and the internal node represents a mixture type based on the type constructor $\cttf$.
\[\begin{array}{@{}c@{\ \ }c@{\ \ }l@{~}l@{}}
  \text{for }\vctt \in \CTT,\ \sem\vctt \in \Type\\
  \sem{\identCTT{\vspe}} \defeq \nu\fspsem{\vspe}\qquad
  \sem{\cttf \cttcompose \vec{\vctt}} \defeq \cttf (\lambda a\in\ctta.\ \sem{\vec{\vctt}(a)})
\end{array}\]

Several examples of $\CTT$s for different type constructors are illustrated in \Cref{CTT examples}.
We refer to
$\identCTT{\vspe}$ as \emph{ident CTT}, 
$\vctt_1 \gop{\times} \vctt_2$ as \emph{product CTT}, 
$\vctt_1 \gop{+} \vctt_2$ as \emph{sum CTT}, 
$\gop{\List}\,\vctt$ as \emph{list CTT}, 
$A\gop{\ra}{\vctt}$ as \emph{degenerate function CTT} (or \emph{exponent CTT}), 
$\gdepfun{a:A}{\vec{\vctt}(a)}$ as \emph{nondegenerate function CTT}, 
$\gdeppair{a:A}{\vec{\vctt}(a)}$ as \emph{dependent pair CTT}, and
$\constCTT{A}$ as \emph{constant CTT}.

\begin{figure}
\begin{equation*}
\begin{array}{@{}r@{\ \ }c@{\ \ }l@{~}l@{}}
  \vctt_1 \gop{\times} \vctt_2  &\defeq& \tconop{\times} \cttcompose (\lambda b\in\Bool.~ \ite{b}{\vctt_1}{\vctt_2}) 
  \quad &\text{ for } \vctt_1,\vctt_2\in\CTT
  \\
  \vctt_1 \gop{+} \vctt_2  &\defeq& \tconop{+} \cttcompose (\lambda b\in\Bool.~ \ite{b}{\vctt_1}{\vctt_2}) 
  \quad &\text{ for } \vctt_1,\vctt_2\in\CTT
  \\
  \gop{\List}\,\vctt  &\defeq& \tconop{\List} \cttcompose (\lambda \_\in\unitset.~ \vctt)
  \quad &\text{ for } \vctt\in\CTT
  \\
  A\gop{\ra}{\vctt}  &\defeq& \tconop{A\ra} \cttcompose (\lambda \_\in\unitset.~\ \vctt)
  \quad &\text{ for } A \in \Type,\ \vctt \in \CTT
  \\
  \gdepfun{a:A}{\vec{\vctt}(a)}  &\defeq& \tconop{\depfun{a:A}{}} \cttcompose (\vec{\vctt})
  \quad &\text{ for } A \in \Type,\ \vec\vctt \in A\ra\CTT
  \\
  \gdeppair{a:A}{\vec{\vctt}(a)}  &\defeq& \tconop{\deppair{a:A}{}} \cttcompose (\vec{\vctt})
  \quad &\text{ for } A \in \Type,\ \vec\vctt \in A\ra\CTT
  \\
  \constCTT{A}  &\defeq& \tconop{A} \cttcompose (\lambda \_\in\Empty.~ \_)
  \quad &\text{ for } A \in \Type
  \\
  \selfCTT{\vspe}  &\defeq& (\vspe.\vspf) \cttcompose (\lambda \_\in\unitset.~ \identCTT{\vspe})
  \quad &\text{ for } \vspe \in \SPE
\end{array}
\end{equation*}
\caption{Examples of CTT}
\label{CTT examples}
\end{figure}

\begin{remark}
  The following equations hold.
  \[
    \sem{\vctt_1 \gop{\times} \vctt_2} = \sem{\vctt_1} \times \sem{\vctt_2} \qquad
    \sem{\vctt_1 \gop{+} \vctt_2} = \sem{\vctt_1} + \sem{\vctt_2} \qquad
    \sem{\gop{\List}\,\vctt} = \List\,\sem{\vctt}
  \]
  \[
    \sem{A\gop{\ra}{\vctt}} = A\ra\sem{\vctt} \qquad
    \sem{\gdepfun{a:A}{\vec{\vctt}(a)}} = \depfun{a:A}{\sem{\vec{\vctt}(a)}} 
  \]
  \[
    \sem{\gdeppair{a:A}{\vec{\vctt}(a)}} = \deppair{a:A}{\sem{\vec{\vctt}(a)}} \qquad
    \sem{\constCTT{B}} = B \qquad
    \sem{\selfCTT{\vspe}} = \fspsem{\vspe}(\nu\fspsem{\vspe})
  \]
\end{remark}

\begin{remark}
  Coco does not yet automatically prove productivity when nondegenerate function CTTs are involved.
\end{remark}

The use of $\selfCTT{\vspe}$ enables us to treat both $\fold_\vspe$ and $\unfold_\vspe$ as functions from a $\CTT$ to a $\CTT$.

\begin{remark}
  $\fold_\vspe : \sem{\selfCTT{\vspe}} \ra \sem{\identCTT{\vspe}} $ and 
  $\unfold_\vspe : \sem{\identCTT{\vspe}} \ra \sem{\selfCTT{\vspe}}$.
\end{remark}

\subsection{Equality-up-to}
\label{Equality-up-to}

This subsection provides formal definitions of \emph{equality-up-to} relations.
We begin by establishing equality-up-to relations for coinductive types,
denoted $\equptoc{n}\ \in\ {\powset{\nu\fspsem{\vspe}\times\nu\fspsem{\vspe}}}$ for level $n \in \nat$. 
We define inductive proposition equality-up-to as follows:
\[
\infer{ }
      { (c_1,c_2) \in {\equptoc{0}\!} }
\qquad
\infer{ (\unfold_\vspe(c_1),\unfold_\vspe(c_2)) \in \vspe.\fspr(\equptoc{n}) }
      { (c_1,c_2) \in {\equptoc{n+1}\!} }
\]

Intuitively, a natural number level for coinductive types corresponds to their \emph{depth},
building on the view of coinductive types as infinite depth trees discussed in \Cref{Coinductive Type}.
Specifically, two elements of a coinductive type are deemed equal up to level $n$ when their \emph{prefix subtrees up to depth $n$}
(\ie the subtrees truncated at depth $n$ from the root) coincide.
For $\stream$, this corresponds to the equality of the initial $n$ segments described in \Cref{Key idea}.

\begin{lemma}\label[lemma]{equptoc_mon}
  $\forall c_1,c_2\in \nu\fspsem{\vspe}$, $\forall n\le m\in\nat,\ c_1\equptoc{m}c_2 \implies c_1\equptoc{n}c_2$.
\end{lemma}

From \Cref{bisim_eq}, we can prove that equality up to all depths is equivalent to actual equality.

\begin{theorem}\label{full_equptoc}
  $\forall c_1,c_2\in \nu\fspsem{\vspe},\ (\forall n\in\nat,\ c_1\equptoc{n}c_2)\implies c_1=c_2$.
\end{theorem}

To extend this relation to mixture types represented by CTTs,
we define the \emph{level set} of a CTT recursively, 
as a generalization of $\nat$ used in $\equptoc{n}$.
Coinductive types have level set $\nat$, while mixture types have level sets that combine the level sets of their child CTTs. 

\[\begin{array}{@{}c@{\ }c@{}}
    \level : \CTT\ra\Type \qquad \\ 
    \level(\identCTT{\vspe}) \defeq \nat \qquad
  \level (\cttf\cttcompose\vec{\vctt}) \defeq \option(\depfun{a:\ctta}{\level(\vec{\vctt}(a))})
  \end{array}\]

We define the equality-up-to relation as an inductive proposition with three cases.
For ident CTTs (leaf case), we directly use the equality-up-to relation defined for coinductive types.
For CTTs with internal nodes, there are two subcases based on the level structure.
When the level is $\some(l)$,
any two elements in $F(\vec{\vctt})$ are equal up to level $\some(l)$
if they are equal up to $l(a)$ with respect to each dimension $a$ in $\ctta$.
When the level is $\none$,
any two elements are equal up to level $\none$.
For $x_1,x_2\in\sem{\vctt}$ and $l\in\level(\vctt)$, we write $x_1 \equpto{l} x_2$ when $x_1$ and $x_2$ are equal up to $l$.
\[\frac{x_1,x_2\in\sem{\identCTT{\vspe}} \quad x_1\equptoc{n} x_2}{x_1\equpto{n} x_2}\quad
  \frac{x_1,x_2\in \sem{\cttf\cttcompose\vec{\vctt}}}{x_1\equpto{\none} x_2}\quad
  \frac{x_1,x_2\in \sem{\cttf\cttcompose\vec{\vctt}} \quad 
    (x_1,x_2) \in \cttf.\fspr(\lambda a.~ \equpto{l'(a)} )}
    {x_1\equpto{\some(l')} x_2}\]

For example, consider $\identCTT{\vspe_\stream} \gop{+} \identCTT{\vspe_\bintree}$,
which has level set $\option(\Bool\ra\nat)$, equivalent to $\option(\nat \times \nat)$.
Here $\inl \ s_1 \equpto{\some(n,\_)} \inl \ s_2$ holds if $s_1\equpto{n}s_2$,
and $\inr \ b_1 \equpto{\some(\_,n)} \inr \ b_2$ holds if $b_1\equpto{n}b_2$.
However, $\inl \ s$ and $\inr \ b$ are only equal up to $\none$,
since $\forall l'\in\Bool\ra\nat, \ (\inl\ s,\inr\ b) \notin \tconop{+}.\fspr(\lambda a.~\equpto{l'(a)})$.
Note that the option type here ensures that any two elements are equal up to at least one level---a 
property leveraged by the CHP framework.

As another example, consider $\identCTT{\vspe_\stream} \gop{\times} \identCTT{\vspe_\bintree}$, which also has a level set equivalent to $\option(\nat \times \nat)$.
Here all elements of $\sem{\identCTT{\vspe_\stream} \gop{\times} \identCTT{\vspe_\bintree}}$ possess the same pair structure, unlike the sum type case.
Consequently, $\some$ contributes no additional structural information.
In this instance, the inclusion of $\option$ creates redundancy within the level set, 
specifically between two levels $\none$ and $\some \ (0, 0)$.
To maintain the generality of our methodology, we universally incorporate $\option$
and address this redundancy through a dedicated mechanism utilizing the preorder within the level set.

\bfparagraph{Order on levels}
We define a minimal element $\bot$ and preorder $\lord$ on level sets.
Equality up to $\bot$ trivially holds for any two elements.
\[\bot \in \level(\identCTT{\vspe}) \defeq 0 \qquad \bot \in \level(\cttf\cttcompose\vec{\vctt}) \defeq \none\]
\begin{remark}
  $\forall x_1,x_2\in\sem{\vctt},\ x_1\equpto{\bot}x_2$.
\end{remark}

\begin{figure}[t]
\[\frac{l_1,l_2\in\level(\identCTT{\vspe}) \qquad l_1\le l_2 \text{ in } \nat}{l_1\lord l_2 \text{ in } \identCTT{\vspe}} \cdots(1)\]
\[\frac{l\in\level (\cttf\cttcompose\vec{\vctt})}{\none\lord l \text{ in } \cttf\cttcompose\vec{\vctt}} \cdots(2)\qquad 
  \frac{l_1',l_2'\in \depfun{a:\ctta}{\level(\vec{\vspe}(a))} \qquad \forall a:\ctta,\ l_1'(a) \lord l_2'(a)} 
    {\some(l_1')\lord\some(l_2') \text{ in } \cttf\cttcompose\vec{\vctt}}\cdots(3)\]
\[\frac{l'\in \depfun{a:\ctta}{\level(\vec{\vspe}(a))} \qquad 
    \forall a:\ctta,\ l'(a) \lord \bot \qquad 
    |\cttf.\vctn.\fctnS| = 1}
    {\some(l')\lord\none \text{ in } \cttf\cttcompose\vec{\vctt}}\cdots(4)\]
\caption{Inductive definition of the order on level sets via four rules}
\label{Definition of order in level set}
\end{figure}

The preorder $\lord$ is defined inductively using four rules in \Cref{Definition of order in level set}.
This ordering captures the relative strength of equivalences: 
$l_1\lord l_2$ means that $\equpto{l_2}$ is a stronger condition than $\equpto{l_1}$, as shown in \Cref{equpto_mon_l}. 
The rules for base cases (1, 2) and pointwise comparison (3) are straightforward in this sense. 
Rule (4) is designed specifically to resolve the $\option$ redundancy problem in product CTT levels discussed above. 
Under the specific condition that a type constructor has only one shape ($|\cttf.\vctn.\fctnS| = 1$)
and all its sub-levels are trivial ($\forall a:\ctta,\ l'(a) \lord \bot$),
this rule collapses $\some(l')$ into $\none$, \ie $\some(l')\lord\none$.
This effectively simplifies the level structure while maintaining soundness, as formalized by \Cref{lord_meaning}.

\begin{remark}
  $\forall l\in\level(\vctt),\ \bot\lord l$.
\end{remark}
\begin{lemma}\label[lemma]{equpto_mon_l}
  $\forall x_1,x_2\in \sem{\vctt}$, $\forall l_1, l_2\in\level(\vctt),\ l_1\lord l_2 \land x_1\equpto{l_2}x_2 \implies x_1\equpto{l_1}x_2$.
\end{lemma}
\begin{lemma}\label[lemma]{lord_meaning}
  $\forall l\in\level(\vctt)$, $l\lord \bot \implies \forall x_1,x_2\in\sem{\vctt}, x_1\equpto{l}x_2$.
\end{lemma}

\bfparagraph{Useful auxiliaries for levels}
To handle levels more intuitively, we define two auxiliaries for levels: 
one generating a level from a natural number, and the other converting a level into a simple function.

First, we define the function $\leveleachtxt : \nat\ra\level(\vctt)$. 
$\leveleachtxt(n)$, or shortly $\leveleach{n}$, represents viewing up to depth $n$ for each dimension.
\[\begin{array}{@{}r@{\ \ }c@{\ \ }l@{~}c@{\ \ }l@{~}} 
  \leveleach{n} &:& \level(\identCTT{\vspe}) &\defeq& n \\
  \leveleach{n} &:& \level(\cttf\cttcompose\vec{\vctt}) &\defeq& \some(\lambda a.~ \leveleach{n} : \level(\vec{\vctt}(a)))
\end{array}\]

Second, we can convert a level $l$ into a simple function $\levelfun{l} : \levelA{\vctt}\ra\nat$, 
by disregarding its $\option$ structure.
$\levelA{\vctt}$ can be understood as the full dimension set of $\vctt$.
\[\levelA{\vctt} \in \Type \quad
  \levelA{\identCTT{\vspe}} \defeq \unitset \quad 
  \levelA{\cttf\cttcompose\vec{\vctt}} \defeq \deppair{a:\ctta}{\levelA{\vec{\vctt}(a)}}\]
\[\begin{array}{@{}r@{\ \ }c@{\ \ }l@{~}l@{}}
  \levelfun{l} &:& \levelA{\vctt}\ra\nat\ &\text{for } l\in\level(\vctt)\\
  \levelfun{l} &\defeq& \lambda \_.~ l\ &\text{for } l\in\level(\identCTT{\vspe})=\nat\\
  \levelfun{l} &\defeq& \lambda \_.~ 0 \ &\text{for } l=\none\in\level(\cttf\cttcompose\vec{\vctt})\\
  \levelfun{l} &\defeq& \lambda (a, a').~ \levelfun{(l'(a))}(a') \qquad &\text{for } l=\some(l')\in\level(\cttf\cttcompose\vec{\vctt})
\end{array}\]

\bfparagraph{Equality from Equality-up-to}

The relationship between equality-up-to and equality is captured by \Cref{full_equptoc} for coinductive types.
We generalize this result to \Cref{full_equpto}, with the stronger version given in \Cref{full_equpto_unif}.

\begin{theorem}\label{full_equpto}
  $\forall x_1,x_2\in \sem{\vctt},\ (\forall l\in\level(\vctt),\ x_1\equpto{l}x_2)\implies x_1=x_2$.
\end{theorem}
\begin{theorem}\label{full_equpto_unif}
  $\forall x_1,x_2\in \sem{\vctt},\ (\forall n\in\nat,\ x_1\equpto{\leveleach{n}}x_2)\implies x_1=x_2$.
\end{theorem}

As a corollary of \Cref{full_equpto_unif}, we obtain a useful result for nondegenerate function CTTs.

\begin{corollary}\label{NonDgnFunCTT_dimeq}
  $\forall x_1,x_2\in \sem{\gdepfun{a:A}{\vec{\vctt}(a)}},\ a\in A,\ \\ 
  \indent\quad (\forall n\in\nat,\ x_1\equpto{l_{n,a}}x_2)  \implies x_1(a)=x_2(a)$ \
  where $l_{n,a}=\some(\lambda a'.~\ite{a\!=\!a'}{\leveleach{n}}{\bot})$
\end{corollary}

\subsection{Productivity}
\label{Productivity}
Now we formally define \emph{productivity} as relations between level sets, based on equality-up-to relations.
We use the notation $\productive{p}(f)$ when $f$ is $p$-productive (\ie $f$ has productivity $p$),
and $\productiveat{p}{l}(f)$ when $f$ is $p$-productive-at level $l$.
We write $p(l_1,l_2)$ for $(l_1,l_2) \in p$.
\[
\begin{array}{@{}r@{\ \ }c@{\ \ }l@{~}l@{}}
  \text{ For } \vctt_1,\vctt_2 \in \CTT,&\ f& : \sem{\vctt_1} \ra \sem{\vctt_2},\ 
  p\in \productivity{\vctt_1}{\vctt_2},\\
  \productiveat{p}{l}(f) &\defeq& \forall x_1,\ x_2,~ (\forall l'\in\level(\vctt_1), p(l',l) \implies x_1 \equpto{l'} x_2) 
  \implies f(x_1) \equpto{l} f(x_2) \\
  \productive{p}(f) &\defeq& \forall\, l\in\level(\vctt_2),~ \productiveat{p}{l}(f)
\end{array}
\]

\begin{lemma}\label[lemma]{productive_inclusion}
  For $\vctt_1,\vctt_2 \in \CTT,\ f \in \sem{\vctt_1} \ra \sem{\vctt_2},\ 
  p_1,p_2\in \productivity{\vctt_1}{\vctt_2}$, \\
\indent\quad if \ $(p_1\subseteq p_2)$ and $\productive{p_1}(f)$ then $\productive{p_2}(f)$.
\end{lemma}

While modeling productivities as relations between level sets offers generality, directly manipulating these relations is often cumbersome.
To make this more manageable, we define two useful productivities: \emph{standard uniform productivity} $\stdunif$ and \emph{standard case productivity} $\stdcase$.

The standard uniform productivity $\stdunif(f)$ is a simple productivity expressed by a function $f : \nat \ra \option(\nat)$ that treats all dimensions identically.
\[\begin{array}{@{}r@{\ \ }r@{\ \ }l@{~}l@{}}
  \stdunif(f)& : \productivity{\vctt_1}{\vctt_2}\defeq \{(l_1,l_2) | \hfill{} &\\
  &(\forall n\in\nat,\ (l_2\lord\leveleach{n}) \implies \exists n_f,\ f(n)=\some(n_f) \land l_1\lord\leveleach{n_f}) \ \land\ \ &\cdots(1)\\
  &\neg(\forall n\in\nat,\ f(n)=\none)\ \land\ \ &\cdots(2)\\
  &(\forall m\in\nat,\ (\forall n,n_f\in\nat,\ f(n)=\some(n_f)\implies n_f\le m) \implies l_1\lord\leveleach{m})\ \} &\cdots(3)\\
  &f\!or\ f : \nat \ra \option(\nat)
\end{array}\]

In the definition of $\stdunif$,
condition (1) means that viewing the domain up to $f(n)$ suffices to determine the codomain up to $n$. 
When $f(n)=\none$, we do not check the domain.
Condition (2) ensures that for constant functions $f$ with value $\none$, domain checking is omitted.
Condition (3) establishes that if $f$ is bounded by $m$, we only consider the domain up to $m$.
We use the notation $\stdunifZ(m)$ for $\stdunif(\lambda n.\ \ite{n\!\!\geq\!\! m}{\some(n\!-\!m)}{\none})$.
The productivities of $\fold$ and $\unfold$ can then be expressed using this notation,
although these results will be refined in \Cref{Slot-leafed Type Tree}.

\begin{theorem}\label{productivity of fold unfold}
  $\productive{\stdunifZ(1)}(\fold_\vspe)$ and 
  $\productive{\stdunifZ(\minus1)}(\unfold_\vspe)$.
\end{theorem}

The standard case productivity $\stdcase(p_1,p_2)$ is a productivity that applies $p_1$ to the left component of the domain and $p_2$ to the right component,
with a shared codomain.
\[\begin{array}{@{}r@{\ \ }l@{\ \ }l@{~}l@{}}
  \stdcase(p_1,p_2) \in& \productivity{\vctt_1\gop{\times}\vctt_2}{\vctt_3}\\ 
  \defeq& \{ (\some(l'_{1,2}),\ l_3)\ |\ (p_1(l'_{1,2}\ \btrue,l_3) \lor l'_{1,2}\ \btrue = \bot)\\
  & \hspace{75pt} \land\ (p_2(l'_{1,2}\ \bfalse,l_3) \lor l'_{1,2}\ \bfalse = \bot) \} \\
  \text{for} &p_1\in\productivity{\vctt_1}{\vctt_3},\ p_2\in\productivity{\vctt_2}{\vctt_3}
\end{array}\]

We also define the trivial productivity $\allproductivity$, which every function satisfies.

\begin{lemma}[Trivial Productivity]\label{productive_cany}\quad\\
  \indent\quad $\forall f : \sem{\vctt_1} \ra \sem{\vctt_2}$, $\productive{\allproductivity}(f)$ where \ 
  $\allproductivity \defeq \{(l_1,l_2)\in \level(\vctt_1)\times\level(\vctt_2) \}$.
\end{lemma}

\subsection{Combination of Productivity}
\label{Combination of Productivity}

This subsection presents several productivity combination principles.
After establishing combination principles for three combinators in \Cref{Key idea},
we present additional examples of combinators along with their respective combination principles.

We first examine the productivity of function composition.
Consider $f : D \ra C$ and $g : C \ra E$, with heterogeneous productivities $p_1$ and $p_2$ respectively.
What is the productivity of $g \fcompose f$?
The values $g(f(x_1))$ and $g(f(x_2))$ are equal up to $l$, if $f(x_1)$ and $f(x_2)$ are equal up to $l'$ for all level $l'$ such that $p_2(l', l)$.
Also, $f(x_1)$ and $f(x_2)$ are equal up to $l'$, if $x_1$ and $x_2$ are equal up to $l''$ for all level $l''$ such that $p_1(l'', l')$.
Altogether, the final results $g(f(x_1))$ and $g(f(x_2))$ are equal up to $l$ if overall input $x_1$ and $x_2$ are equal up to $l''$ whenever $\exists l', p_1(l'', l') \wedge p_2(l', l)$.
It means that $g \fcompose f$ has the productivity $p_1 \pcompose p_2 = \{ (l'', l)\ |\ \exists l', p_1(l'', l') \wedge p_2(l', l) \}$, the relation composition of $p_1$ and $p_2$.
Note that the notation $\fcompose$ is left associative and $\pcompose$ is right associative. 

\begin{theorem}[Productivity of Composition]\label{productive_composition}\quad\\
\indent\quad For $f_1 : \sem{\vctt_1} \ra \sem{\vctt_2}$ and $f_2 : \sem{\vctt_2} \ra \sem{\vctt_3}$
  such that \ $ \productive{p_1}(f_1)$ and $\productive{p_2}(f_2)$,\\ 
\indent\quad $\productive{p_1 \pcompose p_2}(f_2 \fcompose f_1)$ where \\
\indent\quad $p_1 \pcompose p_2 \in \productivity{\vctt_1}{\vctt_3}$, \ 
  $p_1 \pcompose p_2 \defeq \{(l_1,l_3)\ |\ \exists l_2,\ p_1(l_1,l_2) \land p_2(l_2,l_3)\}$.
\end{theorem}

\begin{corollary}[Productivity of Repeated Compositions]\quad\\
\indent\quad For $f : \sem{\vctt} \ra \sem{\vctt}$
  such that \ $ \productive{p}(f)$,\\ 
\indent\quad $\productive{p^n}~(f^n)$ where 
  $p^n \in \productivity{\vctt}{\vctt}$, \ 
  $p^n \defeq p \pcompose p \pcompose \cdots \pcompose p$. 
\end{corollary}

Next, we investigate the function pairing, to find the productivity of $f \fpair g$
where $f : D \ra C_1$ has productivity $p_1$ and $g : D \ra C_2$ has productivity $p_2$.
Pairs $(f(x_1), g(x_1))$ and $(f(x_2), g(x_2))$ are equal up to $l = (l_1, l_2)$ if and only if both
$f(x_1)$ and $f(x_2)$ are equal up to $l_1$ and $g(x_1)$ and $g(x_2)$ are equal up to $l_2$.
$f(x_1)$ and $f(x_2)$ are equal up to $l_1$ if $x_1$ and $x_2$ are equal up to $l'$ for all level $l'$ such that $p_1(l', l_1)$, while
$g(x_1)$ and $g(x_2)$ are equal up to $l_2$ if $x_1$ and $x_2$ are equal up to $l'$ for all level $l'$ such that $p_1(l', l_2)$.
Therefore, the final results $(f \fpair g)(x_1)$ and $(f \fpair g)(x_2)$ are equal up to $l$ if $x_1$ and $x_2$ are equal up to $l'$ whenever $p_1(l', l_1) \vee p_2(l', l_2)$.
We denote this productivity $\{ (l', (l_1, l_2))\ |\ p_1(l', l_1) \vee p_2(l', l_2) \}$ by $p_1 \ppair p_2$;
now $f \fpair g$ has the productivity $p_1 \ppair p_2$.

\begin{lemma}[Productivity of Fpair]\quad\\
  \indent\quad For $f_1 : \sem{\vctt_1} \ra \sem{\vctt_2}$ and $f_2 : \sem{\vctt_1} \ra \sem{\vctt_3}$
    such that $\productive{p_1}(f_1)$ and $\productive{p_2}(f_2)$, \\
  \indent\quad let \ $f_1 \fpair f_2 : \sem{\vctt_1} \ra \sem{\vctt_2\gop{\times}\vctt_3}$, 
    $f_1 \fpair f_2 \defeq \lambda x. (f_1(x), \; f_2(x))$. \\
  \indent\quad Then $\productive{p_1 \ppair p_2}(f_1 \fpair f_2)$ where \\
  \indent\quad $p_1 \ppair p_2 \in \productivity{\vctt_1}{\vctt_2\gop{\times}\vctt_3}$,\\
  \indent\quad $p_1 \ppair p_2 \defeq 
    \{ (l_1,\some(l_{2,3}'))\ |\ 
    p_1(l_1,l_{2,3}'(\btrue)) \lor p_2(l_1,l_{2,3}'(\bfalse)) \}$
\end{lemma}

Similarly, we provide the productivity of function product $f \fproduct g$,
which applies $f$ and $g$ to the respective components of a pair.

\begin{lemma}[Productivity of Fproduct]\quad\\
\indent\quad For $f_1 \in \sem{\vctt_1} \ra \sem{\vctt_2}$ and $f_2 \in \sem{\vctt_3} \ra \sem{\vctt_4}$
  such that \ $\productive{p_1}(f_1)$ and $\productive{p_2}(f_2)$, \\
\indent\quad let \ $f_1 \fproduct f_2 \in \sem{\vctt_{1} \gop{\times} \vctt_{3}} \ra \sem{\vctt_{2} \gop{\times} \vctt_{4}}$, 
  $\ f_1 \fproduct f_2 \defeq \lambda(x_1, x_2). (f_1(x_1),\ f_2(x_2))$. \\
\indent\quad Then $\productive{p_1 \pproduct p_2}(f_1 \fproduct f_2)$ where\\
\indent\quad $p_1 \pproduct p_2 \in \productivity{\vctt_{1} \gop{\times} \vctt_{3}}{\vctt_{2} \gop{\times} \vctt_{4}}$,\\
\indent\quad $p_1 \pproduct p_2 \defeq
  \{ (\some(l'_{1,3}), \some(l'_{2,4}))\ |\ 
  (p_1(l_{1,3}'(\btrue),l_{2,4}'(\btrue))\ \land\ l_{1,3}'(\bfalse)=\bot)\ \lor$\ \\
\indent\hfill $(p_2(l_{1,3}'(\bfalse),l_{2,4}'(\bfalse))\ \land\ l_{1,3}'(\btrue)=\bot) \}$
\end{lemma}

The productivity of $\const(a)$ is simple.
Regardless of the inputs $x_1$ and $x_2$, the outputs of the function $\const(a)(x_1)$ and $\const(a)(x_2)$ are both identical to $a$, thereby equal up to any level.
Since achieving output equivalences require no conditions on the input levels, the productivity is $\Empty$.

\begin{lemma}[Productivity of Constant Function]\quad\\
\indent\quad For $x \in \sem{\vctt_{2}}$, let $\const(x) : \sem{\vctt_{1}} \ra \sem{\vctt_{2}} \defeq \lambda \_.~ x$. 
  Then \ $\productive{\Empty}(\const(x))$.
\end{lemma}

\begin{remark}
  For functions $f : \sem{\vctt_1}\ra\sem{\vctt_2}$ with no expected productivity beyond the trivial $\allproductivity$, 
  we use the notation $\cany{f}{\vctt_1}{\vctt_2}$ (or simply $\cany{f}{}{}$, omitting the $\CTT$s). 
  For example, the tight productivity of a function with constant CTT domain is $\allproductivity$ unless the size of the image is $1$. 
  We regard $\cany{f}{\vctt_1}{\vctt_2}$ as a combinator since we can automatically compute its productivity by \Cref{productive_cany}.
\end{remark}

There are two basic combinators for exponent CTTs:
the function application $\ceapp$ and the currying function $\cecurry$.
Their definitions and combination principles are in \Cref{Combinators related to Exponential CTT}. 
In practice, the generalized forms $\cgeapp$ and $\cgecurry$ are often more convenient:
$\cgeapp$ applies a preprocessing function to an argument before function application,
and $\cgecurry$ copies and reorders arguments during currying.
These are essential for handling compositions involving functions with exponent CTTs as domain or codomain,
which cannot be achieved using only \Cref{productive_composition}.

\begin{lemma}[Productivity of Generalized Exponent Application]\quad\\
\indent\quad For $g: A_1\ra A_2$,\\
\indent\quad let \ $\cgeapp(g) : \sem{(A_2\gop{\ra}\vctt)\gop{\times}\constCTT{A_1}} \ra\sem{\vctt}$, 
  $\cgeapp(g)\defeq \lambda (f,x).~ f(g(x))$.\\
\indent\quad Then $\productive{p'}~(\cgeapp(f))$ for \\
\indent\quad $p'\in \productivity{(A_2\gop{\ra}\vctt)\gop{\times}\constCTT{A_1}}{\vctt}$, \\
\indent\quad $p' = \{(\some(\lambda b.~ \ite{b}{\some(l'_1)}{l_A}),\ l_2)\ |\ l_A\in\level(\constCTT{A_1}),\ l'_1\ \unit=l_2 \}$.
\end{lemma}

\begin{lemma}[Productivity of Generalized Exponent Curry]\quad\\
\indent\quad For $f: \sem{\constCTT{A}\gop{\times}(\vctt_1\gop{\times}\constCTT{A})}\ra\sem{\vctt_2}$
  such that \ $\productive{p}(f)$, \\
\indent\quad let \ $\cgecurry(f) : \sem{\vctt_1}\ra\sem{A\gop{\ra}\vctt_2}$, 
  $\cgecurry(f)\defeq \lambda x.~\lambda a.~ f(a,(x,a))$. \\
\indent\quad Then $\productive{p'}~(\cgecurry(f))$ for \\
\indent\quad $p'\in \productivity{\vctt_1}{A\gop{\ra}\vctt_2}$, \\
\indent\quad $p' = \{(l_1,\some(l'_2))\ |\ \exists l_A,\ l_A'\in\level(\!\!\constCTT{A}\!),\ p(\some(\lambda b.~\ite{b}{l_A}{\some(\lambda b'.~\ite{b'}{l_1}{l_A'})}),\ l'_2\ \unit) \}$.
\end{lemma}

More combination principles with combinators such as commutator or associator are in \Cref{Combination Principles of First-order Productivity}.

\subsection{Fixed Point}
\label{Fixed Point}

Finally, we obtain the fixed point theorem for productive functions.

\begin{theorem}[Fixed Point of 1-Productive Function]\label{Fixed Point of One Productive Function} \quad\\
  \indent\quad For $\vctt\in\CTT$ such that $\sem{\vctt}\neq\Empty$ and $f : \sem{\vctt} \ra \sem{\vctt}$, \\
  \indent\quad if $\ \productive{\stdunifZ(1)}(f)$ then $f$ has a unique fixed point. 
\end{theorem}
\begin{corollary}
  For $A\in\Type$, $\vspe\in\SPE$ such that $A=\Empty\ \lor\ \sem{\identCTT{\vspe}}\neq\Empty$, \\
\indent\quad $\forall f : \sem{A\gop{\ra}\!\identCTT{\vspe}} \ra \sem{A\gop{\ra}\!\identCTT{\vspe}}$, 
  if $\ \productive{\stdunifZ(1)}(f)$ then $f$ has a unique fixed point.
\end{corollary}

It is noteworthy that these theorems require the codomain to be nonempty.
This may initially appear paradoxical, as establishing an example of $\sem{\vctt}$ often depends on a corecursive definition,
which is what this theorem aims to accomplish.
This apparent circularity is resolved by \Cref{coin_prime_moverU}.
In most cases, the conditions of this lemma are trivially satisfied by taking $X = \unitset$. 

\begin{lemma}\label[lemma]{coin_prime_moverU}
  For $\vspe\in\SPE$, if \ $\exists\ X:\Type,\ X\neq\Empty$ and $\exists\alpha : X \ra \fspsem{\vspe}(X)$, then $\sem{\identCTT{\vspe}}\neq\Empty$.
\end{lemma}

Furthermore, we extend the fixed point theorem to encompass functions that do not even have $\stdunifZ(1)$ productivity.
The core intuition underlying this generalization is that the fixed point of the generating function is incrementally constructed ad infinitum,
provided that an infinite number of constructors will eventually be produced by repeated applications of the function.
The most common examples are $1$-productive functions, which introduce at least one constructor in each iteration.
However, a function might produce constructors more slowly, insufficient to classify it as $1$-productive, yet adequate to construct a fixed point.

This relaxed concept of productivity is formalized as \emph{semi-well-foundedness}.
Semi-well-foundedness is a weaker version of well-foundedness, defined by the \emph{semi-accessibility relation} $\sacc$, 
just like well-foundedness defined by the accessibility relation $\acc$. 
Using this idea, even the partial function $\filter(P) : \stream\ra\stream$ can be defined by CHP (see \Cref{example filter}).  
The formal definition and details about semi-accessibility are in \Cref{Well-Founded-S Relation as a Productivity}.

One example of semi-well-founded productivity, except for $\stdunifZ(1)$, is \emph{$1/n$ productivity}.
For $f : \sem{\vctt} \ra \sem{\vctt}$, $f$ is said to be $1/n$-productive if, for some productivity $p$, $\productive{p}(f)$ and $p^n\subseteq \stdunifZ(1)$.
Mutual corecursive functions can be represented by the fixed point of a $1/n$-productive generating function.
See \Cref{example pingpong} for an example.

\subsection{Slot-leafed Type Tree}
\label{Slot-leafed Type Tree}

In this subsection, we establish more precise productivities of constructors and destructors.
We already provided the productivities of $\fold_\vspe$ and $\unfold_\vspe$ in \Cref{productivity of fold unfold}.
However, these productivities are not sufficiently useful in practice.
We will explain the reason and then demonstrate the solution using \emph{slot-leafed type trees}.

The type of $\fold_\vspe$ considered in \Cref{productivity of fold unfold} is $\sem{\selfCTT{\vspe}} \ra \sem{\identCTT{\vspe}}$, 
and that of $\unfold_\vspe$ is $\sem{\identCTT{\vspe}} \ra \sem{\selfCTT{\vspe}}$. 
However, the CTT $\selfCTT{\vspe}$ in this formulation loses important structural information.
For example, consider $\vspe$ such that $\fspsem{\vspe}=\lambda X.~ \nat\times X$, which generates the coinductive type $\stream = \sem{\identCTT{\vspe}}$. 
The desired constructor for $\stream$ is $\cons : \sem{\constCTT{\nat}\gop{\times}\identCTT{\vspe}}\ra\sem{\vspe}$, 
which has a product $\CTT$ domain, allowing us to discuss the levels of $\constCTT{\nat}$ and $\identCTT{\vspe}$ separately. 
In contrast, we cannot consider the level of $\constCTT{\nat}$ for $\fold_\vspe : \sem{\selfCTT{\vspe}} \ra \sem{\identCTT{\vspe}}$.
This motivates the need for appropriate \emph{converters} to transform between $\selfCTT{\vspe}$ and $\constCTT{\nat}\gop{\times}\identCTT{\vspe}$.

To construct these converters, we introduce the concept of \emph{slot-leafed type trees} (STTs).
By constructing SPEs through STTs rather than directly,
we achieve systematic SPE generation while simultaneously obtaining the required converters.
STTs are formally defined inductively as follows:
\[
\begin{array}{@{}r@{\ \ }c@{\ \ }l@{~}l@{}}
  \text{(Leaf case)}& \qquad \identSTT{} &\in& \STT \qquad \\
  \text{(Internal node case)}& \qquad \cttf \sttcompose \vec{\vstt}  &\in& \STT \qquad \text{for } \ctta\in\Type,\ \cttf\in\SPF(\ctta),\ \vec\vstt\in \ctta\ra\STT
\end{array}
\]

The first constructor $\identSTT{}$ represents the leaf case,
while the second constructor $\sttcompose$ represents the internal node case with $\ctta$-indexed children $\vec{\vstt}$.
Note that a slot $\identSTT{}$ in the leaf case contains no further information;
it will later be replaced by the coinductive type (which is the fixed point of the generated $\SPE$)
when we establish the semantics of an STT as a type.

We present several examples of $\STT$ below:
\[
\begin{array}{@{}r@{\ \ }c@{\ \ }l@{~}l@{}}
  \vstt_1 \cop{\times} \vstt_2 &\defeq& \tconop{\times} \sttcompose (\lambda (b\in\Bool).~ \ite{b}{\vstt_1}{\vstt_2}) 
  \quad &\text{ for } \vstt_1,\vstt_2\in\STT
  \\
  \vstt_1 \cop{+} \vstt_2 &\defeq& \tconop{+} \sttcompose (\lambda (b\in\Bool).~ \ite{b}{\vstt_1}{\vstt_2}) 
  \quad &\text{ for } \vstt_1,\vstt_2\in\STT
  \\
  \cop{\List}\,\vstt &\defeq& \tconop{\List} \sttcompose (\lambda (\_\in\unitset).~ \vstt)
  \quad &\text{ for } \vstt\in\STT
  \\
  A\cop{\ra}{\vstt} &\defeq& \tconop{A\ra} \sttcompose (\lambda (\_\in\unitset).~\ \vstt)
  \quad &\text{ for } A\in\Type,\ \vstt\in\STT
  \\
  \constSTT{A} &\defeq& \tconop{A} \sttcompose (\lambda (\_\in\Empty).~ \_)
  \quad &\text{ for } A \in \Type
\end{array}
\]

The construction of an $\SPE$ from an $\STT$ proceeds recursively as follows:
\[\begin{array}{@{}c@{\ }}
  \text{for }\vstt\in\STT,\ \vspe_\vstt \in \SPE\\
  \vspe_{\identSTT{}} \defeq \tconop{\Id}\qquad
  \vspe_{\cttf\sttcompose\vec{\vstt}} \defeq \cttf\ \underline{\fcompose}\ (\lambda (a\in\ctta).\ \vspe_{\vec\vstt(a)})
\end{array}\]

We define a CTT $\selfCTT{\vstt}$ for each $\STT\ \vstt$,
which serves as the proper structural form for constructor domains and destructor codomains.
In the $\stream$ example above, $\selfCTT{\vstt} =\!\!\! \constCTT{\nat}\gop{\times}\identCTT{\vspe_\vstt}$
for $\vstt =\!\!\! \constSTT{\nat} \cop{\times} \identSTT{}$.
Crucially, while providing more structure, 
$\selfCTT{\vstt}$ has the same corresponding type as $\selfCTT{\vspe_\vstt}$, as formalized in \Cref{flatCTTU_eq}.
\[\begin{array}{@{}c@{\ }}
  \selfCTT{\vstt} \defeq \helper(\vstt,\ \vspe_\vstt) \text{ where } \\
  \helper(\identSTT{},\ \vspe)\defeq \identCTT{\vspe}\qquad
  \helper(\cttf \sttcompose \vec{\vstt},\ \vspe)\defeq \cttf \cttcompose (\lambda a.~ \helper(\vec\vstt(a),\ \vspe))
\end{array}\]

\begin{lemma} \label[lemma]{flatCTTU_eq}
  $\forall \vstt,\ \sem{\selfCTT{\vspe_\vstt}} = \sem{\selfCTT{\vstt}}$
\end{lemma}

By \Cref{flatCTTU_eq}, we define two converters $\convone$ and $\convtwo$ with the following properties.
These converters, in turn, are used to define our intended constructors and destructors through $\fold_\vstt$ and $\unfold_\vstt$.

\begin{theorem} 
  Let \ $\convone : \sem{\selfCTT{\vspe_\vstt}} \ra \sem{\selfCTT{\vstt}} \defeq \lambda x.~x$ \ and \   
  $\convtwo : \sem{\selfCTT{\vstt}} \ra \sem{\selfCTT{\vspe_\vstt}} \defeq \lambda x.~x$. \\
\indent\quad Then \ $\productive{\stdunifZ(0)}(\convone) \text{ and } \productive{\stdunifZ(0)}(\convtwo)$.
\end{theorem}
  
\begin{corollary}\label{productivity of folds unfolds}
  Let \ $\unfold_\vstt : \sem{\identCTT{\vspe_\vstt}} \ra \sem{\selfCTT{\vstt}} \defeq \convone \fcompose\ \unfold_{\vspe_\vstt}$ \ and \\  
\indent\quad $\fold_\vstt : \sem{\selfCTT{\vstt}} \ra \sem{\identCTT{\vspe_\vstt}} \defeq \fold_{\vspe_\vstt} \fcompose\ \convtwo$. \\
\indent\quad Then \ $\productive{\stdunifZ(\minus1)}(\unfold_\vstt) \text{ and } \productive{\stdunifZ(1)}(\fold_\vstt)$.
\end{corollary}

\begin{corollary}\label{productivity of stream cons head tail}
  In \ $\stream$, \ $\cons=\fold_\vstt$, \ $\head = \fst \fcompose \unfold_\vstt$ and \\
\indent\quad $\tail= \snd \fcompose \unfold_\vstt$, thereby \\
\indent\quad $\productive{\stdunifZ(1)}(\cons)$, $\productive{\stdunifZ(\minus1)}(\head)$ and $\productive{\stdunifZ(\minus1)}(\tail)$.
\end{corollary}
\section{Second-order Productivity}
\label{Second-order Productivity}

\subsection{Second-order Productivity}

In this section, we introduce \emph{second-order productivity},
a framework that enables automatic derivation of productivity proofs for corecursive functions defined as fixed points.
This addresses a restriction of first-order productivity:
establishing the productivity of component functions that are themselves fixed points (such as $\smapS$ with generating function $F_\smapS$) is challenging.
We illustrate this difficulty in detail through a concrete example, before presenting our solution using second-order productivity.

Consider $\sone = \cons \ 0 \ (\smapS \ \sone)$ from \Cref{Introduction}.
If we know that $\smapS$ is $0$-productive, 
it is straightforward to show that $\Fsone\defeq \lambda x.\ \cons \ 0 \ (\smapS \ x)$ is $1$-productive, 
ensuring that $\sone$ is well-defined as the unique fixed point of $\Fsone$.
However, proving the productivity of $\smapS$ using combination principles is problematic.
When we attempt to apply combination principles to the fixed point equation of $\smapS$,
we encounter a circular dependency:
the productivity of $\smapS$ depends on the productivity of $\smapS$ itself,
yielding the tautology $\productive{\stdunifZ(0)}(\smapS) \ra \productive{\stdunifZ(0)}(\smapS)$.

A possible strategy is to utilize the productivity-at relation and apply induction on the level set of $\stream$ (which in this case is $\nat$).
To prove $\productiveat{\stdunifZ(0)}{0}(\smapS)$,
it suffices to show that $\forall x_1,x_2\in\stream, x_1\equpto{0}x_2$,
which is trivial by definition.
In the inductive case, assuming that $\productiveat{\stdunifZ(0)}{n}(\smapS)$,
it is possible to prove that $\productiveat{\stdunifZ(0)}{n\!+\!1}(F_\smapS\ \smapS)$, 
by slightly modifying the combination principles to handle productivity-at instead of productivity.

We develop this approach into the notion of second-order productivity.
Given a target productivity (such as $\stdunifZ(0)$),
establishing the corresponding second-order productivity for the generating function (such as $F_\smapS$) automatically yields
both the corecursive function as a fixed point (such as $\smapS$)
and the proof that it possesses this target productivity.
We term this concept second-order since it serves as a generative principle for the original productivity,
in contrast to the previously discussed first-order productivity.

Informally, second-order productivity can be understood as the concept of preserving the productivity-at property.
For the formal definition, we introduce the notion of \emph{equal-productivity-at},
an extension of productivity-at.
The notation $f_1\eqpa{l}{p}f_2$ denotes that $f_1$ and $f_2$ are $p$-equally-productive-at level $l$.
\[\begin{array}{@{}c@{}}
  f_1\eqpa{l}{p}f_2 \defeq \forall x_1,x_2\in\sem{\vctt_1},\ (\forall l'\in\level(\vctt_1),\ p(l',l)\!\implies\! x_1\equpto{l'}x_2)\implies f_1(x_1)\equpto{l} f_2(x_2) \\
  \text{for }\ f_1,f_2 : \cFun{1}{2},\ p\in\productivity{\vctt_1}{\vctt_2},\ l\in\level(\vctt_2)
\end{array}\]

\begin{lemma}
  $\forall f_1,f_2 : \cFun{1}{2},\ p\in\productivity{\vctt_1}{\vctt_2},\ l\in\level(\vctt)$,\\
\indent\quad $f_1\eqpa{l}{p}f_2$ if and only if $\productiveat{p}{l}(f_1)$ and $\forall x,\ f_1(x)\equpto{l}f_2(x)$.
\end{lemma}

A generating function $F$ is \emph{second-order productive},
denoted $\soproductive{}$, if it preserves equal-productivity-at relations $\eqpa{l}{p}$.
More specifically, $\soproductive{(p_1,p_2,q)}(F)$ means that
$F(f_1) \eqpa{l_4}{p_2} F(f_2)$ holds for the output whenever $f_1 \eqpa{l_2}{p_1} f_2$ holds for the input
for all $l_2$ related to $l_4$ by $q$.
The definition of $\soproductive{(p_1,p_2,q)}(F)$ is as follows:
\[\begin{array}{@{}r@{\ }l@{}}
  \soproductive{(p_1,p_2,q)}(F) \defeq& \forall l_4\in\level(\vctt_4),\ f_1,f_2 : \cFun{1}{2},\\ 
\indent\quad &(\forall l_2\in\level(\vctt_2),\ q(l_2,l_4)\!\implies\! f_1\eqpa{l_2}{p_1}f_2)\implies F(f_1)\eqpa{l_4}{p_2} F(f_2) \\
\indent\quad \text{for}\ F :& \cFFun{1}{2}{3}{4},\ q\in\productivity{\vctt_2}{\vctt_4},\\
\indent\quad p_1\in& \productivity{\vctt_1}{\vctt_2},\ p_2\in \productivity{\vctt_3}{\vctt_4}
\end{array}\]

Here $p_1$ governs the domain of the generating function,
while $p_2$ governs the codomain of the generating function.
Therefore, in \Cref{so productive inclusion}, they exhibit opposite directions of inclusion.

\begin{lemma} \label[lemma]{so productive inclusion}
  For $F : \cFFun{1}{2}{3}{4}$ such that $\soproductive{(p_1,p_2,q)}(F)$,\ \\ 
\indent\quad$(p_1'\subseteq p_1)\land(p_2\subseteq p_2')\land(q\subseteq q') \implies \soproductive{(p_1',p_2',q')}(F)$.
\end{lemma}

\subsection{Fixed point of Second-order Productive Generating Function}\label{Fixed point of Second-order Productive Generating Function}

In this subsection, we present the fixed point theorem for second-order productivity.
As with the first-order case,
we obtain the unique fixed point of a generating function when it exhibits second-order productivity $\soproductive{(p_1,p_2,q)}$ for specific values of $p_1$, $p_2$ and $q$.
We first introduce the \emph{asymptotic stability condition},
which is one of several requirements that determine these specific values.
We then establish the second-order fixed point theorem.

The \emph{asymptotic stability condition} (ASC) is a property of productivities that ensures eventual stabilization.
Intuitively, this means that for each input level $l_1$,
the productivity $p$ does not distinguish sufficiently large output levels $l_2$.
This intuition is formally captured as follows:
\[\begin{array}{@{}r@{\ \ }c@{\ \ }l@{~}l@{}}
  \text{For } p&\in&\productivity{\vctt_1}{\vctt_2},\\ 
  \asc(p)&\defeq& \forall l_2 \in \level(\vctt_2),\ (\exists m\in\nat,\ l_2\lord\leveleach{m})\ \lor\\ 
  &&\hspace{65pt}(\forall l_1,\ (\exists\ a,\ \levelfun{(l_2)}(a)\!>\!0\ \land\ p(l_1,\ \leveleach{\levelfun{(l_2)}(a)})) \implies p(l_1,l_2))
\end{array}\]

ASC has been proven for standard productivities.

\begin{lemma}
  $\asc(\allproductivity)$. Also \ $\forall f,\ \asc(\stdunif(f))$.
\end{lemma}
\begin{lemma}
  $\forall p_1,p_2$, $(\asc(p_1) \land \asc(p_2)) \implies \asc(\stdcase(p_1,p_2))$.
\end{lemma}

\begin{lemma} \label[lemma]{findim_asc} 
  $\forall p\in\productivity{\vctt_1}{\vctt_2}$, \ $(|A_{\vctt_2}|<\infty) \implies \asc(p)$.
\end{lemma}

Now we present the fixed point theorem.

\begin{theorem}[Fixed Point of Second-order Productive Generating Function] \label{so_fp} \quad\\
  \indent\quad For $\vctt_1,\vctt_2\in \CTT$ such that $\cFun{1}{2}\neq\Empty$ and \\
  \indent\quad $F\in \cFFun{1}{2}{1}{2}$ such that $\soproductive{(p,p,\stdunifZ(1))}(F)$, \\ 
  \indent\quad $F$ has a unique fixed point $f\in\cFun{1}{2}$.\\
  \indent\quad Also $\productive{p}(f)$ if \ $\asc(p)$.
\end{theorem}

Note that even when the productivity of the fixed point is not needed,
second-order productivity can be applied instead of first-order productivity by proving $\soproductive{(\allproductivity,\allproductivity,\stdunifZ(1))}(F)$.
See \Cref{false so} for the formal result.

\begin{corollary}\label{false so}\quad\\
  \indent\quad For $\vctt_1,\vctt_2\in \CTT$ such that $\cFun{1}{2}\neq\Empty$ and $F : \cFFun{1}{2}{1}{2}$, \\
  \indent\quad if $\ \soproductive{(\allproductivity,\allproductivity,\stdunifZ(1))}(F)$ 
  then $F$ has a unique fixed point $f : \cFun{1}{2}$.
\end{corollary}

\subsection{Combination of Second-order Productivity}
\label{Combination of Second-order Productivity}

We present several examples of second-order function combinators, accompanied by their combination lemmas.
By applying these lemmas,
we can automatically derive the second-order productivity of functions constructed from these combinators.
Additional examples are provided in \Cref{Combination Principles of Second-order Productivity}.

\begin{lemma}[Second-order Productivity of Constant Generating Function]\quad\\
\indent\quad For $f : \cFun{3}{4}$ such that $\productive{p}(f)$, \\
\indent\quad let $\sfindep\ f : \cFFun{1}{2}{3}{4},\ \sfindep\ f\defeq \lambda \_.f $. \\
\indent\quad Then $\forall p_0,\ \soproductive{(p_0,p,\Empty)}~(\sfindep\ f)$.
\end{lemma}

\begin{lemma}[Second-order Productivity of Sfself]\quad\\
\indent\quad Let \ $\sfself : \cFFun{1}{2}{1}{2}$, $\sfself \defeq \lambda f.~ f$. \\
\indent\quad Then $\forall p,\ \soproductive{(p,\ p,\ \{(l,l)\ |\ l\in\level(\vctt_2)\})}~(\sfself)$.
\end{lemma}

\begin{lemma}[Second-order Productivity of Composition]\quad\\
\indent\quad For $F_1 : \cFFun{1}{2}{3}{4}$ and $F_2 : \cFFun{1}{2}{4}{5}$\\
\indent\qquad such that $\soproductive{(p_0,p_1,q_1)}(F_1)$ and $\soproductive{(p_0,p_2,q_2)}(F_2)$, \\
\indent\quad let \ $F_2 \sfcompose F_1 : \cFFun{1}{2}{3}{5}$, \\ 
\indent\qquad $F_2 \sfcompose F_1 \defeq \lambda f.~ (F_2\ f) \fcompose (F_1\ f)$. \\
\indent\quad Then $\soproductive{(p_0,\ (p_1\pcompose p_2),\ ((q_1\pcompose p_2)\por q_2))}~(F_2 \sfcompose F_1)$.
\end{lemma}

\begin{lemma}[Second-order Productivity of Sfpair]\quad\\
\indent\quad For $F_1 :\cFFun{1}{2}{3}{4}$ and $F_2 :\cFFun{1}{2}{3}{5}$\\
\indent\qquad such that $\soproductive{(p_0,p_1,q_1)}(F_1)$ and $\soproductive{(p_0,p_2,q_2)}(F_2)$, \\
\indent\quad let \ $F_1 \sfpair F_2 : (\cFun{1}{2})\ra(\sem{\vctt_3}\ra\sem{\vctt_4\gop{\times}\vctt_5})$, \\ 
\indent\qquad $F_1 \sfpair F_2 \defeq \lambda f.~ (F_1\ f)\fpair(F_2\ f)$. \\
\indent\quad Then $\soproductive{(p_0,\ (p_1\ppair p_2),\ (q_1 \ppair q_2))}~(F_1 \sfpair F_2)$.
\end{lemma}
\section{Examples}
\label{Examples}

In this section, we present examples of corecursive definitions formulated within the CHP framework.
After formalizing $\stream$ in \Cref{example stream},
we present five illustrative examples about $\stream$ in \Cref{example zeros}--\Cref{example pingpong}.
We also present coinductive type $\bintree$ and corecursive function $\bfs$ in \Cref{example bintree} and \Cref{example bfs}, 
which are adapted from the AmiCo paper \cite{blanchette17}. 
Additional examples, including the partial corecursive function $\filter$, 
are provided in the appendix (see \Cref{Examples of Defining Coinductive Types and Corecursive Functions}).
For each of these examples, we provide a combination diagram in \Cref{Combination Diagrams for Examples}
to aid understanding.

\subsection{Coinductive Type $\stream$} \label{example stream}

Here we define the coinductive type $\stream$ and outline its key properties.
\[\text{For }T\in\Type,\ \streamP(T)\in\STT\defeq \constSTT{T} \cop{\times} \identSTT{}. \ \ \streamC(T)\in\CTT\defeq {\identCTT{\vspe_{\streamP(T)}}}. \]

\begin{remark}
  $\vspe_{\streamP(T)} = \tconop{\times}\ \spfcompose\ (\lambda b.~ \ite{b}{\tconop{T}}{\tconop{\Id}})$. \ 
  $\fspsem{\vspe}_{\streamP(T)} = \lambda X.~ T\times X$.
\end{remark}
\begin{remark}
  If \ $T\neq\Empty$, then $\sem{\streamC(T)}\neq\Empty$ by \Cref{coin_prime_moverU}.
\end{remark}

We obtain the type by $\stream(T) = \sem{\streamC(T)}$.
We omit the type parameter $T$ when clear from context.
Also, we assume that $T$ is nonempty, ensuring that $\stream(T)$ is nonempty as well.

Next, from the default $\fold$ and $\unfold$ operations,
we derive a more user-friendly set of constructor and destructors.
\[\begin{array}{@{}r@{\ \ }c@{\ \ }l@{~}l@{}}  
  \head &\defeq& \fst \ \fcompose\ \unfold_\streamP &: \sem{\streamC}\ra \sem{\constCTT{T}} \\
  \tail &\defeq& \snd \ \fcompose\ \unfold_\streamP &: \sem{\streamC}\ra\sem{\streamC} \\
  \cons &\defeq& \fold_\streamP &: \sem{\constCTT{T}\gop{\times}\streamC}\ra\sem{\streamC} 
\end{array}\]

The productivities of the constructor and destructors, stated in \Cref{productivity of stream cons head tail},
are easily proved using the combination principles.

\subsection{Corecursive Function $\zeros$} \label{example zeros}
\begin{minipage}{0.65\textwidth}
A simple example of a corecursively defined stream is $\szero$ from \Cref{Introduction},
namely $0 :: 0 :: \cdots$, which consists of an infinite sequence of zeros.
This stream is denoted by $\zeros$, with the generating function $F_\zeros$.
\[F_\zeros : \stream\ \nat \ra \stream\ \nat \defeq 
\lambda x.~\cons(0,x).\]

Using combinators, $F_\zeros$ can be represented as follows.
\[F_\zeros : \sem{\streamC}\ra\sem{\streamC} = \cons\ \fcompose\ (\const(0) \fpair \id)\]

These combinations can be intuitively visualized through \emph{combination diagrams},
as illustrated in the figure (a) on the right.

\end{minipage}
\begin{minipage}{0.34\textwidth}
  \begin{center}
  \begin{tikzpicture}[
      box/.style={draw, rectangle, rounded corners, minimum height=0.8cm, minimum width=1.0cm, align=center},
      arr/.style={->, >=Stealth, thick},
      lbl/.style={font=\small}, 
    ]
    \matrix (m) [matrix of nodes, row sep=20pt, column sep=-10pt, nodes={anchor=center} ]
    {
    & |[box, name=t11]| $\streamC$ &\\
    |[box, name=t21]| $\constCTT{\nat}$ & ${\gop{\times}}$ & |[box, name=t22]| $ \streamC $ \\
    & |[box, name=t31]| $\streamC$ &\\
    };
    \draw [arr] (t11) -- (t21) node [midway, left, yshift=3pt] {$\const(0)$};
    \draw [arr] (t11) -- (t22) node [midway, right, yshift=4pt] {$\id$};
    \draw [arr] (t21) -- (t31);
    \draw [arr] (t22) -- (t31);
    \node [lbl, above of=t31, yshift=-5pt] {$\cons$};
  \end{tikzpicture} \\[-2mm]
  {\small\textsf{(a) Combination diagram of $F_\zeros$}}
\end{center}
\vspace*{1mm}
\end{minipage}

From this combinator representation, we acquire the productivity $(\Empty\ \ppair\ \{(l,l)\})\ \pcompose\ \stdunifZ(1)$ using the combination principles. 
With automatic computation by a single tactic, we can prove that this productivity is a subset of $\stdunifZ(1)$.
By applying \Cref{productive_inclusion} in conjunction with the fixed point theorem (\Cref{Fixed Point of One Productive Function}),
we obtain $\zeros$ as the unique fixed point of $F_\zeros$.

\subsection{Corecursive Function $\smap$} \label{example map}
The function $\smap$ is a map function from $\stream\ T_1$ to $\stream\ T_2$, 
a generalized version of $\smapS$ from \Cref{Introduction} ($\smapS = \smap \ (\lambda n.n\!+\!1)$). 
It is a basic example of a fixed point that can be obtained from a second-order productive generating function $F_\smap$.
Here, we first define $\smap$ using only first-order productivity,
and then redefine it using second-order productivity to obtain its productivity.
\[\begin{array}{@{}r@{\ }c@{~}l@{\ \ }l@{}}  
  \text{For } g : T_1\ra T_2,\ 
  F_\smap(g)& : &(\stream\ T_1\ra\stream\ T_2)\ra(\stream\ T_1\ra\stream\ T_2) \\
  F_\smap(g)&\defeq& 
\lambda f.~\lambda x.~\cons(g(\head\ x),\ f(\tail\ x)).
\end{array}\]

Using first-order combinators, $F_\smap$ can be represented as follows :
\[\begin{array}{@{}r@{\ }c@{~}l@{\ \ }l@{}}  
  F_\smap(g) & : & \sem{\sem{\streamC\ T_1}\gop{\ra}\streamC\ T_2}\ra\sem{\sem{\streamC\ T_1}\gop{\ra}\streamC\ T_2} \\
  F_\smap(g) &=& \cgecurry\ (\cons\ \fcompose\ (
    \cany{g\justcompose\head}{\constCTT{\sem{\streamC\ T_1}}}{\constCTT{T_2}} \fproduct
    \cgeapp(\tail)))
\end{array}\]

Note that the domain $\CTT$ of $\head$ is $\constCTT{\sem{\streamC\ T_1}}$, not $\streamC\ T_1$.
Since it is from the left side of the exponent CTTs, we do not consider its equality-up-to relations.

The productivity of $F_\smap(g)$ is automatically proven to be a subset of $\stdunifZ(1)$.
Therefore, we obtain the unique fixed point $\smap(g)$ of $F_\smap(g)$.

We can also represent $F_\smap$ using second-order combinators:
\[\begin{array}{@{}r@{\ }c@{~}l@{\ \ }l@{}}  
  F_\smap(g) & : & (\sem{\streamC\ T_1}\ra\sem{\streamC\ T_2})\ra(\sem{\streamC\ T_1}\ra\sem{\streamC\ T_2}) \\
  F_\smap(g) &=& (\sfindep\ \cons) \sfcompose
    (\sfindep(\cany{g}{}{}\fcompose\head) \sfpair (\sfself \sfcompose (\sfindep\ \tail)))
\end{array}\]

After specifying the expected productivity $\stdunifZ(0)$ for $\smap(g)$,
we can automatically prove that $\soproductive{(\stdunifZ(0),\stdunifZ(0),\stdunifZ(1))}(F_\smap(g))$
using \Cref{so productive inclusion}.
Then we obtain the unique fixed point of $F_\smap(g)$, which is $\productive{\stdunifZ(0)}$ by \Cref{so_fp}.
Note that we obtain the same $\smap(g)$ from both methods since the fixed point theorems guarantee uniqueness.

\subsection{Corecursive Function $\growing$} \label{example growing}
Now we define the corecursive function $\growing$ using $\smap$, which is the same as $\sone$ from \Cref{Introduction}. 
\[
\begin{array}{@{}r@{\ }c@{~}l@{\ }c@{\ \ }l@{\ \ }}  
  F_\growing &:& \stream\ \nat \ra \stream\ \nat  &\defeq&
    \lambda x.~\cons(0,\ \smap(\lambda n.n\!+\!1)(x)).\\
  F_\growing &:& \sem{\streamC\ \nat}\ra\sem{\streamC\ \nat} &=& \cons\ \fcompose\ 
    (\const(0) \fpair \smap(\lambda n.n\!+\!1))
\end{array}
\]

The productivity of $F_\growing$ is automatically proven to be a subset of $\stdunifZ(1)$.
Therefore, we obtain the unique fixed point $\growing$ of $F_\growing$.

\subsection{Corecursive Function $\zip$} \label{example zip}
$\zip(g)$ is a corecursive function that combines two streams into a single stream via $g: (T\times T) \ra T$. 
\[\begin{array}{@{}r@{\ }c@{~}l@{\ \ }l@{}}  
  F_\zip(g)&:&((\stream\ T\times \stream\ T)\ra\stream\ T)\ra((\stream\ T\times \stream\ T)\ra\stream\ T) \\
  F_\zip(g)&\defeq& 
\lambda f.~\lambda (x,y).~\cons(g(\head\ x,\head\ y),\ f(\tail\ x,\tail\ y)).\\
  F_\zip(g) &:& (\sem{\streamC\gop{\times}\streamC}\ra\sem{\streamC})\ra(\sem{\streamC\gop{\times}\streamC}\ra\sem{\streamC}) \\
  F_\zip(g) &=& \sfindep(\cons) \sfcompose (
    \sfindep(\cany{g}{}{}\fcompose(\head \fproduct \head))\ \sfpair
    (\sfself \sfcompose \sfindep(\tail \fproduct \tail)))
\end{array}\]

After specifying the expected productivity $\stdunifZ(0)$ for $\zip(g)$,
we automatically prove that $\soproductive{(\stdunifZ(0),\stdunifZ(0),\stdunifZ(1))}(F_\zip(g))$.
Then we obtain the unique fixed point $\zip(g)$ of $F_\zip(g)$, which is also $\productive{\stdunifZ(0)}$.

\subsection{Corecursive Function $\pingpong$} \label{example pingpong}
The function $\pingpong(g_1, g_2) : T\ra(\stream\ T\times\stream\ T)$ is an example which shows the usefulness of $1/2$-productivity.
Mathematically, the function is equivalent to a pair of mutual corecursive functions 
$f_1(x)\defeq f_2(g_1(x))$ and $f_2(x)\defeq \cons(x,\ \zip(g_2)(f_1(x),f_2(x)))$,
where $\pingpong(g_1, g_2)(x)=(f_1(x),f_2(x))$. 
\[\begin{array}{@{}r@{\ }c@{~}l@{\ \ }l@{}}  
  F_\pingpong(g_1,g_2)&:& 
  (T\ra(\stream\ T\times\stream\ T))\ra (T\ra(\stream\ T\times\stream\ T)) \\
  F_\pingpong(g_1,g_2)&\defeq& 
  \lambda f.~ \lambda x.~ ( \snd{(f(g_1\ n))},\ \cons(x,\ \zip(g_2)(f\ n) ))\\
  F_\pingpong(g_1,g_2) &:& 
  \sem{T\gop{\ra}(\streamC\gop{\times}\streamC)} \ra \sem{T\gop{\ra}(\streamC\gop{\times}\streamC)} & \\
  F_\pingpong(g_1,g_2) &=& \cgecurry(
    (\snd\fcompose \cgeapp(g_1) \fcompose\snd) \fpair
    (\cons\fcompose(\id \fproduct (\zip(g_2)\fcompose\ceapp))))
\end{array}\]
We can automatically prove that $F_\pingpong(g_1,g_2)$ is $1/2$-productive by a single tactic. 
Therefore we get the unique fixed point $\pingpong(g_1,g_2)$ of $F_\pingpong(g_1,g_2)$.

\subsection{Coinductive Type $\bintree$} \label{example bintree}
We define coinductive type $\bintree$, which is a type of binary trees with infinte depth.
\[\text{For }T\in\Type,\ \bintreeP(T)\in\STT\defeq (\constSTT{T} \cop{\times} \identSTT{}) \cop{\times} \identSTT{}. \ \ \bintreeC(T)\in\CTT\defeq {\identCTT{\vspe_{\bintreeP(T)}}}. \]
\begin{remark}
  $\vspe_{\bintreeP(T)} = \tconop{\times}\ \spfcompose\ (\lambda b.~ \ite{b}
    {\tconop{\times}\ \spfcompose\ (\lambda b'.~ \ite{b'}{\tconop{T}}{\tconop{\Id}})}
    {\tconop{\Id}})$. \\ 
  \indent\quad$\fspsem{\vspe}_{\bintreeP(T)} = \lambda X.~ T\times X\times X$.
\end{remark}
\begin{remark}
  If \ $T\neq\Empty$, then $\sem{\bintreeC(T)}\neq\Empty$ by \Cref{coin_prime_moverU}.
\end{remark}
Then $\bintree(T) = \sem{\bintreeC(T)}$.
We omit the type parameter $T$ when clear from the context. 

Next, from the default $\fold$ and $\unfold$ operations,
we derive a more user-friendly set of constructor and destructors.
\[\begin{array}{@{}r@{\ \ }c@{\ \ }l@{~}l@{}}  
  \bhead &\defeq& \fst \ \fcompose\ \fst \ \fcompose\ \unfold_\bintreeP &: \sem{\bintreeC}\ra \sem{\constCTT{T}} \\
  \bleft &\defeq& \snd \ \fcompose\ \fst \ \fcompose\ \unfold_\bintreeP &: \sem{\bintreeC}\ra\sem{\bintreeC} \\
  \bright &\defeq& \snd \ \fcompose\ \unfold_\bintreeP &: \sem{\bintreeC}\ra\sem{\bintreeC} \\
  \bcons &\defeq& \fold_\bintreeP &: \sem{(\constCTT{T}\gop{\times}\bintreeC)\gop{\times}\bintreeC}\ra\sem{\bintreeC} 
\end{array}\]

The productivities of the constructor and destructors are automatically proved.
\begin{lemma}
  $\productive{\stdunifZ(\minus1)}(\bhead)$, $\productive{\stdunifZ(\minus1)}(\bleft)$,
  $\productive{\stdunifZ(\minus1)}(\bright)$ and
  
  $\productive{\stdunifZ(1)}(\bcons)$.
\end{lemma}

\subsection{Corecursive Function $\bfs$} \label{example bfs}
$\bfs$ is a well-studied corecursive function from Jones and Gibbons \cite{jones93},
which served as an illustrative example in \cite{blanchette17}.
The name stands for breadth-first labeling of nodes in a $\bintree$ using a stream of streams.
Thanks to the heterogeneity of our productivity notion, we can define $\bfs$ directly,
without separating it into $\bfsone$ and $\bfstwo$ as in \cite{blanchette17}.
\[\begin{array}{@{}r@{\ }c@{~}l@{\ \ }l@{}}  
  F_\bfs&:&((\stream\ (\stream\ T)) \ra \bintree\ T\ \times (\stream\ (\stream\ T))) \ra \\
  &&\qquad ((\stream\ (\stream\ T)) \ra \bintree\ T\ \times (\stream\ (\stream\ T)))\\
  F_\bfs&\defeq& 
    \lambda f.~\lambda x.~
      (\bcons(\head^2(x),\ \fst(f(\tail(x))),\ \fst(f(\snd(f(\tail(x)))))),\\
      &&\hspace{35pt} \cons(\tail(\head(x)),\ \snd(f(\snd(f(\tail(x))))))).\\
  F_\bfs &:& (\sem{\streamC\ (\stream\ T)} \ra \sem{(\bintreeC\ T)\ \gop{\times} (\streamC\ (\stream\ T))}) \ra \\
  &&\qquad (\sem{\streamC\ (\stream\ T)} \ra \sem{(\bintreeC\ T)\ \gop{\times} (\streamC\ (\stream\ T))})\\
  F_\bfs &=&
  ((\sfindep\ \bcons) \sfcompose (\\
  &&\quad (\sfindep\ (\cany{\head}{}{} \fcompose \head)\ \sfcompose 
    ((\sfindep\ \fst) \sfcompose \sfself \sfcompose (\sfindep\ \tail)))\ \sfpair \\
  &&\quad ((\sfindep\ \fst) \sfcompose \sfself \sfcompose (\sfindep\ \snd) \sfcompose \sfself \sfcompose (\sfindep\ \tail))\\
  && ))\ \sfpair \\
  && ((\sfindep\ \cons) \sfcompose (\\
  &&\quad \sfindep\ (\cany{\tail}{}{}\fcompose\head)\ \sfpair 
    ((\sfindep\ \snd) \sfcompose \sfself \sfcompose (\sfindep\ \snd) \sfcompose \sfself \sfcompose (\sfindep\ \tail))\\
  &&))
\end{array}\]

After specifying the expected productivity $\stdunifZ(0)$ for $\bfs$,
we automatically prove that \\$\soproductive{(\stdunifZ(0),\stdunifZ(0),\stdunifZ(1))}(F_\bfs)$.
Then we obtain the unique fixed point $\bfs$ of $F_\bfs$, which is also $\productive{\stdunifZ(0)}$.
\section{Implementation and Automation}
\label{Implementation and Automation}

Building upon the CHP theory, we have developed a Rocq library named \emph{Coco}.
This section presents practical code examples demonstrating Coco's capabilities and illustrates how users can extend the library.
Additionally, we discuss the axioms underlying Coco's implementation.

\subsection{Executable Code Example} \label{Executable Code Example}
Below is an executable code example of the corecursive function $\growing$ from \Cref{example growing}.
Note that in the code we use the notation $<\!\![\vctt]\!\!>$ for $\sem{\vctt}$ and $\vctt_1-\!-\!\!>\vctt_2$ for $\sem{\vctt_1}\ra\sem{\vctt_2}$.
$\streamR$ is a record containing the data for $\stream$,
and is implicitly cast to $\streamC$ in the code. 
It would be defined using $\STT$ by the user beforehand (see \Cref{example stream}), but is omitted here for simplicity.

\begin{multicols}{2}
\begin{coqdoccode}
\coqdocnoindent
\coqdockw{Module} \coqdockw{Type} \coqdocvar{StreamGrowingType} <: \coqdocvar{StreamMapType}.\coqdoceol
\coqdocindent{1.00em}
\coqdockw{Include} \coqdocvar{StreamMapType}.\coqdoceol
\coqdocindent{1.00em}
\coqdockw{Definition} \coqdocvar{growingF} (\coqdocvar{self} : \coqdocvar{stream} \coqdocvar{nat}) : \coqdocvar{stream} \coqdocvar{nat} :=\coqdoceol
\coqdocindent{2.00em}
\coqdocvar{cons} (0, \coqdocvar{map} \coqdocvar{S} \coqdocvar{self}).\coqdoceol
\coqdocindent{1.00em}
\coqdockw{Parameter} \coqdocvar{growing} : \coqdocvar{stream} \coqdocvar{nat}.\coqdoceol
\coqdocindent{1.00em}
\coqdockw{Parameter} \coqdocvar{growing\_fix} : \coqdocvar{growingF} \coqdocvar{growing} = \coqdocvar{growing}.\coqdoceol
\coqdocnoindent
\coqdockw{End} \coqdocvar{StreamGrowingType}.\coqdoceol
\coqdocemptyline
\coqdocnoindent
\coqdockw{Module} \coqdocvar{StreamGrowingImpl}.\coqdoceol
\coqdocnoindent
\coqdockw{Import} \coqdocvar{StreamImpl} \coqdocvar{StreamMapImpl}.\coqdoceol
\coqdocemptyline
\coqdocnoindent
\coqdockw{Definition} \coqdocvar{growingF} : \coqdocvar{streamR} \coqdocvar{nat} - -> \coqdocvar{streamR} \coqdocvar{nat} := \coqdoceol
\coqdocindent{1.00em}
\coqdocvar{cons} $\fcompose$ ( (\coqdocvar{cconst} (0 : <[\coqdocvar{nat}]>)) $\fpair$ (\coqdocvar{map} \coqdocvar{S})).\coqdoceol
\coqdocemptyline
\coqdocnoindent
\coqdockw{Program Definition} \coqdocvar{growing} : <[ \coqdocvar{streamR} \coqdocvar{nat} ]> := \coqdoceol
\coqdocindent{1.00em}
\coqdocvar{cfixpoint} \coqdocvar{growingF} \coqdocvar{\_}.\coqdoceol
\coqdocnoindent
\coqdockw{Next} \coqdockw{Obligation}. \coqdocvar{coauto}. \coqdockw{Qed}.\coqdoceol
\coqdocemptyline
\coqdocnoindent
\coqdockw{Lemma} \coqdocvar{growing\_fix} : \coqdocvar{growingF} \coqdocvar{growing} = \coqdocvar{growing}.\coqdoceol
\coqdocnoindent
\coqdockw{Proof}. \coqdocvar{coauto}. \coqdockw{Qed}.\coqdoceol
\coqdocnoindent
\coqdockw{Global Opaque} \coqdocvar{growing}.\coqdoceol
\coqdocemptyline
\coqdocnoindent
\coqdockw{End} \coqdocvar{StreamGrowingImpl}.\coqdoceol
\coqdocemptyline
\coqdocnoindent
\coqdockw{Module} \coqdocvar{StreamGrowing} : \coqdocvar{StreamGrowingType} :=\coqdoceol
\coqdocindent{1.00em}
\coqdocvar{StreamImpl} <+ \coqdocvar{StreamMapImpl} <+ \coqdocvar{StreamGrowingImpl}.\coqdoceol
\end{coqdoccode}
\end{multicols}

First, we define the module type $StreamGrowingType$,
which is an abstraction for users who are not familiar with CHP and the Coco library. 
In this module type, we simply state the generating function $F_\growing$ without combinators, and 
add $\mathtt{Parameter}$s for $\growing$ and its fixed point equation.

Next, we define $StreamGrowingImpl$, the \emph{implementation module}. 
We start by re-defining the generating function, this time using function combinators.
This approach allows us to define $\growing$ and prove its fixed point equation automatically. 
The $\mathtt{Next \ Obligation}$ part provides a default element of $\sem{\streamC\ \nat}$, 
which is also accomplished with a single tactic.

Once we have completed the implementation module, we then check that it conforms to the module type.
If the two definitions of $F_\growing$ satisfy definitional equality
(\ie equality can be proved merely by reduction), 
it will be compiled without error, and we are done. 
It is sometimes difficult to satisfy definitional equality due to the presence of induction or pattern matching
(\eg $F_\filter$ in \Cref{example filter} and $\mathtt{adds}$ in \Cref{example roundrobin}).
In such cases, one can define two generating functions in the implementation module:
one satisfying definitional equality, and the other stated with function combinators.
An equality lemma between them must then be proven.

\subsection{Extending library with new CTT, STT and function combinators} 
\label{Extending library with new CTT, STT and function combinators}

To fully leverage the automation capabilities of the Coco library,
users must employ proper CTTs, STTs and function combinators,
since improper usage may lead to automation failures or even mathematically erroneous subgoals.
For this reason, Coco has been designed for extensibility,
allowing users to define new CTTs, STTs and function combinators without modifying the core library source code.

For example, one might wish to define new coinductive types using a type constructor
$\tconop{\Stream}\defeq\lambda X.~\stream\ (X\ \unit)\in\SPF\unitset$, 
after defining a set of basic corecursive functions on $\stream$.
While proving $\tconop{\mathrm{Stream}}\in\SPF\unitset$ is nontrivial,
the subsequent definitions of a CTT and an STT are straightforward.
\[\begin{array}{@{}r@{\ }c@{\ \ }r@{\ }l@{\ }l@{\ }}
  \gop{\Stream}\,\vctt &\in \CTT, &\gop{\Stream}\,\vctt \defeq& \tconop{\Stream} \cttcompose (\lambda (\_\in\unitset).~ \vctt) \quad &\text{ for } \vctt\in\CTT\\
  \cop{\Stream}\,\vstt &\in \STT, &\cop{\Stream}\,\vstt \defeq& \tconop{\Stream} \sttcompose (\lambda (\_\in\unitset).~ \vstt) \quad &\text{ for } \vstt\in\STT
\end{array}\]

Considering such a CTT, $\smap$ in \Cref{example map} can be considered a function combinator:
\[\begin{array}{@{}l@{}}
  \smap\ (f) : \sem{\gop{\Stream}\vctt_1} \ra \sem{\gop{\Stream}\vctt_2} \hfill\text{ for } f : \sem{\vctt_{1}} \ra \sem{\vctt_{2}} \\
  \mathtt{sfsmap}(F) : (\cFun{1}{2})\ra(\sem{\gop{\Stream}\vctt_3}\ra\sem{\gop{\Stream}\vctt_4}),\ 
  \mathtt{sfsmap}(F)\defeq \lambda f.~ \smap\ (F(f)) \\
  \hfill \text{for } F : \cFFun{1}{2}{3}{4}
\end{array}\]

After proving combination principles for these combinators,
one can easily extend automation tactics to handle these function combinators alongside those in the default Coco library.

\subsection{Axioms}
\label{Axioms}

Coco is implemented using a set of formal axioms.
In addition to the bisimulation axiom presented in \Cref{Bisimulation},
the implementation utilizes \texttt{functional\_extensionality\_dep}, \texttt{\seqsplit{excluded\_middle\_informative}}, and \texttt{dependent\_unique\_choice}. 
For reasoning specifically about semi-accessibility introduced in \Cref{Fixed Point},
\texttt{constructive\_indefinite\_description} is also applied.
\section{Related Work}
\label{Related Work}

The inherent limitations of syntactic guardedness checking have motivated a rich body of work on alternative reasoning frameworks for corecursive definitions.
We have addressed the two most relevant works in the introduction.
Here, we broaden the scope to discuss other notable approaches from the literature.

The notion of achieving self-reference via the \emph{later} modality originates in the work of Nakano \cite{nakano00}.
This enables a principled, type-based form of corecursion, rather than relying on an obscure mechanism within a syntactic checker.
Nevertheless, this method is still fundamentally based on guardedness, albeit in an explicit form.
As a consequence, users must directly handle the modality within the types.
Guatto \cite{guatto18} introduces the concept of \emph{time warps}, a more fine-grained form of temporal modality.
The dependence is expressed with greater precision by means of a cocontinuous function of type $\omega + 1 \ra \omega + 1$.
However, time warps offer a coarser granularity than CHP.
Agda supports corecursion based on \emph{size types}, similarly employing monads to express size information \cite{hughes96,abel14}.
For example, streams are handled using a series of sized types, $ST_i$, which denotes the type of streams with at least $i$ computable elements.
Generally, these approaches involve modifying judgment rules in the type system, which limits their practical utility.

Our notion of equivalences based on levels is related to \emph{converging equivalence relations} (CER) \cite{matthews99} and \emph{complete ordered families of equivalences} (COFE) \cite{gianantonio02,chargueraud10}.
Chargu\'eraud employed these concepts mainly to define the contractivity of functions in order to determine whether a function possesses a fixed point.
The scope of that work, however, did not extend to the relationship between these concepts and modular reasoning about productivity.
Furthermore, while CER combinators were mentioned in \cite{matthews99}, a systematic method for assigning levels was generally underexplored.
Another difference is that the level sets in CHP are not necessarily well-founded, in contrast to the resolution spaces in CER or the carrier types in COFE.

There are other approaches to defining semantic productivity.
Rusu and Nowak \cite{rusu22} utilize \emph{complete partial orders} (CPOs) to establish productiveness.
However, this formulation is not quantitative, and their approach does not support compositionality.
Additionally, the reliance on CPOs typically introduces auxiliary elements like a bottom element $\bot$, which can complicate the reasoning process.
Kozen and Ruozzi \cite{kozen09} employ the framework of \emph{metric spaces} to define contractivity.
However, as Matthews \cite{matthews99} points out, handling continuity can present practical challenges.

\section{Limitations and Future Work}
\label{Future Work}

Several directions could further enhance automation in Coco:

\bfparagraph{Productivity inference for second-order productivity}
Currently, unlike in the first-order case, users must specify the
expected productivity of a fixed point before automatically proving
its second-order productivity. Automating this inference would
eliminate this manual step.

\bfparagraph{Automatic translation to combinator form}
Converting a generating function into composition of combinators
currently requires manual effort. We aim to develop a tool for Rocq
that would automatically translate generating functions to combinator
form.

\bfparagraph{Proof automation for recursion-corecursion}
Mixed recursion-corecursion examples,
such as the filter function in \Cref{example filter},
inherently require manual intervention.
We aim to improve automation to reduce this manual burden.

\begin{acks}
We thank Arthur Chargu\'{e}raud and the anonymous reviewers for their valuable feedback.
We also thank Minki Cho for sharing his prototype implementation of semantic polynomial functors \cite{spf},
upon which our implementation is based.
This work was supported in part by Samsung Research Funding Center of Samsung Electronics
under Project Number SRFC-IT2102-03, 
and by a National Research Foundation of Korea (NRF) grant 
funded by the Korean Ministry of Science and ICT (MSIT) (Grant No. RS-2024-00355459).
Chung-Kil Hur is the corresponding author.
\end{acks}

\bibliography{biblio}

\newpage
\appendix
\addcontentsline{toc}{section}{Appendices}
\section{CHP for Indexed Coinductive Types}
\label{CHP for Indexed Coinductive Types}

In this section, we extend CHP to handle indexed coinductive types.

\subsection{First-order Productivity for Indexed Coinductive Types}
\label{First-order Productivity for Indexed Coinductive Types}
We generalize the definition of CTT to handle general $\SPE(I)$, not just $\SPE\equiv\SPE\unitset$.
\[\begin{array}{@{}r@{}r@{\ \ }c@{\ \ }l@{~}l@{}}
  \text{(Leaf case)}& \qquad \identCTT{\vspe,i} &\in& \CTT \qquad &\text{for } \vspe\in\SPE(I),\ i\in I\\
  \text{(Internal node case)}& \qquad \cttf \cttcompose \vec{\vctt}  &\in& \CTT \qquad &\text{for } \ctta\in\Type,\ \cttf\in\SPF(\ctta),\ \vec\vctt\in \ctta\ra\CTT
\end{array}\]
\[\begin{array}{@{}c@{\ \ }c@{\ \ }l@{~}l@{}}
  \text{for }\vctt\in\CTT,\ \sem\vctt \in \Type\\
  \sem{\identCTT{\vspe,i}} \defeq (\nu\fspsem{\vspe})(i)\qquad
  \sem{\cttf \cttcompose \vec{\vctt}} \defeq \cttf (\lambda (a\in\ctta).\ \sem{\vec{\vctt}(a)})
\end{array}\]

Other CTTs remain the same, but $\selfCTT{\vspe}$ is changed to $\selfCTT{\vspe,i}$.
\[
\begin{array}{@{}r@{\ \ }c@{\ \ }l@{~}l@{}}
  \selfCTT{\vspe,i} \in \CTT &\defeq& (\vspe(i)) \cttcompose (\lambda i'.~ \identCTT{\vspe,i'})
  \quad &\text{ for } \vspe \in \SPE(I),\ i\in I
\end{array}
\]
\begin{remark}
  $\sem{\vspf \cttcompose \vec{\vctt}} = \vspf (\lambda a. \sem{\vec{\vctt} (a)})$. Then 
    $\sem{\selfCTT{\vspe,i}} = \fspsem{\vspe}(\nu\fspsem{\vspe})(i)$.
\end{remark}

The definition of equality-up-to, in coinductive types and in CTTs respectively, is modified as below. 
All the properties about level and equality-up-to still hold for the modified version.
\[
\infer{ c_1,c_2\in (\nu\fspsem{\vspe})(i) }
      { (c_1,c_2) \in {\equptoc{0}\!} }
\qquad
\infer{ (\unfold_\vspe(i)(c_1),\unfold_\vspe(i)(c_2)) \in (\vspe(i)).\fspr(\equptoc{n}) }
      { (c_1,c_2) \in {\equptoc{n+1}\!} }
\]
\[\level : \CTT\ra\Type \qquad
  \level(\identCTT{\vspe,i}) \defeq \nat \qquad
  \level (\cttf\cttcompose\vec{\vctt}) \defeq \option(\depfun{a:\ctta}{\level(\vec{\vctt}(a))})\]
\[\frac{x_1,x_2\!\in\!\sem{\identCTT{\vspe,i}}=(\nu \fspsem{\vspe})(i) \quad x_1\equptoc{n} x_2}{x_1\equpto{n} x_2}\quad
  \frac{x_1,x_2\!\in\! \sem{\cttf\cttcompose\vec{\vctt}}}{x_1\equpto{\none} x_2}\quad
  \frac{x_1,x_2\!\in\! \sem{\cttf\cttcompose\vec{\vctt}} \quad 
    (x_1,x_2) \!\in\! \cttf.\fspr(\lambda a.~ \equpto{l'(a)} )}
    {x_1\equpto{\some(l')} x_2}\]

Also we get the indexed version of \Cref{coin_prime_moverU}.
\begin{lemma}\label[lemma]{coin_prime_mover}
  For $\vspe\in\SPE(I)$ and $i\in I$, if \ $\exists\ X\in I\ra\Type$, \\
  \indent\quad $X(i)\neq\Empty$ and $\exists\alpha \in X \Ra \fspsem{\vspe}(X)$, then $\sem{\identCTT{\vspe,i}}\neq\Empty$.
\end{lemma}

Now all the definitions and properties about productivity are the same as those in \Cref{Productivity}--\Cref{Fixed Point}, except for STT.
Here, a family of coinductive types can be represented by a family of $\STT$s, that is, $I\ra\STT(I)$.
\[\begin{array}{@{}r@{\ \ }c@{\ \ }l@{~}l@{}}
  \identSTT{i~} &\in& \STT(I) \qquad &\text{for } i\in I\\
  \cttf \sttcompose \vec{\vstt}  &\in& \STT(I) \qquad &\text{for } \ctta\in\Type,\ \cttf\in\SPF(\ctta),\ \vec\vstt\in \ctta\ra\STT(I)
\end{array}\]

Below are examples of indexed $\STT$.
\[
\begin{array}{@{}r@{\ \ }c@{\ \ }l@{~}l@{}}
  \vstt_1 \cop{\times} \vstt_2 &\defeq& \tconop{\times} \sttcompose (\lambda (b\in\Bool).~ \ite{b}{\vstt_1}{\vstt_2}) 
  \quad &\text{ for } \vstt_1,\vstt_2\in\STT(I)
  \\
  \vstt_1 \cop{+} \vstt_2 &\defeq& \tconop{+} \sttcompose (\lambda (b\in\Bool).~ \ite{b}{\vstt_1}{\vstt_2}) 
  \quad &\text{ for } \vstt_1,\vstt_2\in\STT(I)
  \\
  \cop{\List}\,\vstt &\defeq& \tconop{\List} \sttcompose (\lambda (\_\in\unitset).~ \vstt)
  \quad &\text{ for } \vstt\in\STT(I)
  \\
  A\cop{\ra}{\vstt} &\defeq& \tconop{A\ra} \sttcompose (\lambda (\_\in\unitset).~\ \vstt)
  \quad &\text{ for } A\in\Type,\ \vstt\in\STT(I)
  \\
  \constSTT{A} &\defeq& \tconop{A} \sttcompose (\lambda (\_\in\Empty).~ \_)
  \quad &\text{ for } A \in \Type
\end{array}
\]

We can recursively generate a $\SPE$ from a set of $\STT$s.
\[\begin{array}{@{}c@{\ }}
  \text{for }\vec\vstt\in I\ra\STT(I),\ \vspe_{\vec\vstt} \in \SPE(I) \defeq \lambda i.~\helper(\vec\vstt(i)) \text{ where}\\
  \helper({\identSTT{i~}}) \defeq \tconop{\Id_{i:I}}\qquad
  \helper({\cttf\sttcompose\vec{\vstt}}) \defeq \cttf\ \underline{\fcompose}\ (\lambda a\in\ctta.\ \helper(\vec\vstt(a)))
\end{array}\]

We can also define a CTT $\selfCTT{\vec\vstt,i}$ of a $\vec\vstt\in I\ra\STT(I)$.
\[\begin{array}{@{}c@{\ }}
  \selfCTT{\vec\vstt,i} \defeq \helper(\vec\vstt(i),\ \vspe_{\vec\vstt}) \text{ where } \\
  \helper(\identSTT{i~},\ \vspe)\defeq \identCTT{\vspe,i}\qquad
  \helper(\cttf \sttcompose \vec{\vstt},\ \vspe)\defeq \cttf \cttcompose (\lambda a.~ \helper(\vec\vstt(a),\ \vspe))
\end{array}\]

\begin{lemma} \label[lemma]{flatCTT_eq}
  $\forall \vec\vstt\in I\ra\STT(I),\ i\in I,\ \sem{\selfCTT{\vspe_{\vec\vstt},i}} = \sem{\selfCTT{\vec\vstt,i}}$
\end{lemma}

By \Cref{flatCTT_eq}, we define two converters $\convone,\ \convtwo$ with following properties.
\begin{theorem} 
  Let \ $\convone(i) : \sem{\selfCTT{\vspe_{\vec\vstt},i}} \ra \sem{\selfCTT{\vec\vstt,i}} \defeq \lambda x.~x$ \ and \\   
\indent\quad $\convtwo(i) : \sem{\selfCTT{\vec\vstt,i}} \ra \sem{\selfCTT{\vspe_{\vec\vstt},i}} \defeq \lambda x.~x$. \\
\indent\quad Then \ $\productive{\stdunifZ(0)}(\convone(i)) \text{ and } \productive{\stdunifZ(0)}(\convtwo(i))$.
\end{theorem}
  
\begin{corollary}
  Let \ $\unfold_{\vec\vstt}(i) : \sem{\identCTT{\vspe_{\vec\vstt},i}} \ra \sem{\selfCTT{\vec\vstt,i}} \defeq \convone(i) \fcompose\ \unfold_{\vspe_{\vec\vstt}}(i)$ \ and \\  
\indent\quad $\fold_{\vec\vstt}(i) : \sem{\selfCTT{\vec\vstt,i}} \ra \sem{\identCTT{\vspe_{\vec\vstt},i}} \defeq \fold_{\vspe_{\vec\vstt}}(i) \fcompose\ \convtwo(i)$. \\
\indent\quad Then \ $\productive{\stdunifZ(\minus1)}(\unfold_{\vec\vstt}(i)) \text{ and } \productive{\stdunifZ(1)}(\fold_{\vec\vstt}(i))$.
\end{corollary}

\begin{remark}
  $\convone(i) \fcompose \convtwo(i) = \lambda x.~x$,\ \ $\convtwo(i) \fcompose \convone(i) = \lambda x.~x$.\\
\indent\quad So \ $\fold_{\vec\vstt}(i) \fcompose \unfold_{\vec\vstt}(i) = \lambda x.~x$,\ \ $\unfold_{\vec\vstt}(i) \fcompose \fold_{\vec\vstt}(i) = \lambda x.~x$.
\end{remark}
\subsection{Second-order Productivity for Indexed Coinductive Types}
\label{Second-order Productivity for Indexed Coinductive Types}

In first-order productivity, handling indexed coinductive types is relatively straightforward, since
most functions can be represented as $f : \depfun{j:J}\sem{\vec\vctt_1(j)}\ra\sem{\vec\vctt_2(j)}$
having the property $\forall j,\ \productive{p}(f(j))$.
However, second-order productivity cannot be directly applied to functions of the form $F : (\depfun{j:J}\sem{\vec\vctt_1(j)}\ra\sem{\vec\vctt_2(j)})\ra(\depfun{j:J}\sem{\vec\vctt_3(j)}\ra\sem{\vec\vctt_4(j)})$.
We therefore define \emph{indexed second-order productivity} as a generalization of second-order productivity.

We first extend the definition of $\eqpa{l}{p}$.
\[\begin{array}{@{}c@{}}
  f_1\eqpa{l}{p}f_2 \defeq \forall x_1,x_2\in\sem{\vec\vctt_1(j)},\ (\forall l'\in\level(\vec\vctt_1(j)),\ p(j)(l',l)\!\!\implies\!\! x_1\!\equpto{l'}\!x_2)\!\!\implies\! f_1(j)(x_1)\equpto{l} f_2(j)(x_2) \\
  \text{for }\ \vec{\vctt_1},\vec{\vctt_2}\in J\ra\CTT,\ j\in J,\ l\in\level(\vec{\vctt_2}(j)),\hfill\\
  \qquad f_1,f_2 : \cFunJ{J}{1}{2},\ p\in\depfun{j\!:\!J}\productivity{\vec{\vctt_1}(j)}{\vec{\vctt_2}(j)}\hfill
\end{array}\]

\begin{lemma}
  $\forall f_1,f_2 : \cFunJ{J}{1}{2}$,\\
\indent\quad $p\in\depfun{j\!:\!J}\productivity{\vec{\vctt_1}(j)}{\vec{\vctt_2}(j)},\ l\in\level(\vec{\vctt_2}(j))$,\\
\indent\quad $f_1\eqpa{l}{p}f_2$ if and only if $\productiveat{(p(j))}{l}(f_1(j))$ and $\forall x,\ f_1(j)(x)\equpto{l}f_2(j)(x)$.
\end{lemma}

We now present the definition of $\soproductiveJ$, the indexed version of $\soproductive$,
along with its associated fixed point theorem.
\[\begin{array}{@{}l@{\ }l@{}}
  \soproductiveJ{(p_1,p_2,q)}(F) \defeq \forall j_2\in J_2,\ l_4\in\level(\vec\vctt_4(j_2)),\ f_1,f_2 : \cFunJ{J_1}{1}{2},\\ 
\indent\quad (\forall j_1\in J_1,\ l_2\in\level(\vec\vctt_2(j_1)),\ q(j_1,j_2)(l_2,l_4)\!\implies\! f_1\eqpa{l_2}{p_1}f_2)\implies F(f_1)\eqpa{l_4}{p_2} F(f_2) \\
\indent\quad \text{for}\ F : \cFFunJ{J_1}{J_2}{1}{2}{3}{4},\\ 
\indent\qquad q\in \depfun{(j_1,j_2)\!:\!J_1\times J_2}\productivity{\vec{\vctt_2}(j_1)}{\vec{\vctt_4}(j_2)},\\
\indent\qquad p_1\in \depfun{j\!:\!J_1}\productivity{\vec{\vctt_1}(j)}{\vec{\vctt_2}(j)},\ p_2\in \depfun{j\!:\!J_2}\productivity{\vec{\vctt_3}(j)}{\vec{\vctt_4}(j)}
\end{array}\]

\begin{theorem}[Fixed Point of Indexed Second-order Productive Generating Function] \label{soj_fp} \quad\\
  \indent\quad For $\vec\vctt_1,\vec\vctt_2\in J\ra\CTT$ such that $\forall j,\ \sem{\vec{\vctt_1}(j)}\ra\sem{\vec{\vctt_2}(j)}\neq\Empty$ and \\
  \indent\quad $F : \cFFunJ{J}{J}{1}{2}{1}{2}$ \\
  \indent\qquad such that $\soproductiveJ{(p,p,\lambda\_.\stdunifZ(1))}(F)$, \\ 
  \indent\quad $F$ has a unique fixed point $f : \cFunJ{J}{1}{2}$.\\
  \indent\quad Also for all $j,\ \productive{p(j)}(f(j))$ if \ $\asc(p(j))$.
\end{theorem}

Although we have a fixed point theorem for indexed second-order productivity, 
the use of combination principles is less convenient for automation,
as it requires instantiating indexes in productivities.
We therefore introduce \emph{partially indexed second-order productivity} as an intermediate notion.

\[\begin{array}{@{}l@{\ }l@{}}
  \soproductivepJ{(p_1,p_2,q)}(F) \defeq \forall l_4\in\level(\vctt_4),\ f_1,f_2 : \cFunJ{J}{1}{2},\\ 
\indent\quad (\forall j\in J,\ l_2\in\level(\vec\vctt_2(j)),\ q(j)(l_2,l_4)\!\implies\! f_1\eqpa{l_2}{p_1}f_2)\implies F(f_1)\eqpa{l_4}{p_2} F(f_2) \\
\indent\quad \text{for}\ F : \cFFunpJ{1}{2}{3}{4},\\ 
\indent\qquad q\in \depfun{j\!:\!J}\productivity{\vec{\vctt_2}(j)}{\vctt_4},\\
\indent\qquad p_1\in \depfun{j\!:\!J}\productivity{\vec{\vctt_1}(j)}{\vec{\vctt_2}(j)},\ p_2\in \productivity{\vctt_3}{\vctt_4}
\end{array}\]

\begin{lemma} \label[lemma]{so_productivepJ_to_so_productiveJ}
For $F : \depfun{j_2:J_2}(\depfun{j_1:J_1}\sem{\vec\vctt_1(j_1)}\ra\sem{\vec\vctt_2(j_1)})\ra(\sem{\vec\vctt_3(j_2)}\ra\sem{\vec\vctt_4(j_2)})$, \\
\indent\quad $p_1\in \depfun{j\!:\!J_1}\productivity{\vec{\vctt_1}(j)}{\vec{\vctt_2}(j)}$,\\ 
\indent\quad $p_2\in \depfun{j\!:\!J_2}\productivity{\vec{\vctt_3}(j)}{\vec{\vctt_4}(j)}$, \\
\indent\quad $q\in \depfun{(j_1,j_2)\!:\!J_1\times J_2}\productivity{\vec{\vctt_2}(j_1)}{\vec{\vctt_4}(j_2)}$,\\
\indent\quad $(\forall j_2,\ \soproductivepJ{(p_1,\ p_2(j_2),\ \lambda j_1.q(j_1,j_2))}(F(j_2))) \iff$ 
  $(\soproductiveJ{(p_1,p_2,q)}(\lambda f.~\lambda j_2.~F(j_2)(f))) $
\end{lemma}

By \Cref{so_productivepJ_to_so_productiveJ}, 
it suffices to employ combination principles for partially indexed second-order productivity, 
which closely resemble those for (non-indexed) second-order productivity.
Examples are provided in \Cref{Combination Principles of Partially Indexed Second-order Productivity}.
\section{Examples of Combination Principles}
\label{Examples of Combination Principles}

Here we present some examples of combination principles, grouped by relevant CTT.

\subsection{Combination Principles of First-order Productivity}
\label{Combination Principles of First-order Productivity}

\subsubsection{Combinators related to General CTT}

\begin{lemma}[Productivity of Identity Function]\quad\\
\indent\quad $\productive{p'}(\id)$ for 
  $\id \defeq \lambda x.x$ and
  $p' = \{(l_1,l_2)\ |\ l_1=l_2 \}$.
\end{lemma}

\begin{lemma}[Productivity of Composition]\quad\\
\indent\quad For $f_1 :\sem{\vctt_1} \ra \sem{\vctt_2}$ and $f_2 : \sem{\vctt_2} \ra \sem{\vctt_3}$
  such that \ $ \productive{p_1}(f_1)$ and $\productive{p_2}(f_2)$,\\ 
\indent\quad $\productive{p_1 \pcompose p_2}(f_2 \fcompose f_1)$ where \\
\indent\quad $p_1 \pcompose p_2 \in \productivity{\vctt_1}{\vctt_3}$, \ 
  $p_1 \pcompose p_2 \defeq \{(l_1,l_3)\ |\ \exists l_2,\ p_1(l_1,l_2) \land p_2(l_2,l_3)\}$. \\
\indent\quad Note that the notation $\fcompose$ is left associative and $\pcompose$ is right associative.
\end{lemma}

\begin{corollary}[Productivity of Repeated Compositions]\quad\\
\indent\quad For $f :\sem{\vctt} \ra \sem{\vctt}$
  such that \ $ \productive{p}(f)$,\\ 
\indent\quad $\productive{p^n}~(f^n)$ where \\
\indent\quad $p^n \in \productivity{\vctt}{\vctt}$, \ 
  $p^n \defeq p \pcompose p \pcompose \cdots \pcompose p$. 
\end{corollary}

\begin{lemma}[Productivity of Map on CTT]\quad\\
\indent\quad For $\vspf \in \SPF(\unitset)$ and $f : \sem{\vctt_{1}} \ra \sem{\vctt_{2}}$ 
  such that \ $\productive{p}(f)$, \\
\indent\quad let \ $\fmap_\vspf(f) : \sem{\vspf\cttcompose(\lambda\_.\vctt_1)} \ra \sem{\vspf\cttcompose(\lambda\_.\vctt_2)}$, 
  $\fmap_\vspf(f) \defeq \vspf.\mathtt{map}\ (\lambda\_.f)$. \\
\indent\quad Then $\productive{p'}~(\vspf.\mathtt{map}\ (\lambda\_.f))$ where \\
\indent\quad $p' \in \productivity{\vspf\cttcompose(\lambda\_.\vctt_1)}{\vspf\cttcompose(\lambda\_.\vctt_2)}$,\\
\indent\quad $p' = \{(\some(l'_1),\ \some(l'_2))\ |\ p(l_1'\ \unit,l_2'\ \unit)\ \lor\ l_1'\ \unit=\bot\}$
\end{lemma}

\begin{lemma}[Productivity of Constant Function]\quad\\
\indent\quad For $x \in \sem{\vctt_{2}}$, let $\const(x): \sem{\vctt_{1}} \ra \sem{\vctt_{2}} \defeq \lambda \_.~ x$. 
  Then \ $\productive{\Empty}(f)$.
\end{lemma}

\begin{lemma}[Productivity of (degenerate) Conditional Function]\quad\\
\indent\quad For $b\in\Bool$ and $f_1,f_2 :\sem{\vctt_1} \ra \sem{\vctt_2}$
  such that \ $ \productive{p_1}(f_1)$ and $\productive{p_2}(f_2)$,\\ 
\indent\quad $\productive{\ite{b}{p_1}{p_2}}(\cbite{b}{f_1}{f_2})$.
\end{lemma}
  
\begin{lemma}[Productivity of Nondegenerate Conditional Function]\quad\\
\indent\quad For $P: A\ra\Bool$ and $f_1,f_2 :\sem{\vctt_1} \ra \sem{\vctt_2}$
  such that \ $ \productive{p_1}(f_1)$ and $\productive{p_2}(f_2)$,\\ 
\indent\quad Let \ $\cnif(P,f_1,f_2) : \sem{\constCTT{A}\gop{\times}\vctt_1}\ra \sem{\vctt_2}$,\ 
  $\cnif(P,f_1,f_2) \defeq \lambda (a,x).~ \ite{P(a)}{f_1(x)}{f_2(x)}$.\\
\indent\quad Then $\productive{p'}(\ite{b}{f_1}{f_2})$ for \\
\indent\quad $p'\in\productivity{\constCTT{A}\gop{\times}\vctt_1}{\vctt_2}$,\\ 
\indent\quad $p'=\{(\some(l_1'),l_2)\ |\ l_1'\ \btrue=\some(\lambda\_.\_)\ \land$\\ 
\indent\hfil $(l_1'\ \bfalse=\bot \lor p_1(l_1'\ \bfalse,l_2) \lor p_2(l_1'\ \bfalse,l_2)) \}$.
\end{lemma}
  
\subsubsection{Combinators related to Product CTT}

\begin{lemma}[Productivity of Fst and Snd]\quad\\
\indent\quad Let $\fst: \sem{\vctt_1\gop{\times}\vctt_2}\ra\sem{\vctt_1}$, $\fst\defeq \lambda (x,y).~x$ and 
  $\snd: \sem{\vctt_1\gop{\times}\vctt_2}\ra\sem{\vctt_2}$, $\snd\defeq \lambda (x,y).~y$.\\
\indent\quad Then $\productive{p_1}{(\fst)}$ and $\productive{p_2}(\snd)$ for \\
\indent\quad $p_1\in\productivity{\vctt_1\gop{\times}\vctt_2}{\vctt_1}$, 
  $p_1 = \{(\some(l'_{1,2}), l_1)\ |\ l'_{1,2}\ \btrue=l_1\ \land\ l'_{1,2}\ \bfalse=\bot \}$,\\
\indent\quad $p_2\in\productivity{\vctt_1\gop{\times}\vctt_2}{\vctt_2}$, 
  $p_2 = \{(\some(l'_{1,2}), l_2)\ |\ l'_{1,2}\ \bfalse=l_2\ \land\ l'_{1,2}\ \btrue=\bot \}$\\
\end{lemma}

\begin{lemma}[Productivity of Pair]\quad\\
\indent\quad For $x_1\in\sem{\vctt_1},x_2\in\sem{\vctt_2}$, 
  let $f_1: \sem{\vctt_2}\ra\sem{\vctt_1\gop{\times}\vctt_2}$, $f_1\defeq \lambda x.~(x_1,x)$ and \\
\indent\quad $f_2: \sem{\vctt_1}\ra\sem{\vctt_1\gop{\times}\vctt_2}$, $f_2\defeq \lambda x.~(x,x_2)$.\\
\indent\quad Then $\productive{p_1}{(f_1)}$ and $\productive{p_2}(f_2)$ for \\
\indent\quad $p_1\in\productivity{\vctt_2}{\vctt_1\gop{\times}\vctt_2}$, 
  $p_1 = \{(l_2, \some(l'_{1,2}))\ |\ l'_{1,2}\ \bfalse=l_2 \}$,\\
\indent\quad $p_2\in\productivity{\vctt_1}{\vctt_1\gop{\times}\vctt_2}$, 
  $p_2 = \{(l_1, \some(l'_{1,2}))\ |\ l'_{1,2}\ \btrue=l_1 \}$\\
\end{lemma}
  
\begin{lemma}[Productivity of Fpair]\quad\\
\indent\quad For $f_1 :\sem{\vctt_1} \ra \sem{\vctt_2}$ and $f_2 :\sem{\vctt_1} \ra \sem{\vctt_3}$
  such that $\productive{p_1}(f_1)$ and $\productive{p_2}(f_2)$, \\
\indent\quad let \ $f_1 \fpair f_2 : \sem{\vctt_1} \ra \sem{\vctt_2\gop{\times}\vctt_3}$, \ 
  $f_1 \fpair f_2 \defeq \lambda x. (f_1(x), \; f_2(x))$. \\
\indent\quad Then $\productive{p_1 \ppair p_2}~(f_1 \fpair f_2)$ where \\
\indent\quad $p_1 \ppair p_2 \in \productivity{\vctt_1}{\vctt_2\gop{\times}\vctt_3}$,\\
\indent\quad $p_1 \ppair p_2 \defeq 
  \{ (l_1,\some(l_{2,3}'))\ |\ 
  p_1(l_1,l_{2,3}'\ \btrue) \lor p_2(l_1,l_{2,3}'\ \bfalse) \}$
\end{lemma}

\begin{lemma}[Productivity of Fproduct]\quad\\
\indent\quad For $f_1 : \sem{\vctt_1} \ra \sem{\vctt_2}$ and $f_2 : \sem{\vctt_3} \ra \sem{\vctt_4}$
  such that \ $\productive{p_1}(f_1)$ and $\productive{p_2}(f_2)$, \\
\indent\quad let \ $f_1 \fproduct f_2 : \sem{\vctt_{1} \gop{\times} \vctt_{3}} \ra \sem{\vctt_{2} \gop{\times} \vctt_{4}}$, 
  $\ f_1 \fproduct f_2 \defeq \lambda(x_1, x_2). (f_1(x_1),\ f_2(x_2))$. \\
\indent\quad Then $\productive{p_1 \pproduct p_2}~(f_1 \fproduct f_2)$ where\\
\indent\quad $p_1 \pproduct p_2 \in \productivity{\vctt_{1} \gop{\times} \vctt_{3}}{\vctt_{2} \gop{\times} \vctt_{4}}$,\\
\indent\quad $p_1 \pproduct p_2 \defeq
  \{ (\some(l'_{1,3}), \some(l'_{2,4}))\ |\ 
  (p_1(l_{1,3}'\ \btrue,l_{2,4}'\ \btrue)\ \land\ l_{1,3}'\ \bfalse=\bot)\ \lor$\ \\
\indent\hfill $(p_2(l_{1,3}'\ \bfalse,l_{2,4}'\ \bfalse)\ \land\ l_{1,3}'\ \btrue=\bot) \}$
\end{lemma}

\begin{lemma}[Productivity of Commutator]\quad\\
  \indent\quad Let $\comm: \sem{\vctt_1\gop{\times}\vctt_2}\ra\sem{\vctt_2\gop{\times}\vctt_1}$, 
  $\comm\defeq \lambda (x,y).~(y,x)$. \\
  \indent\quad Then $\productive{p'}{(\comm)}$ for \\
  \indent\quad $p'\in\productivity{\vctt_1\gop{\times}\vctt_2}{\vctt_2\gop{\times}\vctt_1}$, \\
  \indent\quad $p' = \{(\some(l'_{1,2}), \some(l'_{2,1}))\ |\ l'_{1,2}\ \btrue=l'_{2,1}\ \bfalse\ \land\ l'_{1,2}\ \bfalse=l'_{2,1}\ \btrue \}$
  \end{lemma}

\begin{lemma}[Productivity of Associator and Antiassociator]\quad\\
\indent\quad Let \ $\assoc: \sem{(\vctt_1\gop{\times}\vctt_2)\gop{\times}\vctt_3}\ra\sem{\vctt_1\gop{\times}(\vctt_2\gop{\times}\vctt_3)}$, $\assoc\defeq \lambda ((x,y),z).~(x,(y,z))$ and \\
\indent\qquad $\antiassoc: \sem{\vctt_1\gop{\times}(\vctt_2\gop{\times}\vctt_3)}\ra\sem{(\vctt_1\gop{\times}\vctt_2)\gop{\times}\vctt_3}$, $\antiassoc\defeq \lambda (x,(y,z)).~((x,y),z)$.\\
\indent\quad Then $\productive{p_1}{(\assoc)}$ and $\productive{p_2}(\antiassoc)$ for \\
\indent\quad $p_1\in\productivity{(\vctt_1\gop{\times}\vctt_2)\gop{\times}\vctt_3}{\vctt_1\gop{\times}(\vctt_2\gop{\times}\vctt_3)}$, \\
\indent\quad $p_1 = \{(\some(\lambda b.~ \ite{b}{\some(l'_{1,2})}{l_3}),\ \some(\lambda b.~ \ite{b}{l_1}{\some(l'_{2,3})}))\ |\  $\\
\indent\hfill $l'_{1,2}\ \btrue=l_1\ \land\ l'_{1,2}\ \bfalse=l'_{2,3}\ \btrue\ \land\ l_3=l'_{2,3}\ \bfalse \}$,\\
\indent\quad $p_2\in\productivity{\vctt_1\gop{\times}(\vctt_2\gop{\times}\vctt_3)}{(\vctt_1\gop{\times}\vctt_2)\gop{\times}\vctt_3}$, \\
\indent\quad $p_2 = \{(\some(\lambda b.~ \ite{b}{l_1}{\some(l'_{2,3})}),\ \some(\lambda b.~ \ite{b}{\some(l'_{1,2})}{l_3}))\ |\  $\\
\indent\hfill $l_1=l'_{1,2}\ \btrue\ \land\ l'_{2,3}\ \btrue=l'_{1,2}\ \bfalse\ \land\ l'_{2,3}\ \bfalse=l_3 \}$
\end{lemma}

\begin{lemma}[Productivity of Currying Function]\quad\\
\indent\quad For $f:\sem{\vctt_1\gop{\times}\vctt_2}\ra\sem{\vctt_3}$ such that $\productive{p}(f)$, \\
\indent\quad let \ $\curry(f): \sem{\vctt_1}\ra(\sem{\vctt_2}\ra\sem{\vctt3})$, 
$\curry(f)\defeq \lambda x.~\lambda y.~f(x,y)$. \\
\indent\quad Then $\forall x,\ \productive{p'}{(\curry(f)(x))}$ for \\
\indent\quad $p'\in\productivity{\vctt_2}{\vctt_3}$, \\
\indent\quad $p' = \{(l_2,l_3)\ |\ \exists\ l_1,\ p(\some(\lambda b.~\ite{b}{l_1}{l_2}),\ l_3) \}$
\end{lemma}

\begin{lemma}[Productivity of Uncurrying Function]\quad\\
\indent\quad For $f:\sem{\vctt_1}\ra(\sem{\vctt_2}\ra\sem{\vctt3})$ such that $\forall x,\ \productive{p}(f(x))$, \\
\indent\quad let \ $\uncurry(f): \sem{\vctt_1\gop{\times}\vctt_2}\ra\sem{\vctt_3}$, 
$\uncurry(f)\defeq \lambda (x,y).~f(x)(y)$. \\
\indent\quad Then $\productive{p'}{(\uncurry(f))}$ for \\
\indent\quad $p'\in\productivity{\vctt_1\gop{\times}\vctt_2}{\vctt_3}$, \\
\indent\quad $p' = \{(\some(l'_{1,2}),l_3)\ |\ (p(l'_{1,2}\ \bfalse, l_3)\ \land\ l'_{1,2}\ \btrue=\bot)\ \lor\ (l'_{1,2}\ \bfalse=\bot) \}$
\end{lemma}

\subsubsection{Combinators related to Sum CTT}

\begin{lemma}[Productivity of Inl and Inr]\quad\\
  \indent\quad For $\inl: \sem{\vctt_1}\ra\sem{\vctt_1\gop{+}\vctt_2}$, $\inl\defeq \lambda x.~x$ and 
    $\inr: \sem{\vctt_2}\ra\sem{\vctt_1\gop{+}\vctt_2}$, $\inr\defeq \lambda x.~x$.\\
  \indent\quad Then $\productive{p_1}{(f_1)}$ and $\productive{p_2}(f_2)$ for \\
  \indent\quad $p_1\in\productivity{\vctt_1}{\vctt_1\gop{+}\vctt_2}$, 
    $p_1 = \{(l_1, \some(l'_{1,2}))\ |\ l'_{1,2}\ \btrue=l_1 \}$,\\
  \indent\quad $p_2\in\productivity{\vctt_2}{\vctt_1\gop{+}\vctt_2}$, 
    $p_2 = \{(l_2, \some(l'_{1,2}))\ |\ l'_{1,2}\ \bfalse=l_2 \}$\\
  \end{lemma}  

\begin{lemma}[Productivity of Fcopair]\quad\\
\indent\quad For $f_1 : \sem{\vctt_1} \ra \sem{\vctt_3}$ and $f_2 : \sem{\vctt_2} \ra \sem{\vctt_3}$
  such that \ $\productive{p_1}(f_1)$ and $\productive{p_2}(f_2)$, \\
\indent\quad let \ $f_1 \fcopair f_2 : \sem{\vctt_{1} \gop{+} \vctt_{2}} \ra \sem{\vctt_3}$, \\
\indent\qquad$(f_1 \fcopair f_2) \; (\mathtt{inl} \; x_1) \defeq f_1(x_1)$ and \ 
  $(f_1 \fcopair f_2) \; (\mathtt{inr} \; x_2) \defeq f_2(x_2)$.\\
\indent\quad Then $\productive{p_1 \pcopair p_2}~(f_1 \fcopair f_2)$, where \\
\indent\quad $p_1 \pcopair p_2 \in \productivity{\vctt_{1} \gop{+} \vctt_{2}}{\vctt_3}$,\\
\indent\quad $p_1 \pcopair p_2 \defeq
  \{ (\some(l_{1,2}'), l_3)\ |\ $
  $(p_1(l'_{1,2}\ \btrue, l_3)\ \land\ l'_{1,2}\ \bfalse=\bot)\ \lor$\ \\
\indent\hfill$(p_2(l'_{1,2}\ \bfalse, l_3)\ \land\ l'_{1,2}\ \btrue=\bot)\ \lor$
  $(l'_{1,2}\ \btrue=\bot\ \land\ l'_{1,2}\ \bfalse=\bot) \}$
\end{lemma}

\begin{lemma}[Productivity of Fcoproduct]\quad\\
\indent\quad For $f_1 : \sem{\vctt_1} \ra \sem{\vctt_2}$ and $f_2 : \sem{\vctt_3} \ra \sem{\vctt_4}$
  such that \ $\productive{p_1}(f_1)$ and $\productive{p_2}(f_2)$, \\
\indent\quad let \ $f_1 \fcoproduct f_2 : \sem{\vctt_{1} \gop{+} \vctt_{3}} \ra \sem{\vctt_{2} \gop{+} \vctt_{4}}$, \\
\indent\qquad $(f_1 \fcoproduct f_2) \; (\mathtt{inl} \; x_1) \defeq \mathtt{inl} \; (f_1(x_1))$ and \ 
$(f_1 \fcoproduct f_2) \; (\mathtt{inr} \; x_3) \defeq \mathtt{inr} \; (f_2(x_2))$.\\
\indent\quad Then $\productive{p_1 \pcoproduct p_2}~(f_1 \fcoproduct f_2)$, where \\
\indent\quad $p_1 \pcoproduct p_2 \in \productivity{\vctt_{1} \gop{+} \vctt_{3}}{\vctt_{2} \gop{+} \vctt_{4}}$,\\
\indent\quad $p_1 \pcoproduct p_2 \defeq
  \{ (\some(l'_{1,3}), \some(l'_{2,4}))\ |\ $
  $(p_1(l'_{1,3}\ \btrue, l'_{2,4}\ \btrue)\ \land\ l'_{1,3}\ \bfalse=\bot)\ \lor$\ \\
\indent\hfill$(p_2(l'_{1,3}\ \bfalse, l'_{2,4}\ \bfalse)\ \land\ l'_{1,3}\ \btrue=\bot)\ \lor$\ 
  $(l'_{1,3}\ \btrue=\bot\ \land\ l'_{1,3}\ \bfalse=\bot) \}$
\end{lemma}

\subsubsection{Combinators related to Exponential CTT} \label{Combinators related to Exponential CTT}

\begin{lemma}[Productivity of Exponent Application]\quad\\
\indent\quad Let \ $\ceapp : \sem{(A\gop{\ra}\vctt)\gop{\times}\constCTT{A}} \ra\sem{\vctt}$, 
  $\ceapp\defeq \lambda (f,x).~ f(x)$.\\
\indent\quad Then $\productive{p'}~(\ceapp(f))$ for \\
\indent\quad $p'\in \productivity{(A\gop{\ra}\vctt)\gop{\times}\constCTT{A}}{\vctt}$, \\
\indent\quad $p' = \{(\some(\lambda b.~ \ite{b}{\some(l'_1)}{l_A}),\ l_2)\ |\ l_A\in\level(\constCTT{A}),\ l'_1\ \unit=l_2 \}$.
\end{lemma}

\begin{corollary}[Productivity of Generalized Exponent Application]\quad\\
\indent\quad For $g: A_1\ra A_2$,\\
\indent\quad let \ $\cgeapp(g) : \sem{(A_2\gop{\ra}\vctt)\gop{\times}\constCTT{A_1}} \ra\sem{\vctt}$, 
  $\cgeapp(g)\defeq \lambda (f,x).~ f(g(x))$.\\
\indent\quad Then $\productive{p'}~(\cgeapp(f))$ for \\
\indent\quad $p'\in \productivity{(A_2\gop{\ra}\vctt)\gop{\times}\constCTT{A_1}}{\vctt}$, \\
\indent\quad $p' = \{(\some(\lambda b.~ \ite{b}{\some(l'_1)}{l_A}),\ l_2)\ |\ l_A\in\level(\constCTT{A_1}),\ l'_1\ \unit=l_2 \}$.
\end{corollary}

\begin{remark}
  $\forall g: \sem{\constCTT{A_1}}\ra\sem{\constCTT{A_2}}$,\ 
  $\cgeapp(g) = \ceapp(g)\ \fcompose\ (\id \fproduct g)$.
\end{remark}
  
\begin{lemma}[Productivity of Exponent Curry]\quad\\
\indent\quad For $f: \sem{\vctt_1\gop{\times}\constCTT{A}}\ra\sem{\vctt_2}$
  such that \ $\productive{p}(f)$, \\
\indent\quad let \ $\cecurry(f) : \sem{\vctt_1}\ra\sem{A\gop{\ra}\vctt_2}$, 
  $\cecurry(f)\defeq \lambda x.~\lambda a.~ f(x,a)$. \\
\indent\quad Then $\productive{p'}~(\cecurry(f))$ for \\
\indent\quad $p'\in \productivity{\vctt_1}{A\gop{\ra}\vctt_2}$, \\
\indent\quad $p' = \{(l_1,\some(l'_2))\ |\ \exists l_A\in\level(\constCTT{A}),\ p(\some(\lambda b.~\ite{b}{l_1}{l_A}),\ l'_2\ \unit) \}$.
\end{lemma}

\begin{corollary}[Productivity of Generalized Exponent Curry]\quad\\
\indent\quad For $f: \sem{\constCTT{A}\gop{\times}(\vctt_1\gop{\times}\constCTT{A})}\ra\sem{\vctt_2}$
  such that \ $\productive{p}(f)$, \\
\indent\quad let \ $\cgecurry(f) : \sem{\vctt_1}\ra\sem{A\gop{\ra}\vctt_2}$, 
  $\cgecurry(f)\defeq \lambda x.~\lambda a.~ f(a,(x,a))$. \\
\indent\quad Then $\productive{p'}~(\cgecurry(f))$ for \\
\indent\quad $p'\in \productivity{\vctt_1}{A\gop{\ra}\vctt_2}$, \\
\indent\quad $p' = \{(l_1,\some(l'_2))\ |\ \exists l_A,\ l_A'\in\level(\!\!\constCTT{A}\!),\ p(\some(\lambda b.~\ite{b}{l_A}{\some(\lambda b'.~\ite{b'}{l_1}{l_A'})}),\ l'_2\ \unit) \}$.
\end{corollary}
  
\begin{remark}
  $\forall f: \sem{\constCTT{A}\gop{\times}(\vctt_1\gop{\times}\constCTT{A})}\ra\sem{\vctt_2}$,\\ 
  \indent\quad $\cgecurry(f) = \cecurry(f)\ \fcompose\ 
  \comm\ \fcompose\ \antiassoc\ \fcompose\ (\id \fproduct (\id \fpair \id))$.
\end{remark}
  
In cases where level dependency on the curried argument is not desired,
we can use $\ceswap$ and $\ceproj$ instead.

\begin{lemma}[Productivity of Exponent Swap]\quad\\
\indent\quad For $f: A\ra{\sem{\vctt_1}\ra\sem{\vctt_2}}$
  such that \ $\forall a,\ \productive{p(a)}(f(a))$, \\
\indent\quad Let \ $\ceswap: \sem{\vctt_1}\ra\sem{A\gop{\ra}\vctt_2}$, \ 
  $\ceswap\defeq \lambda x.~ \lambda a.~ f(a)(x)$.\\
\indent\quad Then $\productive{p'}~(\ceswap(f))$ for \\
\indent\quad $p'\in \productivity{\vctt_1}{A\gop{\ra}\vctt_2}$, \ 
  $p' = \{ (l_1,\some(l'_2))\ |\ \exists a,\ p(a)(l_1,l_2\ \unit) \}$.
\end{lemma}
  
\begin{lemma}[Productivity of Exponent Projection]\quad\\
\indent\quad Let \ $\ceproj: A\ra{\sem{A\gop{\ra}\vctt}\ra\sem{\vctt}}$, 
  $\ceproj\defeq \lambda a.~ \lambda f.~ f(a)$. \\
\indent\quad Then $\productive{p'}~(\ceproj(a))$ for \\
\indent\quad $p'\in \productivity{A\gop{\ra}\vctt}{\vctt}$, \
  $p' = \{(\some(l'_1),l_2)\ |\ l'_1\ \unit = l_2 \}$.
\end{lemma}

\begin{lemma}[Productivity of Exponent Paring]\quad\\
\indent\quad Let \ $\cepair: \sem{(A\gop{\ra}\vctt_1)\gop{\times}(A\gop{\ra}\vctt_2)}\ra \sem{A\gop{\ra}(\vctt_1\gop{\times}\vctt_2)}$, 
  $\cepair\defeq \lambda f.~ \lambda a.~ (\fst(f)(a),\snd(f)(a))$. \\
\indent\quad Then $\productive{p'}~(\cepair)$ for \\
\indent\quad $p'\in \productivity{(A\gop{\ra}\vctt_1)\gop{\times}(A\gop{\ra}\vctt_2)}{A\gop{\ra}(\vctt_1\gop{\times}\vctt_2)}$, \\
\indent\quad$p' = \{(\some(l'_1), \some(\lambda\_.\some(l_2')))\ |\ 
  \forall b\in\Bool,\ l'_1\ b = \some(\lambda\_.l_2'\ b) \}$.
\end{lemma}

\subsubsection{Combinators related to Nondegenerate Function CTT}

\begin{lemma}[Productivity of Nondegenerate Function Swap]\quad\\
\indent\quad For $f: \depfun{a:A}{\sem{\vctt_1}\ra\sem{\vec{\vctt_2}(a)}}$
  such that \ $\forall a,\ \productive{p(a)}(f(a))$, \\
\indent\quad Let \ $\cnswap: \sem{\vctt_1}\ra\sem{\gdepfun{a:A}{\vec{\vctt_2}(a)}}$, \ 
  $\cnswap\defeq \lambda x.~ \lambda a.~ f(a)(x)$.\\
\indent\quad Then $\productive{p'}~(\cnswap(f))$ for \\
\indent\quad $p'\in \productivity{\vctt_1}{\gdepfun{a:A}{\vec{\vctt_2}(a)}}$, \ 
  $p' = \{ (l_1,\some(l'_2))\ |\ \exists a,\ p(a)(l_1,l_2'\ a) \}$.
\end{lemma}
  
\begin{lemma}[Productivity of Nondegenerate Function Projection]\quad\\
\indent\quad Let \ $\cnproj: \depfun{a:A}{\sem{\gdepfun{a':A}{\vec\vctt(a')}}\ra\sem{\vec\vctt(a)}}$, 
  $\cnproj\defeq \lambda a.~ \lambda f.~ f(a)$. \\
\indent\quad Then $\productive{p'(a)}~(\cnproj(a))$ for \\
\indent\quad $p'\in \depfun{a:A}{\productivity{\gdepfun{a':A}{\vec\vctt(a')}}{\vec\vctt(a)}}$, \\
\indent\quad $p' = \lambda a.~ \{(\some(l'_1),l_2)\ |\ l'_1\ a = l_2 \land (\forall a'\neq a,\ l'_1\ a'=\bot) \}$.
\end{lemma}

\subsection{Combination Principles of Second-order Productivity}
\label{Combination Principles of Second-order Productivity}

\subsubsection{Combinators related to General CTT}

\begin{lemma}[Second-order Productivity of Constant Generating Function]\quad\\
\indent\quad For $f :\cFun{3}{4}$ such that $\productive{p}(f)$, \\
\indent\quad let $\sfindep\ f : \cFFun{1}{2}{3}{4},\ \sfindep\ f\defeq \lambda \_.f $. \\
\indent\quad Then $\forall p_0,\ \soproductive{(p_0,p,\Empty)}~(\sfindep\ f)$.
\end{lemma}

\begin{lemma}[Second-order Productivity of Sfself]\quad\\
\indent\quad let \ $\sfself : \cFFun{1}{2}{1}{2}$, $\sfself \defeq \lambda f.~ f$. \\
\indent\quad Then $\forall p,\ \soproductive{(p,\ p,\ \{(l,l)\ |\ l\in\level(\vctt_2)\})}~(\sfself)$.
\end{lemma}

\begin{lemma}[Second-order Productivity of Sfcompose]\quad\\
\indent\quad For $F_1 :\cFFun{1}{2}{3}{4}$ and $F_2 :\cFFun{1}{2}{4}{5}$\\
\indent\qquad such that $\soproductive{(p_0,p_1,q_1)}(F_1)$ and $\soproductive{(p_0,p_2,q_2)}(F_2)$, \\
\indent\quad let \ $F_2\sfcompose F_1 : \cFFun{1}{2}{3}{5}$, \\ 
\indent\qquad $F_2\sfcompose F_1 \defeq \lambda f.~ (F_2\ f)\fcompose (F_1\ f)$. \\
\indent\quad Then $\soproductive{(p_0,\ (p_1\pcompose p_2),\ ((q_1\pcompose p_2)\por q_2))}~(F_2\sfcompose F_1)$.
\end{lemma}

\begin{lemma}[Second-order Productivity of Sfmap]\quad\\
\indent\quad For $\vspf \in \SPF(\unitset)$ and $F :\cFFun{1}{2}{3}{4}$ 
  such that $\soproductive{(p_0,p_1,q_1)}(F)$, \\
\indent\quad Let \ $\sfmap_\vspf(F) : (\cFun{1}{2})\ra(\sem{\vspf\cttcompose(\lambda\_.\vctt_3)}\ra\sem{\vspf\cttcompose(\lambda\_.\vctt_4)})$,\\
\indent\qquad $\sfmap_\vspf(F)\defeq \lambda f.~ \vspf.\mathtt{map}\ (\lambda\_.F\ f)$.\\
\indent\quad Then $\soproductive{(p_0,p_1',q')}~(\sfmap_\vspf(F))$ where \\
\indent\quad $p' \in \productivity{\vspf\cttcompose(\lambda\_.\vctt_3)}{\vspf\cttcompose(\lambda\_.\vctt_4)}$,\\
\indent\quad $p' = \{(\some(l'_3),\ \some(l'_4))\ |\ p(l_3'\ \unit,l_4'\ \unit)\ \lor\ l_3'\ \unit=\bot\}$ and \\
\indent\quad $q'\in\productivity{\vctt_2}{\vspf\cttcompose(\lambda\_.\vctt_4)}$, \\
\indent\quad $q' = \{(l_2,\some(l'_4))\ |\ q(l_2,l'_4\ \unit) \}$
\end{lemma}

\subsubsection{Combinators related to Product CTT}

\begin{lemma}[Second-order Productivity of Sfpair]\quad\\
\indent\quad For $F_1 :\cFFun{1}{2}{3}{4}$ and $F_2 :\cFFun{1}{2}{3}{5}$\\
\indent\qquad such that $\soproductive{(p_0,p_1,q_1)}(F_1)$ and $\soproductive{(p_0,p_2,q_2)}(F_2)$, \\
\indent\quad let \ $F_1\sfpair F_2 : (\cFun{1}{2})\ra(\sem{\vctt_3}\ra\sem{\vctt_4\gop{\times}\vctt_5})$, \\ 
\indent\qquad $F_1\sfpair F_2 \defeq \lambda f.~ (F_1\ f)\fpair(F_2\ f)$. \\
\indent\quad Then $\soproductive{(p_0,\ (p_1\ppair p_2),\ (q_1\ppair q_2))}~(F_1\sfpair F_2)$.
\end{lemma}
  
\begin{lemma}[Second-order Productivity of Sfproduct]\quad\\
\indent\quad For $F_1 :\cFFun{1}{2}{3}{4}$ and $F_2 :\cFFun{1}{2}{5}{6}$\\
\indent\qquad such that $\soproductive{(p_0,p_1,q_1)}(F_1)$ and $\soproductive{(p_0,p_2,q_2)}(F_2)$, \\
\indent\quad let \ $F_1\sfproduct F_2 : (\cFun{1}{2})\ra(\sem{\vctt_3\gop{\times}\vctt_5}\ra\sem{\vctt_4\gop{\times}\vctt_6})$, \\ 
\indent\qquad $F_1\sfproduct F_2 \defeq \lambda f.~ (F_1\ f)\fproduct(F_2\ f)$. \\
\indent\quad Then $\soproductive{(p_0,\ (p_1\pproduct p_2),\ (q_1\ppair q_2))}~(F_1\sfproduct F_2)$.
\end{lemma}

\subsubsection{Combinators related to Sum CTT}

\begin{lemma}[Second-order Productivity of Sfcopair]\quad\\
\indent\quad For $F_1 :\cFFun{1}{2}{3}{5}$ and $F_2 :\cFFun{1}{2}{4}{5}$\\
\indent\qquad such that $\soproductive{(p_0,p_1,q_1)}(F_1)$ and $\soproductive{(p_0,p_2,q_2)}(F_2)$, \\
\indent\quad let \ $F_1\sfcopair F_2 : (\cFun{1}{2})\ra(\sem{\vctt_3\gop{+}\vctt_4}\ra\sem{\vctt_5})$, \\ 
\indent\qquad $F_1\sfcopair F_2 \defeq \lambda f.~ (F_1\ f)\fcopair(F_2\ f)$. \\
\indent\quad Then $\soproductive{(p_0,\ (p_1\pcopair p_2),\ (q_1\por q_2))}~(F_1\sfcopair F_2)$.
\end{lemma}

\begin{lemma}[Second-order Productivity of Sfcoproduct]\quad\\
\indent\quad For $F_1 :\cFFun{1}{2}{3}{4}$ and $F_2 :\cFFun{1}{2}{5}{6}$\\
\indent\qquad such that $\soproductive{(p_0,p_1,q_1)}(F_1)$ and $\soproductive{(p_0,p_2,q_2)}(F_2)$, \\
\indent\quad let \ $F_1\sfcoproduct F_2 : (\cFun{1}{2})\ra(\sem{\vctt_3\gop{+}\vctt_5}\ra\sem{\vctt_4\gop{+}\vctt_6})$, \\ 
\indent\qquad $F_1\sfcoproduct F_2 \defeq \lambda f.~ (F_1\ f)\fcoproduct(F_2\ f)$. \\
\indent\quad Then $\soproductive{(p_0,\ (p_1\pcoproduct p_2),\ (q_1\ppair q_2))}~(F_1\sfcoproduct F_2)$.
\end{lemma}

\subsubsection{Combinators related to Exponential CTT} 

\begin{lemma}[Second-order Productivity of Sfecurry]\quad\\
\indent\quad For $F_1 :(\cFun{1}{2})\ra(\sem{\vctt_3\gop{\times}\constCTT{A}}\ra\sem{\vctt_4})$
  such that $\soproductive{(p_0,p,q)}(F_1)$, \\
\indent\quad let \ $\sfecurry(F) : (\cFun{1}{2})\ra(\sem{\vctt_3}\ra\sem{A\gop{\ra}\vctt_4})$, \\ 
\indent\qquad $\sfecurry(F) \defeq \lambda f.~ \cecurry(F\ f)$. \\
\indent\quad Then $\soproductive{(p_0,\ p',\ q')}~(\sfecurry(F))$where \\
\indent\quad $p' \in \productivity{\vctt_3}{A\gop{\ra}\vctt_4}$,\\
\indent\quad $p' = \{(l_3,\some(l'_4))\ |\ \exists l_A\in\level(\constCTT{A}),\ p(\some(\lambda b.~\ite{b}{l_3}{l_A}),\ l'_4\ \unit) \}$ and \\
\indent\quad $q'\in\productivity{\vctt_2}{A\gop{\ra}\vctt_4}$, \\
\indent\quad $q' = \{(l_2,\some(l'_4))\ |\ q(l_2,l'_4\ \unit) \}$.
\end{lemma}

\subsection{Combination Principles of Partially Indexed Second-order Productivity}
\label{Combination Principles of Partially Indexed Second-order Productivity}

\subsubsection{Combinators related to General CTT}

\begin{lemma}[Partially Indexed Second-order Productivity of Constant Generating Function]\quad\\
\indent\quad For $f :\cFun{3}{4}$ such that $\productive{p}(f)$, \\
\indent\quad let $\spifindep(f) : \depfun{j\!:\!J}(\depfun{j'\!:\!J}\sem{\vec{\vctt_1}(j')}\ra\sem{\vec{\vctt_2}(j')})\ra(\cFun{3}{4})$, \\
\indent\quad $\spifindep(f) \defeq \lambda \_.~f$. \\
\indent\quad Then $\forall p_0,\ \soproductivepJ{(p_0,p,\Empty)}~(\spifindep(f))$.
\end{lemma}

\begin{lemma}[Partially Indexed Second-order Productivity of Spifself]\quad\\
\indent\quad let \ $\spifself : \depfun{j\!:\!J}(\depfun{j'\!:\!J}\sem{\vec{\vctt_1}(j')}\ra\sem{\vec{\vctt_2}(j')})\ra(\sem{\vec{\vctt_1}(j)}\ra\sem{\vec{\vctt_2}(j)})$, \\
\indent\quad $\spifself \defeq \lambda j.~ \lambda f.~ f(j)$. \\
\indent\quad Then $\forall p,j,\ \soproductivepJ{(p,\ p(j),\ \lambda j.\{(l,l)\ |\ l\in\level(\vctt_2)\})}~(\spifself(j))$.
\end{lemma}

\begin{lemma}[Partially Indexed Second-order Productivity of Spifcompose]\quad\\
\indent\quad For $F_1 :\cFFunpJ{1}{2}{3}{4}$ and \\
\indent\quad $F_2 :\cFFunpJ{1}{2}{4}{5}$\\
\indent\qquad such that $\soproductivepJ{(p_0,p_1,q_1)}(F_1)$ and $\soproductivepJ{(p_0,p_2,q_2)}(F_2)$, \\
\indent\quad let \ $F_2\spifcompose F_1 : \cFFunpJ{1}{2}{3}{5}$, \\ 
\indent\qquad $F_2\spifcompose F_1 \defeq \lambda f.~ (F_2\ f)\fcompose (F_1\ f)$. \\
\indent\quad Then $\soproductivepJ{(p_0,\ (p_1\pcompose p_2),\ (\lambda j.((q_1(j))\pcompose p_2)\por q_2(j)))}~(F_2\spifcompose F_1)$.
\end{lemma}

\begin{lemma}[Partially Indexed Second-order Productivity of Spifmap]\quad\\
\indent\quad For $\vspf \in \SPF(\unitset)$ and $F :\cFFunpJ{1}{2}{3}{4}$ \\
\indent\qquad such that $\soproductivepJ{(p_0,p_1,q_1)}(F)$, \\
\indent\quad Let \ $\spifmap_\vspf(F) : (\cFunJ{J}{1}{2})\ra(\sem{\vspf\cttcompose(\lambda\_.\vctt_3)}\ra\sem{\vspf\cttcompose(\lambda\_.\vctt_4)})$,\\
\indent\qquad $\spifmap_\vspf(F)\defeq \lambda f.~ \vspf.\mathtt{map}\ (\lambda\_.F\ f)$.\\
\indent\quad Then $\soproductive{(p_0,p_1',q')}~(\spifmap_\vspf(F))$ where \\
\indent\quad $p' \in \productivity{\vspf\cttcompose(\lambda\_.\vctt_3)}{\vspf\cttcompose(\lambda\_.\vctt_4)}$,\\
\indent\quad $p' = \{(\some(l'_3),\ \some(l'_4))\ |\ p(l_3'\ \unit,l_4'\ \unit)\ \lor\ l_3'\ \unit=\bot\}$ and \\
\indent\quad $q'\in\depfun{j\!:\!J}\productivity{\vctt_2}{\vspf\cttcompose(\lambda\_.\vctt_4)}$, \\
\indent\quad $q' = \lambda j.~\{(l_2,\some(l'_4))\ |\ q(j)(l_2,l'_4\ \unit) \}$
\end{lemma}

\subsubsection{Combinators related to Product CTT}

\begin{lemma}[Partially Indexed Second-order Productivity of Spifpair]\quad\\
\indent\quad For $F_1 :\cFFunpJ{1}{2}{3}{4}$ and \\
\indent\quad $F_2 :\cFFunpJ{1}{2}{3}{5}$\\
\indent\qquad such that $\soproductivepJ{(p_0,p_1,q_1)}(F_1)$ and $\soproductivepJ{(p_0,p_2,q_2)}(F_2)$, \\
\indent\quad let \ $F_1\spifpair F_2 : (\cFunJ{J}{1}{2})\ra(\sem{\vctt_3}\ra\sem{\vctt_4\gop{\times}\vctt_5})$, \\ 
\indent\qquad $F_1\spifpair F_2 \defeq \lambda f.~ (F_1\ f)\fpair(F_2\ f)$. \\
\indent\quad Then $\soproductive{(p_0,\ (p_1\ppair p_2),\ (\lambda j.~(q_1(j))\ppair(q_2(j))))}~(F_1\spifpair F_2)$.
\end{lemma}

\begin{lemma}[Partially Indexed Second-order Productivity of Spifproduct]\quad\\
\indent\quad For $F_1 :\cFFunpJ{1}{2}{3}{4}$ and \\
\indent\quad $F_2 :\cFFunpJ{1}{2}{5}{6}$\\
\indent\qquad such that $\soproductivepJ{(p_0,p_1,q_1)}(F_1)$ and $\soproductivepJ{(p_0,p_2,q_2)}(F_2)$, \\
\indent\quad let \ $F_1\spifproduct F_2 : (\cFunJ{J}{1}{2})\ra(\sem{\vctt_3\gop{\times}\vctt_5}\ra\sem{\vctt_4\gop{\times}\vctt_6})$, \\ 
\indent\qquad $F_1\spifproduct F_2 \defeq \lambda f.~ (F_1\ f)\fproduct (F_2\ f)$. \\
\indent\quad Then $\soproductivepJ{(p_0,\ (p_1\pproduct p_2),\ (\lambda j.~(q_1(j))\ppair(q_2(j))))}~(F_1\spifproduct F_2)$.
\end{lemma}

\subsubsection{Combinators related to Sum CTT}

\begin{lemma}[Partially Indexed Second-order Productivity of Spifcopair]\quad\\
\indent\quad For $F_1 :\cFFunpJ{1}{2}{3}{5}$ and \\
\indent\quad $F_2 :\cFFunpJ{1}{2}{4}{5}$\\
\indent\qquad such that $\soproductivepJ{(p_0,p_1,q_1)}(F_1)$ and $\soproductivepJ{(p_0,p_2,q_2)}(F_2)$, \\
\indent\quad let \ $F_1\spifcopair F_2 : (\cFunJ{J}{1}{2})\ra(\sem{\vctt_3\gop{+}\vctt_4}\ra\sem{\vctt_5})$, \\ 
\indent\qquad $F_1\spifcopair F_2 \defeq \lambda f.~ (F_1\ f)\fcopair(F_2\ f)$. \\
\indent\quad Then $\soproductivepJ{(p_0,\ (p_1\pcopair p_2),\ (\lambda j.(q_1(j))\por (q_2(j))))}~(F_1\spifcopair F_2)$.
\end{lemma}

\begin{lemma}[Partially Indexed Second-order Productivity of Spifcoproduct]\quad\\
\indent\quad For $F_1 :\cFFunpJ{1}{2}{3}{4}$ and \\
\indent\quad $F_2 :\cFFunpJ{1}{2}{5}{6}$\\
\indent\qquad such that $\soproductivepJ{(p_0,p_1,q_1)}(F_1)$ and $\soproductivepJ{(p_0,p_2,q_2)}(F_2)$, \\
\indent\quad let \ $F_1\spifcoproduct F_2 : (\cFunJ{J}{1}{2})\ra(\sem{\vctt_3\gop{+}\vctt_5}\ra\sem{\vctt_4\gop{+}\vctt_6})$, \\ 
\indent\qquad $F_1\spifcoproduct F_2 \defeq \lambda f.~ (F_1\ f)\fcoproduct(F_2\ f)$. \\
\indent\quad Then $\soproductivepJ{(p_0,\ (p_1\pcoproduct p_2),\ (\lambda j.~(q_1(j))\ppair(q_2(j))))}~(F_1\spifcoproduct F_2)$.
\end{lemma}

\section{Examples of Defining Coinductive Types and Corecursive Functions}
\label{Examples of Defining Coinductive Types and Corecursive Functions}

In the first subsection, we provide combination diagrams for the examples presented in \Cref{Examples}.
The subsequent subsections present new examples not covered in \Cref{Examples}.

\subsection{Combination Diagrams for Examples}
\label{Combination Diagrams for Examples}
Here we provide combination diagrams for $F_\smap$: one using first-order function combinators and another using second-order function combinators.
We also provide combination diagrams for $F_\growing$, $F_\zip$, $F_\pingpong$ and $F_\bfs$. 

\begin{center}
\begin{figure}
\begin{tikzpicture}[
    box/.style={draw, rectangle, rounded corners, minimum height=0.8cm, minimum width=1.0cm, align=center},
    arr/.style={->, >=Stealth, thick},
    lbl/.style={font=\small}, 
  ]
  \matrix (m) [matrix of nodes, row sep=20pt, column sep=-30pt, nodes={anchor=center} ]
  {
  |[box, name=t01]| $\sem{\streamC\ T_1} \gop{\ra} \streamC\ T_2$ &\ \hspace{20pt}\ &\ \hspace{55pt}\ &\ \hspace{65pt}\ & \\
  &&|[box, name=t11]| $\constCTT{\sem{\streamC\ T_1}}$ & ${\gop{\times}}$ & |[box, name=t12]| $ (\sem{\streamC\ T_1} \gop{\ra} \streamC\ T_2) \gop{\times} \constCTT{\sem{\streamC\ T_1}} $ \\
  &&|[box, name=t31]| $\constCTT{T_2}$ & ${\gop{\times}}$ & |[box, name=t32]| $ \streamC\ T_2 $ \\
  &&& |[box, name=t41]| $\streamC\ T_2$ &\\
  |[box, name=t51]| $\sem{\streamC\ T_1} \gop{\ra} \streamC\ T_2$ &&&& \\
  };
  \draw [arr] (t11) -- (t31) node [midway, right] {$\cany{g\justcompose\head}{}{}$};
  \draw [arr] (t12) -- (t32) node [midway, right] {$\cgeapp (\cany{\tail}{}{})$};
  \draw [arr] (t31) -- (t41) node [midway, right, xshift=5pt] {$\cons$};
  \draw [arr] (t32) -- (t41);
  \coordinate (arr_start) at ($(t01.south west) + (40pt, 0pt)$);
  \coordinate (arr_end) at ($(t51.north west) + (40pt, 0pt)$);
  \coordinate (brace1_start) at ($(t11.north west) + (-2pt, 5pt)$);
  \coordinate (brace1_end) at ($(t11.north west) + (-2pt, -115pt)$);
  \coordinate (brace2_start) at ($(t12.north east) + (0pt, 5pt)$);
  \coordinate (brace2_end) at ($(t12.north east) + (0pt, -115pt)$);
  \draw [arr] (arr_start) -- (arr_end) node [midway, right] {$\cgecurry$};
  \draw[thick] (brace1_start) to[bend right=10] (brace1_end);
  \draw[thick] (brace2_start) to[bend left=10] (brace2_end);
\end{tikzpicture}
\captionof{figure}{Combination diagram of $F_\smap(g)$}\label{map diagram}
\end{figure}
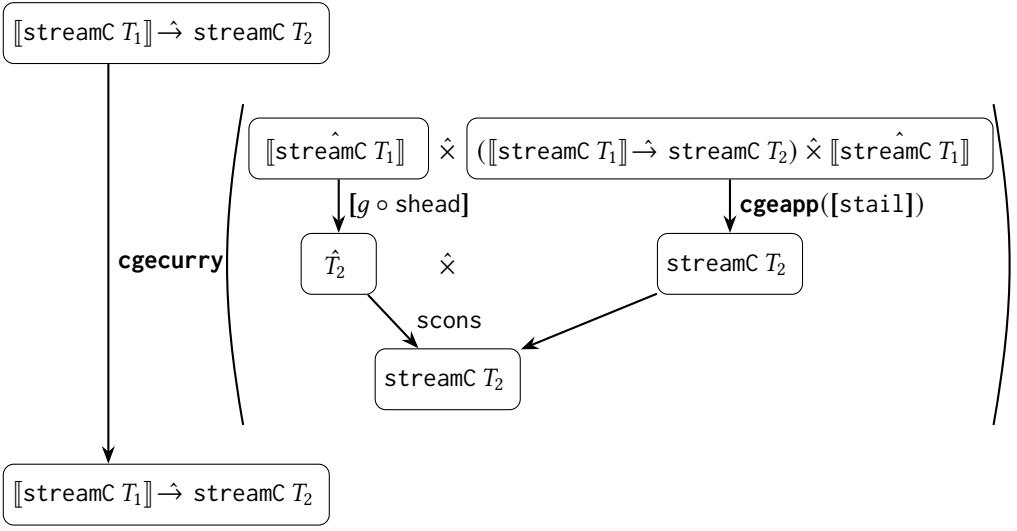
\end{center}

\begin{center}
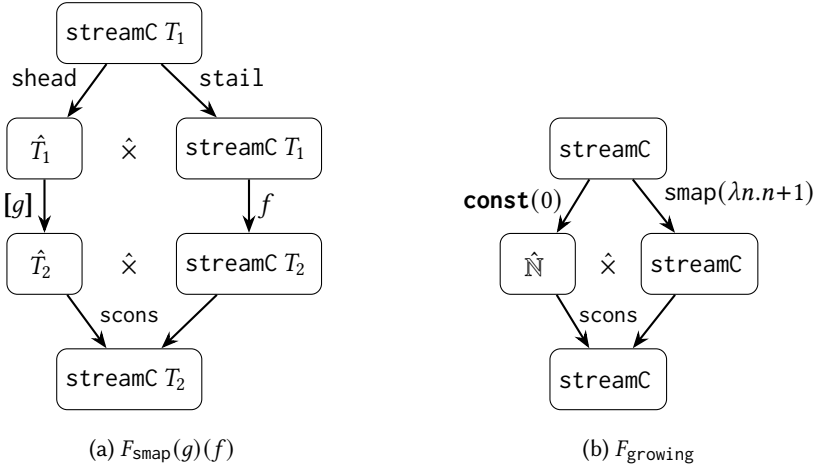
\begin{figure}
\begin{subfigure}{0.45\textwidth}
\begin{center}
  \begin{tikzpicture}[
      box/.style={draw, rectangle, rounded corners, minimum height=0.8cm, minimum width=1.0cm, align=center},
      arr/.style={->, >=Stealth, thick},
      lbl/.style={font=\small}, 
    ]
    \matrix (m) [matrix of nodes, row sep=20pt, column sep=-10pt, nodes={anchor=center} ]
    {
    & |[box, name=t11]| $\streamC\ T_1$ &\\
    |[box, name=t21]| $\constCTT{T_1}$ & ${\gop{\times}}$ & |[box, name=t22]| $ \streamC\ T_1 $ \\
    |[box, name=t31]| $\constCTT{T_2}$ & ${\gop{\times}}$ & |[box, name=t32]| $ \streamC\ T_2 $ \\
    & |[box, name=t41]| $\streamC\ T_2$ &\\
    };
    \draw [arr] (t11) -- (t21) node [midway, left, yshift=5pt] {$\head$};
    \draw [arr] (t11) -- (t22) node [midway, right, yshift=5pt] {$\tail$};
    \draw [arr] (t21) -- (t31) node [midway, left] {$\cany{g}{}{}$};
    \draw [arr] (t22) -- (t32) node [midway, right] {$f$};
    \draw [arr] (t31) -- (t41);
    \draw [arr] (t32) -- (t41);
    \node [lbl, above of=t41, yshift=-5pt] {$\cons$};
\end{tikzpicture}
\caption{$F_\smap(g)(f)$}\label{map so diagram}
\end{center}
\end{subfigure}
\begin{subfigure}{0.45\textwidth}
\begin{center}
  \begin{tikzpicture}[
      box/.style={draw, rectangle, rounded corners, minimum height=0.8cm, minimum width=1.0cm, align=center},
      arr/.style={->, >=Stealth, thick},
      lbl/.style={font=\small}, 
    ]
    \matrix (m) [matrix of nodes, row sep=20pt, column sep=-10pt, nodes={anchor=center} ]
    {
    & |[box, name=t11]| $\streamC$ &\\
    |[box, name=t21]| $\constCTT{\nat}$ & ${\gop{\times}}$ & |[box, name=t22]| $ \streamC $ \\
    & |[box, name=t31]| $\streamC$ &\\
    };
    \draw [arr] (t11) -- (t21) node [midway, left, yshift=2pt] {$\const(0)$};
    \draw [arr] (t11) -- (t22) node [midway, right, yshift=5pt] {$\smap(\lambda n.n\!+\!1)$};
    \draw [arr] (t21) -- (t31);
    \draw [arr] (t22) -- (t31);
    \node [lbl, above of=t31, yshift=-5pt] {$\cons$};
\end{tikzpicture}
\caption{$F_\growing$}\label{growing diagram}
\end{center}
\end{subfigure}
\caption{Combination diagrams of $F_\smap(g)(f)$ and $F_\growing$}
\end{figure}
\end{center}

\begin{center}
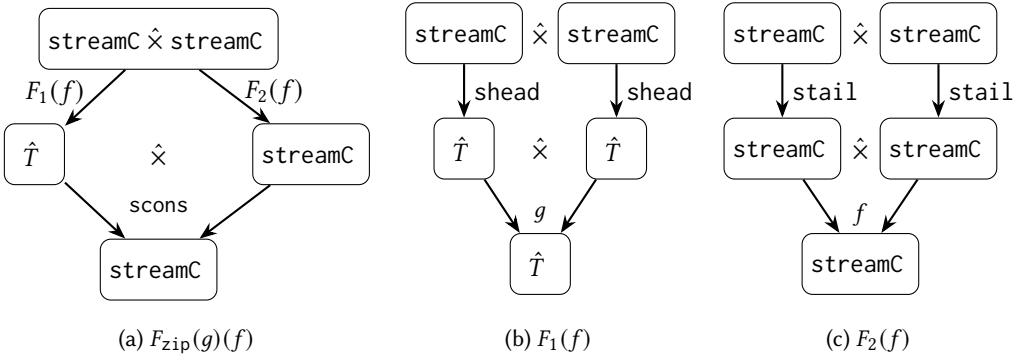
\begin{figure}
\begin{subfigure}{0.38\textwidth}
\begin{center}
\begin{tikzpicture}[
    box/.style={draw, rectangle, rounded corners, minimum height=0.8cm, minimum width=0.8cm, align=center},
    arr/.style={->, >=Stealth, thick},
    lbl/.style={font=\small}, 
  ]
  \matrix (m) [matrix of nodes, row sep=20pt, column sep=-10pt, nodes={anchor=center} ]
  {
  & |[box, name=t11]| $\streamC \gop{\times} \streamC$ &\\
  |[box, name=t21]| $\constCTT{T}$ & ${\gop{\times}}$ & |[box, name=t22]| $ \streamC $ \\
  & |[box, name=t31]| $\streamC$ &\\
  };
  \draw [arr] (t11) -- (t21) node [midway, left, yshift=2pt] {$F_1(f)$};
  \draw [arr] (t11) -- (t22) node [midway, right, yshift=2pt] {$F_2(f)$};
  \draw [arr] (t21) -- (t31);
  \draw [arr] (t22) -- (t31);
  \node [lbl, above of=t31, yshift=-5pt] {$\cons$};
\end{tikzpicture}
\end{center}
\caption{$F_\zip(g)(f)$}\label{zip diagram(a)}
\end{subfigure}
\begin{subfigure}{0.3\textwidth}
\begin{center}
\begin{tikzpicture}[
    box/.style={draw, rectangle, rounded corners, minimum height=0.8cm, minimum width=0.8cm, align=center},
    arr/.style={->, >=Stealth, thick},
    lbl/.style={font=\small}, 
  ]
  \matrix (m) [matrix of nodes, row sep=20pt, column sep=-5pt, nodes={anchor=center} ]
  {
  |[box, name=t21]| $\streamC$ & ${\gop{\times}}$ & |[box, name=t22]| $ \streamC $ \\
  |[box, name=t31]| $\constCTT{T}$ & ${\gop{\times}}$ & |[box, name=t32]| $ \constCTT{T} $ \\
  & |[box, name=t41]| $\constCTT{T}$ &\\
  };
  \draw [arr] (t21) -- (t31) node [midway, right] {$\head$};
  \draw [arr] (t22) -- (t32) node [midway, right] {$\head$};
  \draw [arr] (t31) -- (t41);
  \draw [arr] (t32) -- (t41);
  \node [lbl, above of=t41, yshift=-10pt] {$g$};
\end{tikzpicture}
\end{center}
\caption{$F_1(f)$} \label{zip diagram(b)}
\end{subfigure}
\begin{subfigure}{0.3\textwidth}
\begin{center}
\begin{tikzpicture}[
    box/.style={draw, rectangle, rounded corners, minimum height=0.8cm, minimum width=0.8cm, align=center},
    arr/.style={->, >=Stealth, thick},
    lbl/.style={font=\small}, 
  ]
  \matrix (m) [matrix of nodes, row sep=20pt, column sep=-15pt, nodes={anchor=center} ]
  {
  |[box, name=t21]| $\streamC$ & ${\gop{\times}}$ & |[box, name=t22]| $ \streamC $ \\
  |[box, name=t31]| $\streamC$ & ${\gop{\times}}$ & |[box, name=t32]| $ \streamC $ \\
  & |[box, name=t41]| $\streamC$ &\\
  };
  \draw [arr] (t21) -- (t31) node [midway, right] {$\tail$};
  \draw [arr] (t22) -- (t32) node [midway, right] {$\tail$};
  \draw [arr] (t31) -- (t41);
  \draw [arr] (t32) -- (t41);
  \node [lbl, above of=t41, yshift=-10pt] {$f$};
\end{tikzpicture}
\end{center}
\caption{$F_2(f)$} \label{zip diagram(c)}
\end{subfigure}
\caption{Combination diagram of $F_\zip(g_1,g_2)(f)$} \label{zip diagram}
\end{figure}
\end{center}

\begin{center}
\begin{figure}
\begin{subfigure}{0.8\textwidth}
\begin{tikzpicture}[
    box/.style={draw, rectangle, rounded corners, minimum height=0.8cm, minimum width=1.2cm, align=center},
    arr/.style={->, >=Stealth, thick},
    lbl/.style={font=\small}, 
  ]
    \matrix (m) [matrix of nodes, row sep=20pt, column sep=-50pt, nodes={anchor=center} ]
    {
    |[box, name=t01]| $T \gop{\ra} (\streamC \gop{\times} \streamC)$ &\ \hspace{95pt}\ &&& \\
    &&&|[box, name=t11]| $\constCTT{T} \gop{\times} ((T \gop{\ra} (\streamC \gop{\times} \streamC)) \gop{\times} \constCTT{T})$ & \\
    && |[box, name=t21]| $\streamC$ & ${\gop{\times}}$ & |[box, name=t22]| $\streamC$\\
    |[box, name=t31]| $T \gop{\ra} (\streamC \gop{\times} \streamC)$ &&&& \\
    };
    \draw [arr] (t01) -- (t31) node [midway, right] {$\cgecurry$};
    \draw [arr] (t11) -- (t21) node [midway, right] {$F_1$};
    \draw [arr] (t11) -- (t22) node [midway, right] {$F_2$};
    \coordinate (brace1_start) at ($(t11.north west) + (-5pt, 5pt)$);
    \coordinate (brace1_end) at ($(t11.north west) + (-5pt, -75pt)$);
    \coordinate (brace2_start) at ($(t11.north east) + (5pt, 5pt)$);
    \coordinate (brace2_end) at ($(t11.north east) + (5pt, -75pt)$);
    \draw[thick] (brace1_start) to[bend right=10] (brace1_end);
    \draw[thick] (brace2_start) to[bend left=10] (brace2_end);
\end{tikzpicture}
\caption{$F_\pingpong(g_1,g_2)$}\label{pingpong diagram(a)}
\end{subfigure}
\begin{subfigure}{0.48\textwidth}
\begin{tikzpicture}[
    box/.style={draw, rectangle, rounded corners, minimum height=0.8cm, minimum width=0.7cm, align=center},
    arr/.style={->, >=Stealth, thick},
    lbl/.style={font=\small}, 
  ]
  \matrix (m) [matrix of nodes, row sep=20pt, column sep=-15pt, nodes={anchor=center} ]
  {
  |[box, name=t11]| $\!\constCTT{T} \gop{\times} ((T \gop{\ra}\! (\streamC \gop{\times} \streamC)) \gop{\times} \constCTT{T})\!$\\
  |[box, name=t21]| $(T \gop{\ra} (\streamC \gop{\times} \streamC)) \gop{\times} \constCTT{T}$\\
  |[box, name=t31]| $\streamC \gop{\times} \streamC$\\
  |[box, name=t41]| $\streamC$\\
  };
  \draw [arr] (t11) -- (t21) node [midway, right] {$\snd$};
  \draw [arr] (t21) -- (t31) node [midway, right] {$\mathtt{cegapp}(g_1)$};
  \draw [arr] (t31) -- (t41) node [midway, right] {$\snd$};
\end{tikzpicture}
\caption{$F_1$} \label{pingpong diagram(b)}
\end{subfigure}
\begin{subfigure}{0.48\textwidth}
\begin{tikzpicture}[
    box/.style={draw, rectangle, rounded corners, minimum height=0.8cm, minimum width=0.8cm, align=center},
    arr/.style={->, >=Stealth, thick},
    lbl/.style={font=\small}, 
  ]
  \matrix (m) [matrix of nodes, row sep=20pt, column sep=-15pt, nodes={anchor=center} ]
  {
  |[box, name=t11]| $\!\constCTT{T}\!\!$ & ${\gop{\times}}$ & |[box, name=t12]| $ (T \gop{\ra} (\streamC \gop{\times} \streamC)) \gop{\times} \constCTT{T} \!$ \\
    &  & |[box, name=t21]| $ \streamC \gop{\times} \streamC $ \\
    |[box, name=t31]| $\!\constCTT{T}\!\!$ & \hspace{20pt}${\gop{\times}}$\hspace{-20pt} & |[box, name=t32]| $\streamC$ \\
  &|[box, name=t41]| $\streamC$ & \\
  };
  \draw [arr] (t11) -- (t31) node [midway, right] {$\id$};
  \draw [arr] (t12) -- (t21) node [midway, right] {$\ceapp$};
  \draw [arr] (t21) -- (t32) node [midway, right] {$\zip(g_2)$};
  \coordinate (arr_end1) at ($(t41.north west) + (10pt, 0pt)$);
  \draw [arr] (t31) -- (arr_end1) node [midway, right, xshift=2pt] {$\cons$};
  \coordinate (arr_end2) at ($(t41.north east) + (-10pt, 0pt)$);
  \draw [arr] (t32) -- (arr_end2);
\end{tikzpicture}
\caption{$F_2$} \label{pingpong diagram(c)}
\end{subfigure}
\caption{Combination diagram of $F_\pingpong(g_1,g_2)$} \label{pingpong diagram}
\end{figure}
\end{center}

\begin{center}
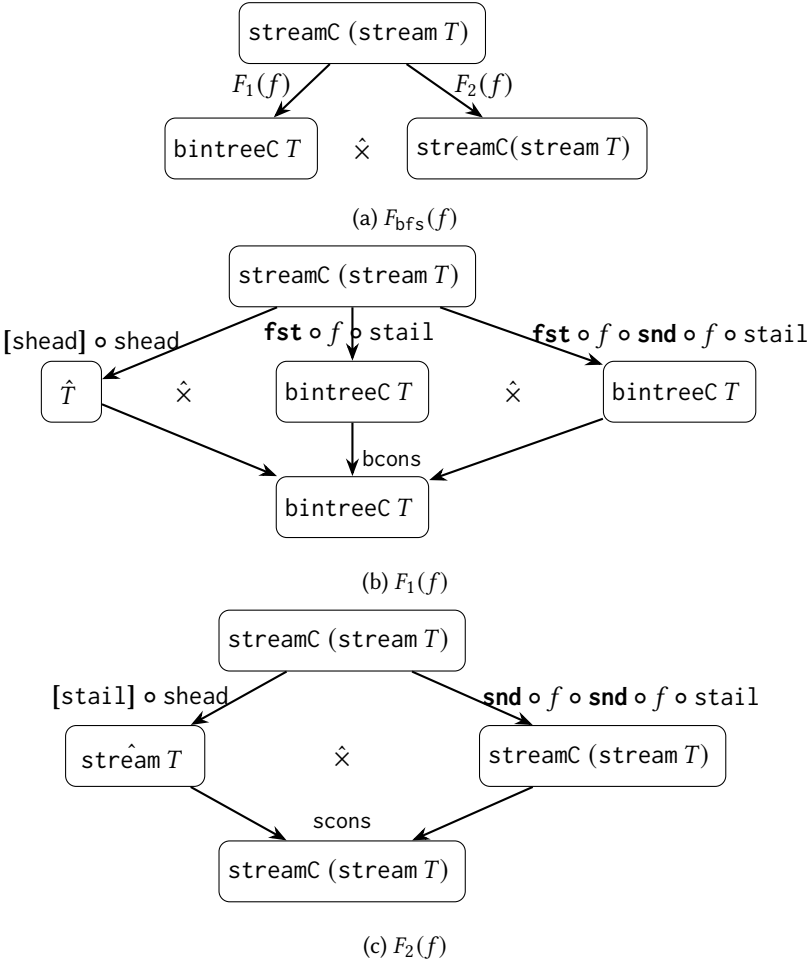
\begin{figure}
\begin{subfigure}{0.99\textwidth}
\begin{center}
\begin{tikzpicture}[
    box/.style={draw, rectangle, rounded corners, minimum height=0.8cm, minimum width=0.8cm, align=center},
    arr/.style={->, >=Stealth, thick},
    lbl/.style={font=\small}, 
  ]
  \matrix (m) [matrix of nodes, row sep=20pt, column sep=-30pt, nodes={anchor=center} ]
  {
  & |[box, name=t11]| $\streamC\ (\stream\ T)$ &\\
  |[box, name=t21]| $\bintreeC\ T$ & ${\gop{\times}}$ & |[box, name=t22]| $ \streamC (\stream\ T)$ \\
  };
  \draw [arr] (t11) -- (t21) node [midway, left, yshift=2pt] {$F_1(f)$};
  \draw [arr] (t11) -- (t22) node [midway, right, yshift=2pt] {$F_2(f)$};
\end{tikzpicture}
\end{center}
\caption{$F_\bfs(f)$}\label{bfs diagram(a)}
\end{subfigure}
\begin{subfigure}{0.99\textwidth}
\begin{center}
\begin{tikzpicture}[
    box/.style={draw, rectangle, rounded corners, minimum height=0.8cm, minimum width=0.8cm, align=center},
    arr/.style={->, >=Stealth, thick},
    lbl/.style={font=\small}, 
  ]
  \matrix (m) [matrix of nodes, row sep=20pt, column sep=15pt, nodes={anchor=center} ]
  {
  && |[box, name=t11]| $\streamC\ (\stream\ T)$ &\\
  |[box, name=t21]| $\constCTT{T}$ & \hspace{7pt}${\gop{\times}}$\hspace{-7pt} & |[box, name=t22]| $\bintreeC\ T$ & \hspace{-10pt}${\gop{\times}}$\hspace{10pt} & |[box, name=t23]| $ \bintreeC\ T$ \\
  && |[box, name=t31]| $\bintreeC\ T$ &\\
  };
  \draw [arr] (t11) -- (t21) node [midway, left] {$\cany{\head}{}{}\fcompose\head$};
  \draw [arr] (t11) -- (t22) node [midway, left, xshift=35pt] {$\fst\fcompose f\fcompose\tail$};
  \draw [arr] (t11) -- (t23) node [midway, right] {$\fst\fcompose f\fcompose\snd\fcompose f\fcompose\tail$};
  \draw [arr] (t21) -- (t31);
  \draw [arr] (t22) -- (t31);
  \draw [arr] (t23) -- (t31);
  \node [lbl, above of=t31, xshift=15pt, yshift=-10pt] {$\bcons$};
\end{tikzpicture}
\end{center}
\caption{$F_1(f)$} \label{bfs diagram(b)}
\end{subfigure}
\begin{subfigure}{0.99\textwidth}
\begin{center}
\begin{tikzpicture}[
    box/.style={draw, rectangle, rounded corners, minimum height=0.8cm, minimum width=0.8cm, align=center},
    arr/.style={->, >=Stealth, thick},
    lbl/.style={font=\small}, 
  ]
  \matrix (m) [matrix of nodes, row sep=20pt, column sep=5pt, nodes={anchor=center} ]
  {
  & |[box, name=t11]| $\streamC\ (\stream\ T)$ &\\
  |[box, name=t22]| $\constCTT{\stream\ T}$ & ${\gop{\times}}$ & |[box, name=t23]| $ \streamC\ (\stream\ T)$ \\
  & |[box, name=t31]| $\streamC\ (\stream\ T)$ &\\
  };
  \draw [arr] (t11) -- (t22) node [midway, left] {$\cany{\tail}{}{}\fcompose\head$};
  \draw [arr] (t11) -- (t23) node [midway, right] {$\snd\fcompose f\fcompose\snd\fcompose f\fcompose\tail$};
  \draw [arr] (t22) -- (t31);
  \draw [arr] (t23) -- (t31);
  \node [lbl, above of=t31, yshift=-10pt] {$\cons$};
\end{tikzpicture}
\end{center}
\caption{$F_2(f)$} \label{bfs diagram(c)}
\end{subfigure}
\caption{Combination diagram of $F_\bfs(f)$} \label{bfs diagram}
\end{figure}
\end{center}

\subsection{Corecursive Function $\fib$} \label{example fib}
$\fib$ is a corecursively defined stream whose $i$-th element is the $i$-th fibonacci number.
The productivity of $F_\fib$ is automatically proven to be a subset of $\stdunifZ(1)$.
Therefore, we obtain the unique fixed point $\fib$ of $F_\fib$.
\[\begin{array}{@{}r@{\ }c@{~}l@{\ \ }l@{}}  
  F_\fib&:&\stream\ \nat \ra \stream\ \nat \\
  F_\fib&\defeq& 
\lambda x.~ \cons \ 0 \ (\cons \ 1 \ (\zip \ (\lambda (n_1,n_2). n_1+n_2)\ (x,\ \tail \ x))).
\end{array}\]
\[\begin{array}{@{}r@{\ }c@{~}l@{\ \ }l@{}}  
  F_\fib &:& \sem{\streamC}\ra\sem{\streamC} \\
  F_\fib &=& \cons \fcompose (\const(0) \fpair (\cons \fcompose (\const(1) \fpair (
    (\zip \ (\lambda (n_1,n_2). n_1+n_2)) \fcompose (\id \fpair \tail)))))
\end{array}\]

\begin{center}
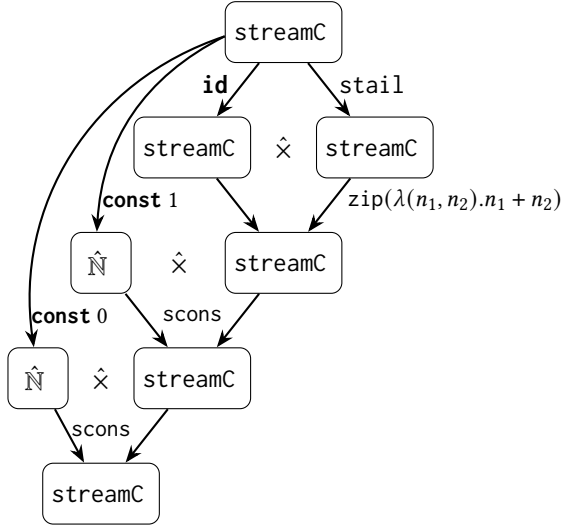
\begin{figure}
\begin{tikzpicture}[
    box/.style={draw, rectangle, rounded corners, minimum height=0.8cm, minimum width=0.8cm, align=center},
    arr/.style={->, >=Stealth, thick},
    lbl/.style={font=\small}, 
  ]
  \matrix (m) [matrix of nodes, row sep=20pt, column sep=-10pt, nodes={anchor=center} ]
  {
  &&& |[box, name=t11]| $\streamC $ &\\
  &&|[box, name=t21]| $\streamC$ & ${\gop{\times}}$ & |[box, name=t22]| $ \streamC $ \\
  & |[box, name=t31]| $\constCTT{\nat}$ & \hspace{-5pt}${\gop{\times}}$\hspace{5pt} & |[box, name=t32]| $ \streamC $ \\
  |[box, name=t41]| $\constCTT{\nat}$ & ${\gop{\times}}$ & |[box, name=t42]| $ \streamC $ \\
  & |[box, name=t51]| $\streamC $ &\\
  };
  \draw [arr] (t11) -- (t21) node [midway, left, yshift=2pt] {$\id$};
  \draw [arr] (t11) -- (t22) node [midway, right, yshift=2pt] {$\tail$};
  \draw [arr] (t21) -- (t32);
  \draw [arr] (t22) -- (t32);
  \node [lbl, above of=t32, xshift=65pt, yshift=-5pt] {$\zip(\lambda (n_1,n_2). n_1+n_2)$};
  \draw [arr] (t31) -- (t42);
  \draw [arr] (t32) -- (t42);
  \node [lbl, above of=t42, yshift=-5pt] {$\cons$};
  \draw [arr] (t41) -- (t51);
  \draw [arr] (t42) -- (t51);
  \node [lbl, above of=t51, yshift=-5pt] {$\cons$};
  \coordinate (brace1_start) at ($(t11.south west) + (0pt, 10pt)$);
  \coordinate (brace1_end) at ($(t31.north west) + (10pt, 0pt)$);
  \coordinate (brace2_end) at ($(t41.north west) + (10pt, 0pt)$);
  \draw[arr] (brace1_start) to[bend right=30] (brace1_end);
  \draw[arr] (brace1_start) to[bend right=40] (brace2_end);
  \node [lbl, above of=t31, xshift=15pt, yshift=-5pt] {$\const\ 1$};
  \node [lbl, above of=t41, xshift=12pt, yshift=-5pt] {$\const\ 0$};
\end{tikzpicture}
\caption{Combination diagram of $F_\fib$} \label{fib diagram}
\end{figure}
\end{center}

\subsection{Corecursive Function $\roundrobin$} \label{example roundrobin}
We first define the recursive function $\mathtt{adds}$ with inductive domain $\nat$. 
Given a function $l$ and a stream $x$,
$\mathtt{adds}(m)(l,x)$ produces a stream by prepending $m$ elements $l(0),\ l(1),\ ...,\ l(m-1)$ to $x$.
\[\begin{array}{@{}r@{\ }c@{~}l@{\ \ }l@{}}  
  \text{For } m&\in& \nat,\ l\in \nat\ra T, \\ 
  \mathtt{adds}(m)(l) &:& \stream\ T \ra \stream\ T\\
  \mathtt{adds}(0)(l) &\defeq& \lambda x.~x\\
  \mathtt{adds}(m+1)(l) &\defeq& \lambda x.~ \mathtt{adds}(m)(l)(\cons(l(m), x))
\end{array}\]
\[\begin{array}{@{}r@{\ }c@{~}l@{\ \ }l@{}}  
  \mathtt{adds}(m)(l) &:& \sem{\streamC\ T} \ra \sem{\streamC\ T}\\
  \mathtt{adds}(0)(l) &=& \id\\
  \mathtt{adds}(m+1)(l) &=& \mathtt{adds}(m)(l) \fcompose \cons \fcompose 
  (\const(l(m):\sem{\constCTT{T}}) \fpair \id)
\end{array}\]
We then obtain $\productive{\stdunifZ(m)}(\mathtt{adds}(m)(l))$ by induction on $m$.

Using $\mathtt{adds}$, we define the corecursive function $\roundrobin(m)$.
Given $xs\in\nat\ra\stream$, 
$\roundrobin(m)(xs)$ is a stream that picks elements one by one from $xs(0),\ xs(1), ...,\ xs(m-1)$ in round-robin order.
\[\begin{array}{@{}r@{\ }c@{~}l@{\ \ }l@{}}  
  F_\roundrobin(m)&:&((\nat\ra \stream\ T)\ra\stream\ T)\ra((\nat\ra \stream\ T)\ra\stream\ T) \\
  F_\roundrobin(m)&\defeq& 
\lambda f.~\lambda xs.~\mathtt{adds}(m)(\lambda n.~\head(xs(n)))(f(\lambda n.~\tail(xs(n)))).
\end{array}\]
\[\begin{array}{@{}r@{\ }c@{~}l@{\ \ }l@{}}  
  F_\roundrobin(m) &:& (\sem{\nat\gop{\ra}\streamC}\ra\sem{\streamC})\ra(\sem{\nat\gop{\ra}\streamC}\ra\sem{\streamC}) \\
  F_\roundrobin(m) &=& (\sfindep\ \mathtt{cuncurry}(\mathtt{adds}(m)))\ \sfcompose \\
    &&\qquad(\sfindep(\tconop{\nat\ra}.\mathtt{map}\ (\lambda\_.\head)) \sfpair
      (\sfself \sfcompose \sfindep(\tconop{\nat\ra}.\mathtt{map}\ (\lambda\_.\tail))))
\end{array}\]

\begin{center}
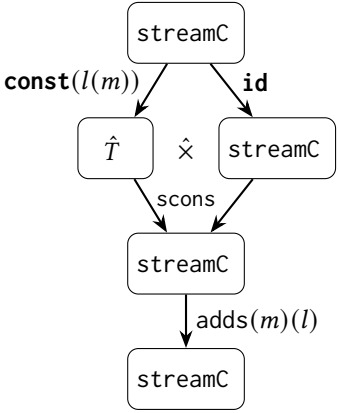
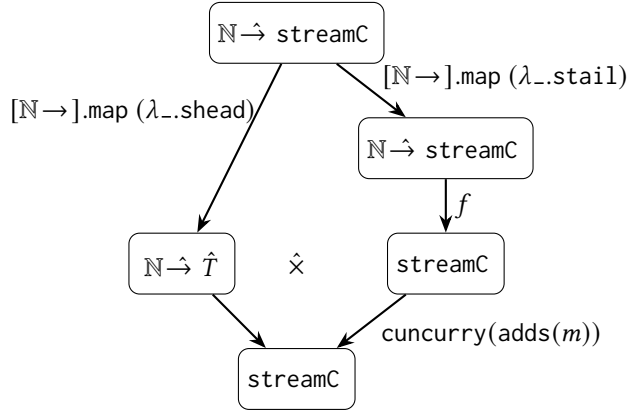
\begin{figure}
\begin{subfigure}{0.35\textwidth}
\begin{center}
  \begin{tikzpicture}[
      box/.style={draw, rectangle, rounded corners, minimum height=0.8cm, minimum width=1.0cm, align=center},
      arr/.style={->, >=Stealth, thick},
      lbl/.style={font=\small}, 
    ]
    \matrix (m) [matrix of nodes, row sep=20pt, column sep=-10pt, nodes={anchor=center} ]
    {
    & |[box, name=t11]| $\streamC$ &\\
    |[box, name=t21]| $\constCTT{T}$ & ${\gop{\times}}$ & |[box, name=t22]| $ \streamC $ \\
    & |[box, name=t31]| $\streamC$ &\\
    & |[box, name=t41]| $\streamC$ &\\
    };
    \draw [arr] (t11) -- (t21) node [midway, left, yshift=3pt] {$\const(l(m))$};
    \draw [arr] (t11) -- (t22) node [midway, right, yshift=3pt] {$\id$};
    \draw [arr] (t21) -- (t31);
    \draw [arr] (t22) -- (t31);
    \draw [arr] (t31) -- (t41) node [midway, right] {$\mathtt{adds}(m)(l)$};
    \node [lbl, above of=t31, yshift=-5pt] {$\cons$};
\end{tikzpicture}
\caption{$\mathtt{adds}(m+1)(l)$}\label{adds diagram}
\end{center}
\end{subfigure}
\begin{subfigure}{0.60\textwidth}
\begin{center}
\begin{tikzpicture}[
    box/.style={draw, rectangle, rounded corners, minimum height=0.8cm, minimum width=1.0cm, align=center},
    arr/.style={->, >=Stealth, thick},
    lbl/.style={font=\small}, 
  ]
  \matrix (m) [matrix of nodes, row sep=20pt, column sep=-10pt, nodes={anchor=center} ]
  {
  & |[box, name=t11]| $\nat \gop{\ra} \streamC$ &\\
  & & |[box, name=t22]| $ \nat \gop{\ra} \streamC$ \\
  |[box, name=t31]| \ $\nat \gop{\ra} \constCTT{T}$\hspace{-2pt} & ${\gop{\times}}$ & |[box, name=t32]| $ \streamC $ \\
  & |[box, name=t41]| $\streamC$ &\\
  };
  \draw [arr] (t11) -- (t22) node [midway, right, yshift=5pt] {$\tconop{\nat\ra}.\mathtt{map}\ (\lambda\_.\tail)$};
  \draw [arr] (t11) -- (t31) node [midway, left, yshift=14pt, xshift=10pt] {$\tconop{\nat\ra}.\mathtt{map}\ (\lambda\_.\head)$};
  \draw [arr] (t22) -- (t32) node [midway, right] {$f$};
  \draw [arr] (t31) -- (t41);
  \draw[arr] (t32) -- (t41) node [midway, right, yshift=-5pt, xshift=0pt] {$\mathtt{cuncurry}(\mathtt{adds}(m))$};
\end{tikzpicture}
\caption{$F_\roundrobin(m)(f)$}\label{roundrobin diagram}
\end{center}
\end{subfigure}
\caption{Combination diagrams of $\mathtt{adds}(m+1)(l)$ and $F_\roundrobin(m)(f)$}
\end{figure}
\end{center}

After specifying the expected productivity $p_m=\stdunif(\lambda n.~\some((n+m\!-\!1)/m))$ for $\roundrobin(m)$,
we can automatically prove that $\soproductive{(p_m,p_m,\stdunifZ(1))}(F_\roundrobin(m))$. 
Then we obtain the unique fixed point $\roundrobin(m)$ of $F_\roundrobin(m)$, which is also $\productive{p_m}$. 

\subsection{Corecursive Function $\distribute$} \label{example distribute}
We first define the corecursive function $\mathtt{nkth}(m)$, 
which extracts the $(n\times m)$-th elements from the input stream for each $n\in\nat$.
Note that we use 0-indexing, where a stream starts with its 0th element.
\[\begin{array}{@{}r@{\ }c@{~}l@{\ \ }l@{}}  
  \text{For } 0<m&\in& \nat,\\ 
  F_\mathtt{nkth}(m) &:& (\stream\ T \ra \stream\ T)\ra(\stream\ T \ra \stream\ T)\\
  F_\mathtt{nkth}(m) &\defeq& \lambda f.~\lambda x.~\cons(\head\ x,\ f(\tail^m(x)))
\end{array}\]
\[\begin{array}{@{}r@{\ }c@{~}l@{\ \ }l@{}}  
  F_\mathtt{nkth}(m) &:& (\sem{\streamC} \ra \sem{\streamC})\ra(\sem{\streamC} \ra \sem{\streamC})\\
  F_\mathtt{nkth}(m) &=& (\sfindep\ \cons)\sfcompose
    ((\sfindep\ \head) \sfpair (\sfself \sfcompose (\sfindep\ \tail^m)) )
\end{array}\]

After specifying the expected productivity $p_m=\stdunif(\lambda n.~\some(n~\times~m))$ for $\mathtt{nkth}(m)$,
we can automatically prove that $\soproductive{(p_m,p_m,\stdunifZ(1))}(F_\mathtt{nkth}(m))$. 
Then we obtain the unique fixed point $\mathtt{nkth}(m)$ of $F_\mathtt{nkth}(m)$, which is also $\productive{p_m}$.

Now we can define $\distribute$ using $\mathtt{nkth}$.
\[\begin{array}{@{}r@{\ }c@{~}l@{\ \ }l@{}}  
  \text{For } 0<m&\in& \nat,\\ 
  \distribute(m) &:& (\stream\ T \ra (\nat\ra \stream\ T))\ra(\stream\ T \ra (\nat\ra \stream\ T))\\
  \distribute(m) &\defeq& \lambda x.~\lambda i.~\ite{i<m}{\mathtt{nkth}(m)(\tail^i(x))}{\mathtt{nkth}(m)(x)}
\end{array}\]
\[\begin{array}{@{}r@{\ }c@{~}l@{\ \ }l@{}}  
  \distribute(m) &:& (\sem{\streamC} \ra \sem{\nat\gop{\ra}\streamC})\ra(\sem{\streamC} \ra \sem{\nat\gop{\ra}\streamC})\\
  \distribute(m) &=& \ceswap(\lambda i.~\ite{i<m}{\mathtt{nkth}(m) \fcompose \tail^i}{\mathtt{nkth}(m)})
\end{array}\]
Then we get $\productive{\stdunif(\lambda n.~\some(m\times n+m-1))}(\distribute(m))$ for $m>0$.

\begin{center}
\begin{figure}
\begin{subfigure}{0.28\textwidth}
\begin{center}
\begin{tikzpicture}[
    box/.style={draw, rectangle, rounded corners, minimum height=0.8cm, minimum width=0.8cm, align=center},
    arr/.style={->, >=Stealth, thick},
    lbl/.style={font=\small}, 
  ]
  \matrix (m) [matrix of nodes, row sep=20pt, column sep=-10pt, nodes={anchor=center} ]
  {
  & |[box, name=t11]| $\streamC$ &\\
  & & |[box, name=t22]| $ \streamC $ \\
  |[box, name=t21]| $\constCTT{T}$ & ${\gop{\times}}$ & |[box, name=t32]| $ \streamC $ \\
  & |[box, name=t41]| $\streamC$ &\\
  };
  \draw [arr] (t11) -- (t21) node [midway, left, yshift=3pt] {$\head$};
  \draw [arr] (t11) -- (t22) node [midway, right, yshift=3pt] {$\tail^m$};
  \draw [arr] (t22) -- (t32) node [midway, right] {$f$};
  \draw [arr] (t21) -- (t41);
  \draw [arr] (t32) -- (t41);
  \node [lbl, above of=t41, yshift=-5pt] {$\cons$};
\end{tikzpicture}
\caption{$\mathtt{nkth}(m)(f)$}\label{nkth diagram}
\end{center}
\end{subfigure}
\begin{subfigure}{0.7\textwidth}
\begin{tikzpicture}[
  box/.style={draw, rectangle, rounded corners, minimum height=0.8cm, minimum width=1.2cm, align=center},
  arr/.style={->, >=Stealth, thick},
  lbl/.style={font=\small}, 
]
  \matrix (m) [matrix of nodes, row sep=20pt, column sep=-10pt, nodes={anchor=center} ]
  {
  |[box, name=t01]| $\streamC$ &\ \hspace{50pt}\ & \\
  &&|[box, name=t11]| $\streamC$  \\
  && |[box, name=t21]| $\streamC$ \\
  |[box, name=t31]| $\nat\gop{\ra}{\streamC}$ && \\
  };
  \draw [arr] (t01) -- (t31) node [midway, right] {$\ceswap\ \ \lambda (i\in\nat).$};
  \draw [arr] (t11) -- (t21) node [midway, right] {$\ite{i\!<\!m}{(\mathtt{nkth}(m) \fcompose \tail^i)}{\mathtt{nkth}(m)}$};
  \coordinate (brace1_start) at ($(t11.north west) + (-32pt, 5pt)$);
  \coordinate (brace1_end) at ($(t11.north west) + (-32pt, -75pt)$);
  \coordinate (brace2_start) at ($(t11.north east) + (132pt, 5pt)$);
  \coordinate (brace2_end) at ($(t11.north east) + (132pt, -75pt)$);
  \draw[thick] (brace1_start) to[bend right=10] (brace1_end);
  \draw[thick] (brace2_start) to[bend left=10] (brace2_end);
\end{tikzpicture}
\caption{$\distribute(m)$}\label{distribute diagram}
\end{subfigure}
\caption{Combination diagrams of $\mathtt{nkth}(m)(f)$ and $\distribute(m)$}
\end{figure}
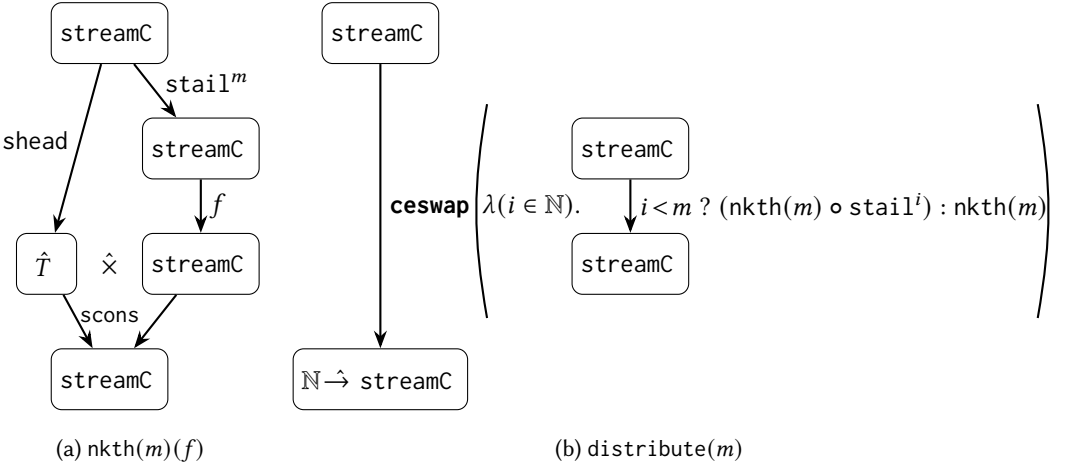
\end{center}

\subsection{Corecursive Function $\filter$} \label{example filter}
To illustrate the application of \Cref{wf_fp},
we define the partial corecursive function
$\filter(P) : \stream\ra\stream$ with generating function $F_\filter(P)$ as follows.
\[\begin{array}{@{}l@{\ }l@{\ \ }l@{}}  
  \text{For } P : T\ra\Bool,\ 
  F_\filter(P) : \stream\ra\stream  \\
  F_\filter(P) = \lambda f.~\lambda x.~ \ite{P(\head\ x)}{\cons(\head\ x,\ f(\tail\ x))}{f(\tail\ x)}
\end{array}\]
Using combinators, $F_\filter(P)$ can be represented as follows.
We employ nondegenerate function CTT since we need to consider whether $P(\head \ x)$ holds for the argument $x\in\sem{\streamC}$.
\[\begin{array}{@{}r@{\ }c@{~}l@{\ \ }l@{}}  
  F_\filter(P) &:& 
  \sem{\gdepfun{\_:\sem{\streamC}}{\streamC}} \ra \sem{\gdepfun{\_:\sem{\streamC}}{\streamC}} & \\
  F_\filter(P) &=& \cnswap (\lambda x.~ \ite{P(\head\ x)}{\\
  &&\quad \cons\fcompose (\const(\head\ x) \fpair \cnproj(\tail\ x)\ )}{\\
  &&\quad \cnproj(\tail\ x)} \\
  &&)
\end{array}\]

\begin{center}
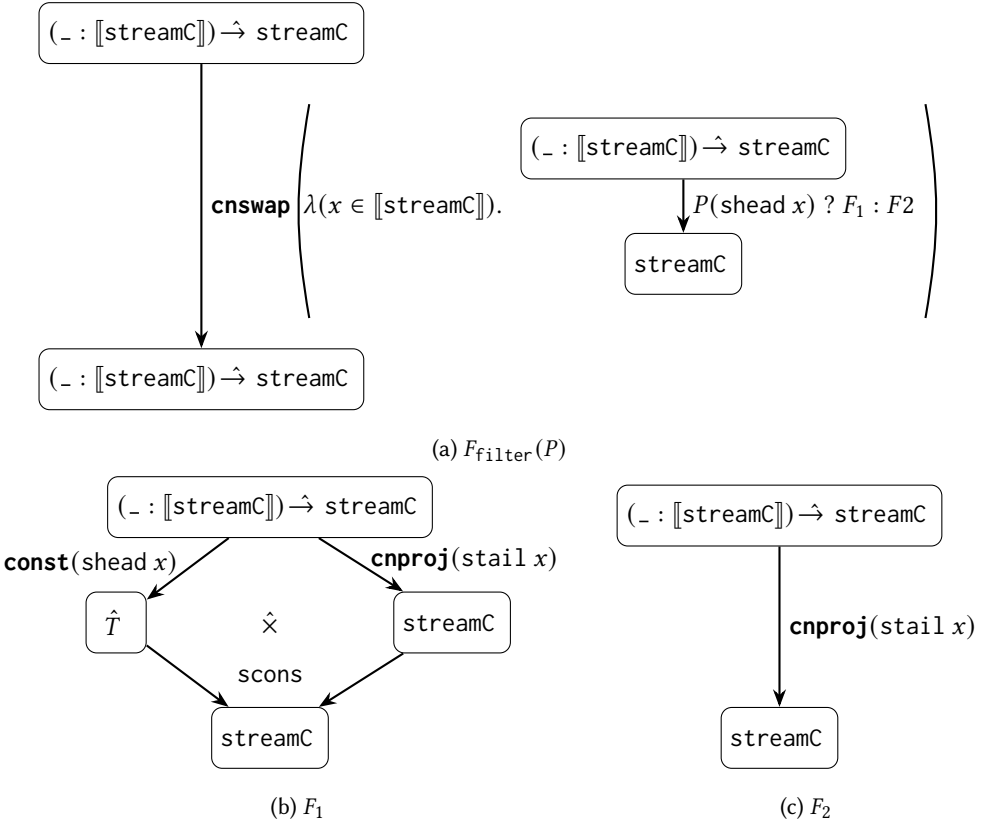
\begin{figure}
\begin{subfigure}{0.9\textwidth}
\begin{tikzpicture}[
    box/.style={draw, rectangle, rounded corners, minimum height=0.8cm, minimum width=1.2cm, align=center},
    arr/.style={->, >=Stealth, thick},
    lbl/.style={font=\small}, 
  ]
    \matrix (m) [matrix of nodes, row sep=20pt, column sep=-10pt, nodes={anchor=center} ]
    {
    |[box, name=t01]| $\gdepfun{\_:\sem{\streamC}}{\streamC}$ &\ \hspace{70pt}\ & \\
    &&|[box, name=t11]| $\gdepfun{\_:\sem{\streamC}}{\streamC}$  \\
    && |[box, name=t21]| $\streamC$ \\
    |[box, name=t31]| $\gdepfun{\_:\sem{\streamC}}{\streamC}$ && \\
    };
    \draw [arr] (t01) -- (t31) node [midway, right] {$\cnswap\ \ \lambda (x\in\sem{\streamC}).$};
    \draw [arr] (t11) -- (t21) node [midway, right] {$\ite{P(\head\ x)}{F_1}{F2}$};
    \coordinate (brace1_start) at ($(t11.north west) + (-80pt, 5pt)$);
    \coordinate (brace1_end) at ($(t11.north west) + (-80pt, -75pt)$);
    \coordinate (brace2_start) at ($(t11.north east) + (30pt, 5pt)$);
    \coordinate (brace2_end) at ($(t11.north east) + (30pt, -75pt)$);
    \draw[thick] (brace1_start) to[bend right=10] (brace1_end);
    \draw[thick] (brace2_start) to[bend left=10] (brace2_end);
\end{tikzpicture}
\caption{$F_\filter(P)$}\label{filter diagram(a)}
\end{subfigure}
\begin{subfigure}{0.58\textwidth}
\begin{tikzpicture}[
    box/.style={draw, rectangle, rounded corners, minimum height=0.8cm, minimum width=0.8cm, align=center},
    arr/.style={->, >=Stealth, thick},
    lbl/.style={font=\small}, 
  ]
  \matrix (m) [matrix of nodes, row sep=20pt, column sep=-15pt, nodes={anchor=center} ]
  {
  &|[box, name=t11]| $\gdepfun{\_:\sem{\streamC}}{\streamC}$ & \\
    |[box, name=t31]| $\constCTT{T}$ & ${\gop{\times}}$ & |[box, name=t32]| $\streamC$ \\
  &|[box, name=t41]| $\streamC$ & \\
  };
  \draw [arr] (t11) -- (t31) node [midway, left, yshift=2pt] {$\const(\head\ x)$};
  \draw [arr] (t11) -- (t32) node [midway, right, yshift=2pt] {$\cnproj(\tail\ x)$};
  \draw [arr] (t31) -- (t41) node [midway, right, xshift=15pt] {$\cons$};
  \draw [arr] (t32) -- (t41);
\end{tikzpicture}
\caption{$F_1$} \label{filter diagram(b)}
\end{subfigure}
\begin{subfigure}{0.38\textwidth}
\begin{tikzpicture}[
    box/.style={draw, rectangle, rounded corners, minimum height=0.8cm, minimum width=0.8cm, align=center},
    arr/.style={->, >=Stealth, thick},
    lbl/.style={font=\small}, 
  ]
    \matrix (m) [matrix of nodes, row sep=20pt, column sep=-10pt, nodes={anchor=center} ]
    {
    |[box, name=t11]| $\gdepfun{\_:\sem{\streamC}}{\streamC}$  \\
    \\
    \\
    |[box, name=t21]| $\streamC$  \\
    };
  \draw [arr] (t11) -- (t21) node [midway, right] {$\cnproj(\tail\ x)$};
\end{tikzpicture}
\caption{$F_2$} \label{filter diagram(c)}
\end{subfigure}
\caption{Combination diagram of $F_\filter(P)$} \label{filter diagram}
\end{figure}
\end{center}

The well-definedness of $\filter(P)(x)$ requires the condition $\infsat{P}{x}$,
a coinductive property defined below.
This property asserts that the stream $x$ has infinitely many elements satisfying $P$,
ensuring that retaining only these elements suffices to define a stream.
\[ \frac{x\in\stream\ T\quad n\in\nat\quad \tail^n(x)=\cons(h,t)\quad P(h)=\btrue\quad \infsat{P}{t}}{\infsat{P}{x}} \]

We cannot directly apply \Cref{wf_fp}, as the productivity of $F_\filter(P)$ is not well-founded in general.
However, when restricting attention to streams satisfying $\infsat{P}{x}$,
the following property holds:

\begin{lemma}\label[lemma]{filterF_prod}
  For the canonical productivity $p$ of $F_\filter(P)$ and $x\in\stream\ T$ such that $\infsat{P}{x}$, \\
  \indent\quad $\forall n\in\nat,\ \sacc(p)(\some(\lambda x'.~\ite{x\!=\!x'}{\leveleach{n}}{\bot}))$.
\end{lemma}

By combining \Cref{wf_fp_pre}, \Cref{NonDgnFunCTT_dimeq} and \Cref{filterF_prod}, 
we obtain $\filter(P)$ and its fixed point equation. 
\begin{lemma}\label[lemma]{filter_fix}
  For $P : T\ra\Bool$, $\exists~\filter(P) : \stream\ T\ra\stream\ T$, \\
  \indent\quad $\forall x\in\stream\ T,\ \infsat{P}{x} \implies F_\filter(P)(\filter(P))(x) = \filter(P)(x)$.
\end{lemma}
Note that the only manual steps for the user are representing $F_\filter(P)$ using combinators and proving \Cref{filterF_prod}.
All other steps are automated by the framework.

\subsection{Corecursive Function $\bmap$} \label{example bmap}
The function $\bmap$ is a map function from $\bintree\ T_1$ to $\bintree\ T_2$, 

\[\begin{array}{@{}l@{\ }l@{\ \ }l@{}}  
  \text{For } g &:& T_1\ra T_2,\\ 
  F_\bmap(g) &:& (\bintree\ T_1\ra\bintree\ T_2)\ra(\bintree\ T_1\ra\bintree\ T_2) \\
  F_\bmap(g) &=& \lambda f.~\lambda x.~ \bcons\ (g(\bhead\ x),\ f(\bleft\ x),\ f(\bright\ x))
\end{array}\]
\[\begin{array}{@{}r@{\ }c@{~}l@{\ \ }l@{}}  
  F_\bmap(g) & : & (\sem{\bintreeC\ T_1}\ra\sem{\bintreeC\ T_2})\ra(\sem{\bintreeC\ T_1}\ra\sem{\bintreeC\ T_2}) \\
  F_\bmap(g) &=& (\sfindep\ \bcons)\ \sfcompose 
    (\sfindep(\cany{g}{}{}\fcompose\bhead)\ \sfpair 
    (\sfself \sfcompose (\sfindep\ \bleft))\ \sfpair 
    (\sfself \sfcompose (\sfindep\ \bright)))
\end{array}\]

\begin{center}
\begin{figure}
\begin{tikzpicture}[
    box/.style={draw, rectangle, rounded corners, minimum height=0.8cm, minimum width=1.0cm, align=center},
    arr/.style={->, >=Stealth, thick},
    lbl/.style={font=\small}, 
  ]
  \matrix (m) [matrix of nodes, row sep=20pt, column sep=0pt, nodes={anchor=center} ]
  {
  && |[box, name=t11]| $\bintreeC\ T_1$ &\\
  |[box, name=t21]| $\constCTT{T_1}$ & ${\gop{\times}}$ & |[box, name=t22]| $ \bintreeC\ T_1 $ & ${\gop{\times}}$ & |[box, name=t23]| $ \bintreeC\ T_1 $ \\
  |[box, name=t31]| $\constCTT{T_2}$ & ${\gop{\times}}$ & |[box, name=t32]| $ \bintreeC\ T_2 $ & ${\gop{\times}}$ & |[box, name=t33]| $ \bintreeC\ T_2 $ \\
  && |[box, name=t41]| $\bintreeC\ T_2$ &\\
  };
  \draw [arr] (t11) -- (t21) node [midway, left, yshift=2pt] {$\bhead$};
  \draw [arr] (t11) -- (t22) node [midway, right, yshift=0pt] {$\bleft$};
  \draw [arr] (t11) -- (t23) node [midway, right, yshift=2pt] {$\bright$};
  \draw [arr] (t21) -- (t31) node [midway, left] {$\cany{g}{}{}$};
  \draw [arr] (t22) -- (t32) node [midway, right] {$f$};
  \draw [arr] (t23) -- (t33) node [midway, right] {$f$};
  \draw [arr] (t31) -- (t41);
  \draw [arr] (t32) -- (t41);
  \draw [arr] (t33) -- (t41);
  \node [lbl, above of=t41, xshift=15pt, yshift=-5pt] {$\bcons$};
\end{tikzpicture}
\caption{Combination diagram of $F_\bmap(g)(f)$}\label{bmap diagram}
\end{figure}
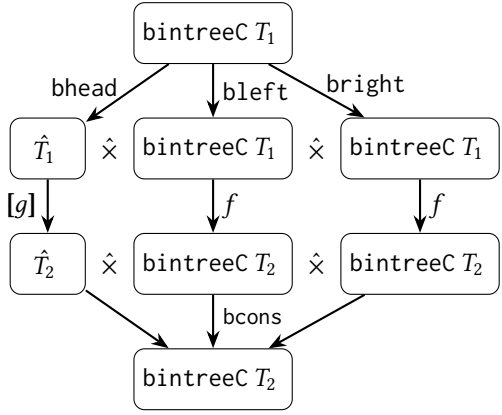
\end{center}

After specifying the expected productivity $\stdunifZ(0)$ for $\bmap(g)$,
we can automatically prove that $\soproductive{(\stdunifZ(0),\stdunifZ(0),\stdunifZ(1))}(F_\bmap(g))$. 
Then we obtain the unique fixed point $\bmap(g)$ of $F_\bmap(g)$, which is also $\productive{\stdunifZ(0)}$. 

\subsection{Corecursive Function $\sbt$} \label{example sbt}
$\sbt$ is the Stern-Brocot tree, an infinite binary tree with fraction nodes presented in \cite{hinze09}.
It served as an illustrative example in \cite{blanchette17}.

$\sbtsuc \defeq \bmap(\lambda\ (n_1,n_2).~ (n_1+n_2,n_2))$,\ 
$\sbtinv \defeq \bmap(\lambda\ (n_1,n_2).~ (n_2,n_1))$

\[\begin{array}{@{}l@{\ }l@{\ \ }l@{}}  
  F_\sbt &:& \bintree(\nat*\nat)\ra\bintree(\nat*\nat) \\
  F_\sbt &=& \lambda x.~ \bcons\ ((1,1),\ \sbtinv(\sbtsuc(\sbtinv(x))),\ \sbtsuc(x))
\end{array}\]
\[\begin{array}{@{}r@{\ }c@{~}l@{\ \ }l@{}}  
  F_\sbt &:& \sem{\bintreeC(\nat*\nat)}\ra\sem{\bintreeC(\nat*\nat)}\\
  F_\sbt &=& \cons\ \fcompose\ ((\const(0,0)) \fpair (\sbtinv\fcompose\sbtsuc\fcompose\sbtinv) \fpair \sbtsuc)
\end{array}\]

\begin{center}
\begin{figure}
\begin{tikzpicture}[
    box/.style={draw, rectangle, rounded corners, minimum height=0.8cm, minimum width=1.0cm, align=center},
    arr/.style={->, >=Stealth, thick},
    lbl/.style={font=\small}, 
  ]
  \matrix (m) [matrix of nodes, row sep=20pt, column sep=0pt, nodes={anchor=center} ]
  {
  && |[box, name=t11]| $\bintreeC$ &\\
  |[box, name=t21]| $\constCTT{\nat}$ & ${\gop{\times}}$ & |[box, name=t22]| $ \bintreeC $ & ${\gop{\times}}$ & |[box, name=t23]| $ \bintreeC $ \\
  && |[box, name=t31]| $\bintreeC$ &\\
  };
  \draw [arr] (t11) -- (t21) node [midway, left, yshift=2pt] {$\const(0,0)$};
  \draw [arr] (t11) -- (t22);
  \draw [arr] (t11) -- (t23) node [midway, right, yshift=2pt] {$\sbtsuc$};
  \draw [arr] (t21) -- (t31);
  \draw [arr] (t22) -- (t31);
  \draw [arr] (t23) -- (t31);
  \node [lbl, above of=t31, xshift=15pt, yshift=-5pt] {$\bcons$};
  \coordinate (brace1_start) at ($(t11.north east) + (20pt, -10pt)$);
  \coordinate (brace1_end) at ($(t11.south east) + (-22pt, -10pt)$);
  \draw[thick] (brace1_start) to[bend left=20] (brace1_end);
  \node [lbl, above of=t11, xshift=95pt, yshift=-25pt] {$\sbtinv\fcompose\sbtsuc\fcompose\sbtinv$};
\end{tikzpicture}
\caption{Combination diagram of $F_\sbt$}\label{sbt diagram}
\end{figure}
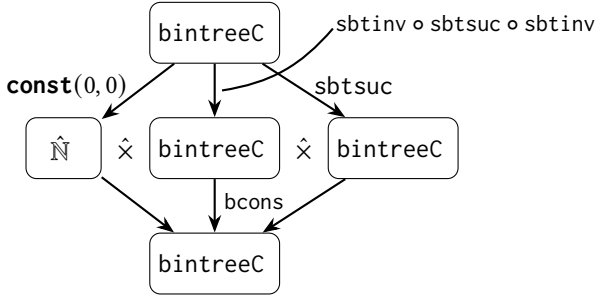
\end{center}

The productivity of $F_\sbt$ is automatically proven to be a subset of $\stdunifZ(1)$.
Therefore, we obtain the unique fixed point $\sbt$ of $F_\sbt$.

\subsection{Coinductive Type $\lang$} \label{example lang}
We define the coinductive type $\lang$ for formal languages following \cite{rutten98automata}.
This type was used to demonstrate AmiCo's capabilities in \cite{blanchette17}.
\[\text{For }T\in\Type,\ \langP(T)\in\STT\defeq \constSTT{\Bool} \cop{\times} (T \cop{\ra} \identSTT{}). \ \ \langC(T)\in\CTT\defeq {\identCTT{\vspe_{\langP(T)}}}. \]
\begin{remark}
  $\vspe_{\langP(T)} = \tconop{\times}\ \spfcompose\ (\lambda b.~ \ite{b}
    {\tconop{\Bool}}
    {\tconop{T\ra}\ \spfcompose\ (\lambda\_.~\tconop{\Id})})$. \\ 
  \indent\quad$\fspsem{\vspe}_{\langP(T)} = \lambda X.~ \Bool\times (T\ra X)$.
\end{remark}
\begin{remark}
  If $T\neq\Empty$, then $\sem{\langC(T)}\neq\Empty$ by \Cref{coin_prime_moverU}.
\end{remark}
Then $\lang(T) = \sem{\langC(T)}$.
We omit the type parameter $T$ when clear from context.

Next, from the default $\fold$ and $\unfold$ operations,
we derive a more user-friendly set of constructor and destructors.
\[\begin{array}{@{}r@{\ \ }c@{\ \ }l@{~}l@{}}  
  \llhead &\defeq& \fst \ \fcompose\ \unfold_\langP &: \sem{\langC}\ra \sem{\constCTT{\Bool}} \\
  \ltail &\defeq& \snd \ \fcompose\ \unfold_\langP &: \sem{\langC}\ra\sem{T\gop{\ra}\langC} \\
  \lcons &\defeq& \fold_\langP &: \sem{\constCTT{\Bool}\gop{\times}(T\gop{\ra}\langC)}\ra\sem{\langC} 
\end{array}\]

The productivities of the constructor and destructors are automatically proved.
\begin{lemma}
  $\productive{\stdunifZ(\minus1)}(\llhead)$, $\productive{\stdunifZ(\minus1)}(\ltail)$, $\productive{\stdunifZ(1)}(\lcons)$.
\end{lemma}

\subsection{Corecursive Function $\lplus$} \label{example lplus}
The function $\lplus$ is alternation, the union of two languages.
It served as an illustrative example in \cite{blanchette17}.
\[\begin{array}{@{}l@{\ }l@{\ \ }l@{}}  
  F_\lplus &:& ((\lang\ T\ \times\ \lang\ T)\ra \lang\ T) \ra ((\lang\ T\ \times\ \lang\ T)\ra \lang\ T) \\
  F_\lplus &=& \lambda f.~\lambda (x_1,x_2).~ \lcons\ ( \llhead(x_1)\ \bor\ \llhead(x_2),\ \lambda a.f(\ltail(x_1),\ltail(x_2)))
\end{array}\]
\[\begin{array}{@{}r@{\ }c@{~}l@{\ \ }l@{}}  
  F_\lplus & : & (\sem{\langP(T)\cop{\times}\langP(T)}\ra\sem{\langP(T)})\ra (\sem{\langP(T)\cop{\times}\langP(T)}\ra\sem{\langP(T)}) \\
  F_\lplus &=& (\sfindep\ \lcons)\ \sfcompose\\
  && (\sfindep(\cany{\lambda(b_1,b_2).b_1\bor b_2}{}{}\fcompose (\llhead\fproduct \llhead))\ \sfproduct\\
  &&\quad (\sfecurry(\sfself\sfcompose (\sfindep\ \ceapp)) \sfcompose
    \sfindep(\cepair\fcompose(\ltail\fproduct \ltail)))\\
  &&)\ \sfcompose\\
  &&(\sfindep(\id\fpair\id))
\end{array}\]

\begin{center}
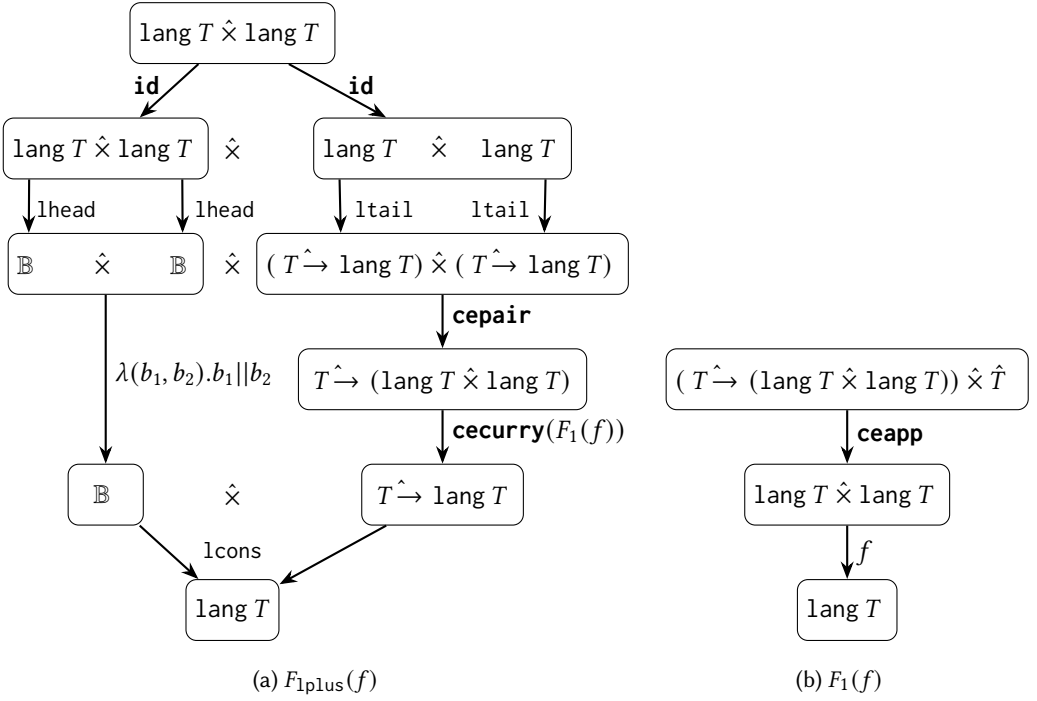
\begin{figure}
\begin{subfigure}{0.64\textwidth}
\begin{center}
  \begin{tikzpicture}[
      box/.style={draw, rectangle, rounded corners, minimum height=0.8cm, minimum width=1.0cm, align=center},
      arr/.style={->, >=Stealth, thick},
      lbl/.style={font=\small}, 
    ]
    \matrix (m) [matrix of nodes, row sep=20pt, column sep=-30pt, nodes={anchor=center} ]
    {
    & |[box, name=t11]| $\lang\ T \gop{\times} \lang\ T$ &\\
    |[box, name=t21]| $\lang\ T \gop{\times} \lang\ T$ & ${\gop{\times}}$ & |[box, name=t22]| $\lang\ T\quad\ \gop{\times}\quad \lang\ T$ \\
    |[box, name=t31]| $\Bool\qquad\ \gop{\times}\qquad \Bool$ & ${\gop{\times}}$ & |[box, name=t32]| $(\gop{T\ra}\lang\ T) \gop{\times} (\gop{T\ra}\lang\ T)$ \\
    && |[box, name=t42pre]| $\gop{T\ra}(\lang\ T \gop{\times} \lang\ T)$ \\
    |[box, name=t41]| $\Bool$ & ${\gop{\times}}$ & |[box, name=t42]| $\gop{T\ra}\lang\ T$ \\
    & |[box, name=t51]| $\lang\ T$\\
    };
    \draw [arr] (t11) -- (t21) node [midway, left, yshift=2pt] {$\id$};
    \draw [arr] (t11) -- (t22) node [midway, right, yshift=2pt] {$\id$};
    \coordinate (arr1_start) at ($(t21.south west) + (10pt, 0pt)$);
    \coordinate (arr1_end) at ($(t31.north west) + (8pt, 0pt)$);
    \draw[arr] (arr1_start) to (arr1_end);
    \node [lbl, above of=t31, xshift=-15pt, yshift=-8pt] {$\llhead$};
    \coordinate (arr2_start) at ($(t21.south east) + (-10pt, 0pt)$);
    \coordinate (arr2_end) at ($(t31.north east) + (-8pt, 0pt)$);
    \draw[arr] (arr2_start) to (arr2_end);
    \node [lbl, above of=t31, xshift=45pt, yshift=-8pt] {$\llhead$};
    \coordinate (arr3_start) at ($(t22.south west) + (10pt, 0pt)$);
    \coordinate (arr3_end) at ($(t32.north west) + (32pt, 0pt)$);
    \draw[arr] (arr3_start) to (arr3_end);
    \node [lbl, above of=t32, xshift=-22pt, yshift=-8pt] {$\ltail$};
    \coordinate (arr4_start) at ($(t22.south east) + (-10pt, 0pt)$);
    \coordinate (arr4_end) at ($(t32.north east) + (-32pt, 0pt)$);
    \draw[arr] (arr4_start) to (arr4_end);
    \node [lbl, above of=t32, xshift=22pt, yshift=-8pt] {$\ltail$};
    \draw [arr] (t31) -- (t41) node [midway, right, yshift=2pt] {$\lambda(b_1,b_2).b_1\bor b_2$};
    \draw [arr] (t32) -- (t42pre) node [midway, right, yshift=2pt] {$\cepair$};
    \draw [arr] (t42pre) -- (t42) node [midway, right, yshift=2pt] {$\cecurry(F_1(f))$};
    \draw [arr] (t41) -- (t51);
    \draw [arr] (t42) -- (t51);
    \node [lbl, above of=t51, yshift=-6pt] {$\lcons$};
\end{tikzpicture}
\caption{$F_\lplus(f)$}\label{lplus diagram 1}
\end{center}
\end{subfigure}
\begin{subfigure}{0.35\textwidth}
\begin{center}
  \begin{tikzpicture}[
      box/.style={draw, rectangle, rounded corners, minimum height=0.8cm, minimum width=1.2cm, align=center},
      arr/.style={->, >=Stealth, thick},
      lbl/.style={font=\small}, 
    ]
    \matrix (m) [matrix of nodes, row sep=20pt, column sep=-15pt, nodes={anchor=center} ]
    {
    |[box, name=t11]| $(\gop{T\ra}(\lang\ T \gop{\times} \lang\ T)) \gop{\times} \constCTT{T}$ \\
    |[box, name=t21]| $\lang\ T \gop{\times} \lang\ T$  \\
    |[box, name=t31]| $\lang\ T$ \\
    };
    \draw [arr] (t11) -- (t21) node [midway, right] {$\ceapp$};
    \draw [arr] (t21) -- (t31) node [midway, right] {$f$};
\end{tikzpicture}
\caption{$F_1(f)$}\label{lplus diagram 2}
\end{center}
\end{subfigure}
\caption{Combination diagrams of $F_\lplus(f)$}\label{lplus diagram}
\end{figure}
\end{center}

After specifying the expected productivity $\stdunifZ(0)$ for $\lplus$,
we can automatically prove that $\soproductive{(\stdunifZ(0),\stdunifZ(0),\stdunifZ(1))}(F_\lplus)$. 
Then we obtain the unique fixed point $\lplus$ of $F_\lplus$, which is also $\productive{\stdunifZ(0)}$. 

\subsection{Indexed Coinductive Type $\covector$} \label{example covector}
We define the indexed coinductive type $\covector(T):\hnat\ra\Type$,
where $\hnat \defeq \Nat \sqcup \{ \infty \}$ denotes the hyper-natural numbers.
$\covector(T)(h)$ is a type of vectors whose elements have type $T$ and whose length is $h$.
Note that we use standard addition and subtraction on $\hnat$.
\[\begin{array}{@{}r@{\ }c@{~}l@{\ \ }l@{}}  
\covectorP(T) &\in& \hnat\ra\STT(\hnat) \hspace{50pt} &\text{for } T\in\Type\\
\covectorP(T)(0) &\defeq& \constSTT{\unitset}&\\
\covectorP(T)(1+n) &\defeq& \constSTT{\unitset} \cop{\times} \identSTT{_n\ } \hspace{50pt} &\text{for } n\in\nat\\
\covectorP(T)(\infty) &\defeq& \constSTT{\unitset} \cop{\times} \identSTT{_\infty\ } 
\end{array}\]
\[\begin{array}{@{}r@{\ }l@{\ \ }l@{\ }}  
\covectorC(T)&\defeq \lambda h.~{\identCTT{\vspe_{\covectorP(T)},h}} & : \hnat\ra\CTT\\
\covector(T)&\defeq \lambda h.~\sem{\covectorC(h)} & : \hnat\ra\Type
\end{array}\]

\begin{remark}
  $\vspe_{\covectorP(T)}(0) = \tconop{\unitset}$, 
  $\vspe_{\covectorP(T)}(1+n) = \tconop{\times}\ \spfcompose\ (\lambda b.~ \ite{b}{\tconop{T}}{\tconop{\Id_{n:\hnat}}})$, \\ 
  \indent\quad$\vspe_{\covectorP(T)}(\infty) = \tconop{\times}\ \spfcompose\ (\lambda b.~ \ite{b}{\tconop{T}}{\tconop{\Id_{\infty:\hnat}}})$. \\
  \indent\quad$\fspsem{\vspe}_{\covectorP(T)}\in (\hnat\!\!\ra\Type)\ra(\hnat\!\!\ra\Type),\ 
  \fspsem{\vspe}_{\covectorP(T)} = \lambda X.~\lambda h.~ \ite{h\!=\!0}{\unitset}{T\times X(h-1)}$.
\end{remark}
\begin{remark}
  $\forall h,\ \sem{\covectorC(T)(h)}\neq\Empty$ by \Cref{coin_prime_mover} (with $X=\lambda h.~\unitset$).
\end{remark}
We omit the type parameter $T$ when clear from context.

Next, we define from $\fold$ and $\unfold$ more familiar constructors and destructors.
\[\begin{array}{@{}r@{\ }c@{\ }l@{~}c@{\ \ }l@{}}  
  \cvhead &:& \depfun{h\!:\!\hnat}\sem{\covectorC(1\!+\!h)}\ra \sem{\constCTT{T}} \\
  \cvhead &\defeq& \lambda h.~ \fst\ \fcompose\ \unfold_\covectorP(1\!+\!h) \\
  \cvtail &:& \depfun{h\!:\!\hnat}\sem{\covectorC(1\!+\!h)}\ra \sem{\covectorC(h)} \\
  \cvtail &\defeq& \lambda h.~ \snd\ \fcompose\ \unfold_\covectorP(1\!+\!h) \\
  \cvcons &:& \depfun{h\!:\!\hnat}\sem{\constCTT{T}\gop{\times}\covectorC(h)}\ra\sem{\covectorC(1\!+\!h)} \\
  \cvcons &\defeq& \lambda h.~ \fold_\covectorP(1\!+\!h)\\
  \cvnil &:& \sem{\covectorC(0)}\\
  \cvnil &\defeq& \fold_\covectorP(0)(\unit)
\end{array}\]
Note that we omit a destructor using $\unfold_\covectorP(0) : \sem{\covectorC(0)}\ra\sem{\constCTT{\unitset}}$, which is not useful in practice.

The productivity of these constructors and destructors can be easily proved by the combination principles.
\begin{lemma}
  $\forall h\in\hnat,\ \productive{\stdunifZ(\minus1)}(\cvhead(h))$, $\productive{\stdunifZ(\minus1)}(\cvtail(h))$, \\
  \indent\quad and $\productive{\stdunifZ(1)}(\cvcons(h))$.
\end{lemma}

\subsection{Indexed Corecursive Function $\append$} \label{example append}
The function $\append: \depfun{(h_1,h_2)\!:\!\hnat\!\!\times\!\hnat} \sem{(\covectorC(T)(h_1))\gop{\times}(\covectorC(T)(h_2))}\ra$
$\sem{\covectorC(T)(h_1+h_2)}$ concatenates two input covectors into a single covector. 
If the first input has infinite length, then the second input is ignored. 
Note that the index type of $\append$ differs from that of $\covector$. 
\[\begin{array}{@{}r@{\ }c@{~}l@{\ \ }l@{}}  
  F_\append &:& 
  (\ \depfun{(h_1,h_2)\!:\!\hnat\!\!\times\!\hnat} (\covector(T)(h_1)\times\covector(T)(h_2))\hspace{20pt}\\ 
  &&\hfill\ra(\covector(T)(h_1+h_2))\ ) \ra \\
  &&(\ \depfun{(h_1,h_2)\!:\!\hnat\!\!\times\!\hnat} (\covector(T)(h_1)\times\covector(T)(h_2))\\ 
  &&\hfill\ra(\covector(T)(h_1+h_2))\ ) \hspace{12pt}\\
  F_\append &\defeq& \lambda f.~\lambda (h_1,h_2).~\lambda (x_1,x_2). \\
  &&x_2 \hfill \text{for } h_1=0\\
  &&x_1 \hfill \text{for } h_1=\infty\\
  &&\cvcons(h_1'\!+\!h_2)(\cvhead(h_1')(x_1),
  \hfill \text{for } h_1=h_1'\!+\!1\in\nat\\
  &&\hspace{68pt} f(h_1',h_2)(\cvtail(h_1')(x_1),\ x_2)\ ) \hspace{130pt}
\end{array}\]
\[\begin{array}{@{}r@{\ }c@{~}l@{\ \ }l@{}}  
  F_\append &:&  
  (\ \depfun{(h_1,h_2)\!:\!\hnat\!\!\times\!\hnat} \sem{(\covectorC(h_1))\gop{\times}(\covectorC(h_2))}\hspace{20pt}\\ 
  &&\hfill\ra\sem{\covectorC(h_1+h_2)}\ ) \ra \\
  &&(\ \depfun{(h_1,h_2)\!:\!\hnat\!\!\times\!\hnat} \sem{(\covectorC(h_1))\gop{\times}(\covectorC(h_2))}\\ 
  &&\hfill\ra\sem{\covectorC(h_1+h_2)}\ ) \hspace{12pt}\\
  F_\append &=& \lambda f.~\lambda (h_1,h_2).~F_\mathtt{append,pi}(h_1,h_2)(f) \text{ where }\\
  F_\mathtt{append,pi}(h_1,h_2)&\defeq& \spifindep\ \snd \hfill \text{for } h_1=0\\
  &&\spifindep\ \fst \hfill \text{for } h_1=\infty\\
  &&\spifindep(\cvcons(h_1'\!+\!h_2)) \spifcompose (
  \hfill \text{for } h_1=h_1'\!+\!1\in\nat\\
  &&\hspace{10pt} \spifindep(\cvhead(h_1')\fcompose \fst)\ \spifpair \\
  &&\hspace{10pt} (\spifself(h_1',h_2) \spifcompose \spifindep(
    (\cvtail(h_1') \fproduct \id))) \hspace{10pt} \\
  &&)
\end{array}\] 

\begin{center}
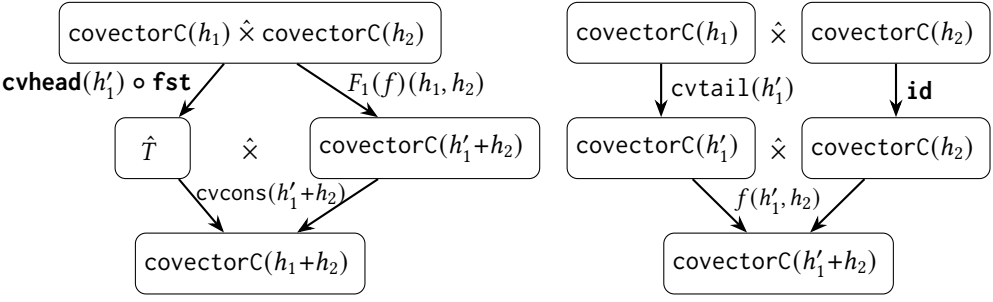
\begin{figure}
\begin{subfigure}{0.5\textwidth}
\begin{center}
  \begin{tikzpicture}[
      box/.style={draw, rectangle, rounded corners, minimum height=0.8cm, minimum width=1.0cm, align=center},
      arr/.style={->, >=Stealth, thick},
      lbl/.style={font=\small}, 
    ]
    \matrix (m) [matrix of nodes, row sep=20pt, column sep=-50pt, nodes={anchor=center} ]
    {
    & |[box, name=t11]| $\covectorC(h_1) \gop{\times} \covectorC(h_2)$ &\\
    |[box, name=t21]| $\constCTT{T}$ & ${\gop{\times}}$ & |[box, name=t22]| $ \covectorC(h_1'\!+\!h_2) $ \\
    & |[box, name=t31]| $\covectorC(h_1\!+\!h_2)$ &\\
    };
    \draw [arr] (t11) -- (t21) node [midway, left, yshift=2pt] {$\cvhead(h_1')\fcompose\fst$};
    \draw [arr] (t11) -- (t22) node [midway, right, yshift=2pt] {$F_1(f)(h_1,h_2)$};
    \draw [arr] (t21) -- (t31);
    \draw [arr] (t22) -- (t31);
    \node [lbl, above of=t31, yshift=-3pt, xshift=8pt] {$\cvcons(h_1'\!+\!h_2)$};
\end{tikzpicture}
\caption{$F_\append(f)(h_1,h_2)$, where $h_1=h_1'+1$}\label{append diagram 1}
\end{center}
\end{subfigure}
\begin{subfigure}{0.49\textwidth}
\begin{center}
  \begin{tikzpicture}[
      box/.style={draw, rectangle, rounded corners, minimum height=0.8cm, minimum width=1.2cm, align=center},
      arr/.style={->, >=Stealth, thick},
      lbl/.style={font=\small},
    ]
    \matrix (m) [matrix of nodes, row sep=20pt, column sep=-35pt, nodes={anchor=center} ]
    {
    |[box, name=t11]| $\covectorC(h_1)$ & ${\gop{\times}}$ & |[box, name=t12]| $\covectorC(h_2)$ \\
    |[box, name=t21]| $\covectorC(h_1')$ & ${\gop{\times}}$ & |[box, name=t22]| $\covectorC(h_2)$ \\
    & |[box, name=t31]| $\covectorC(h_1'\!+\!h_2)$ &\\
    };
    \draw [arr] (t11) -- (t21) node [midway, right] {$\cvtail(h_1')$};
    \draw [arr] (t12) -- (t22) node [midway, right] {$\id$};
    \draw [arr] (t21) -- (t31);
    \draw [arr] (t22) -- (t31);
    \node [lbl, above of=t31, yshift=-5pt] {$f(h_1',h_2)$};
\end{tikzpicture}
\caption{$F_1(f)(h_1,h_2)$}\label{append diagram 2}
\end{center}
\end{subfigure}
\caption{Combination diagrams of $F_\append(f)(h_1,h_2)$ and $F_1(f)(h_1,h_2)$}\label{append diagram}
\end{figure}
\end{center}

After specifying the expected productivities $\lambda\_.\stdunifZ(0)$ for $\append$,
we can automatically prove that
$\forall h_1,h_2\in\hnat,\ 
\soproductivepJ{(\lambda\_.\stdunifZ(0),\stdunifZ(0),\lambda\_.\stdunifZ(1))} (F_\mathtt{append,pi}(h_1,h_2))$.
Then we obtain the unique fixed point $\append$ of $F_\append$,
with $\forall h_1,h_2\in\hnat,\ \productive{\stdunifZ(0)}(\append(h_1,h_2))$.
\section{Well-Founded-S Relation as a Productivity}
\label{Well-Founded-S Relation as a Productivity}

We define the \emph{semi-accessibility relation} $\sacc$ mentioned in \Cref{Fixed Point}.
This is a variant of the accessibility relation $\acc$, adapted to the context of levels.
Before formalizing $\sacc$, we define the hyper-level $\hlevel$,
which is the hyper-natural variant of $\level$,
and then define two new orderings $\hord$ and $\hpord$ on the set of hyper-levels.

The definition of $\hlevel$ is similar to that of $\level$,
except it uses $\hnat$ in the base case.
\[\begin{array}{@{}c@{\ }c@{}}
    \hlevel : \CTT\ra\Type \qquad \\ 
    \hlevel(\identCTT{\vspe}) \defeq \hnat \qquad
  \hlevel (\cttf\cttcompose\vec{\vctt}) \defeq \option(\depfun{a:\ctta}{\hlevel(\vec{\vctt}(a))})
  \end{array}\]
Every level can also be considered a hyper-level via the natural embedding $\iota : \level(\vctt) \ra \hlevel(\vctt)$,
whose definition we omit.

The preorders $\hord$ and $\hpord$ are defined inductively in \Cref{Definition of hord in hyper-level set} and \Cref{Definition of hpord in hyper-level set},
respectively, using four and three rules.
The relation $\hord$ corresponds directly to $\lord$, with $\hlevel$ replacing $\level$.
The only difference between $\hord$ and $\hpord$ is that $\hpord$ omits the fourth rule of $\hord$.
Hence, if $l_1 \hpord l_2$, then $l_1 \hord l_2$.

\begin{figure}[t]
  \[\frac{l_1,l_2\in\hlevel(\identCTT{\vspe}) \qquad l_1\le l_2 \text{ in } \hnat}{l_1\hord l_2 \text{ in } \identCTT{\vspe}} \cdots(1)\]
  \[\frac{l\in\hlevel (\cttf\cttcompose\vec{\vctt})}{\none\hord l \text{ in } \cttf\cttcompose\vec{\vctt}} \cdots(2)\qquad 
    \frac{l_1',l_2'\in \depfun{a:\ctta}{\hlevel(\vec{\vspe}(a))} \qquad \forall a:\ctta,\ l_1'(a) \hord l_2'(a)} 
      {\some(l_1')\hord\some(l_2') \text{ in } \cttf\cttcompose\vec{\vctt}}\cdots(3)\]
  \[\frac{l'\in \depfun{a:\ctta}{\hlevel(\vec{\vspe}(a))} \qquad 
      \forall a:\ctta,\ l'(a) \lord \bot \qquad 
      |\cttf.\vctn.\fctnS| = 1}
      {\some(l')\hord\none \text{ in } \cttf\cttcompose\vec{\vctt}}\cdots(4)\]
  \caption{Inductive definition of $\hord$ on hyper-level sets via four rules}
  \label{Definition of hord in hyper-level set}
\end{figure}

\begin{figure}[t]
  \[\frac{l_1,l_2\in\hlevel(\identCTT{\vspe}) \qquad l_1\le l_2 \text{ in } \hnat}{l_1\hpord l_2 \text{ in } \identCTT{\vspe}} \cdots(1)\]
  \[\frac{l\in\hlevel (\cttf\cttcompose\vec{\vctt})}{\none\hpord l \text{ in } \cttf\cttcompose\vec{\vctt}} \cdots(2)\qquad 
    \frac{l_1',l_2'\in \depfun{a:\ctta}{\hlevel(\vec{\vspe}(a))} \qquad \forall a:\ctta,\ l_1'(a) \hpord l_2'(a)} 
      {\some(l_1')\hpord\some(l_2') \text{ in } \cttf\cttcompose\vec{\vctt}}\cdots(3)\]
  \caption{Inductive definition of $\hpord$ on hyper-level sets via three rules}
  \label{Definition of hpord in hyper-level set}
\end{figure}
  
The semi-accessibility relation $\sacc$ is defined inductively using four rules, as shown in \Cref{Definition of sacc}.
Intuitively, $\sacc(p)(l)$ means that the level $l$ is accessible in the context of generating a fixed point,
that is, we can extract information up to level $l$ from the productivity $p$.

\begin{figure}[t]
  \[\frac{}{\sacc(p)(\bot)} \cdots(1) \qquad
    \frac{\sacc(p)(l') \qquad l \le l'}{\sacc(p)(l)} \cdots(2)
  \]
  \[\frac{\forall l', p(l', l) \implies \sacc(p)(l')} 
      {\sacc(p)(l)}\cdots(3)\]
  \[\frac{q : \level(\vctt) \ra \Prop \quad
      \forall l', q(l') \!\!\implies\!\! \sacc(p)(l') \quad \forall l_{bd}, (\forall l', q(l') \!\!\implies\!\! \iota(l') \hpord l_{bd}) \!\!\implies\!\! \iota(l) \hord l_{bd}}
      {\hfill\sacc(p)(l)\hfill\cdots(4)}\]
  \caption{Inductive definition of $\sacc$ via four rules}
  \label{Definition of sacc}
\end{figure}

We define \emph{semi-well-foundedness} in terms of $\sacc$ as follows.
\[\text{For } p\in\productivity{\vctt}{\vctt},\ \wfs(p) \defeq (\forall l\in\level (\vctt),\ \sacc(p)(l))\]

We now present the fixed point theorems involving semi-accessibility and semi-well-foundedness.

\begin{lemma}\label[lemma]{wf_fp_pre} \quad\\
  \indent\quad For $\vctt\in\CTT$ such that $\sem{\vctt}\neq\Empty$ and $p\in \productivity{\vctt}{\vctt}$, $f : \sem{\vctt} \ra \sem{\vctt}$, \\
  \indent\quad if $\ \productive{p}(f)$, then $\exists x,\ \forall l,\ \sacc(p)(l) \implies f(x)\equpto{l}x$. 
\end{lemma}

\begin{theorem}[Fixed Point of Well-Founded Productive Function]\label{wf_fp} \quad\\
  \indent\quad For $\vctt\in\CTT$ such that $\sem{\vctt}\neq\Empty$ and $p\in \productivity{\vctt}{\vctt}$, $f : \sem{\vctt} \ra \sem{\vctt}$, \\
  \indent\quad if $\ \productive{p}(f)$ and $\wfs(p)$, then $f$ has a unique fixed point. 
\end{theorem}

\begin{corollary}[Fixed Point of (1/n)-Productive Function] \quad\\
  \indent\quad For $n\in\nat$, $\vctt\in\CTT$ such that $\sem{\vctt}\neq\Empty$ and $p\in \productivity{\vctt}{\vctt}$, $f : \sem{\vctt} \ra \sem{\vctt}$, \\
  \indent\quad if $\ \productive{p}(f)$ and \ $p^n\!\subseteq \stdunifZ(1)$, then $f$ has a unique fixed point. 
\end{corollary}

\end{document}